\pdfoutput=1

\documentclass{aastex631}
\usepackage{amsmath}
\usepackage{appendix}
\usepackage{longtable}
\usepackage{url}
\shorttitle{SLF Variability of Galactic O-type Stars}
\shortauthors{D.-X. Shen et al.}
\graphicspath{{./}{figures/}}

\begin{document}

\title{A Study of Stochastic Low-Frequency Variability for Galactic O-type Stars}

\author[0000-0002-1926-8248]{Dong-Xiang, Shen}
\affiliation{School of Physical Science and Technology, Xinjiang University, Urumqi 830046, People's Republic of China; {\it chunhuazhu@sina.cn}}

\author{Chun-Hua, Zhu\footnotemark[1]}
\affiliation{School of Physical Science and Technology, Xinjiang University, Urumqi 830046, People's Republic of China; {\it chunhuazhu@sina.cn}}

\author{Guo-Liang, L$\ddot{\rm u}$\footnotemark[1]}
\affiliation{Xinjiang Astronomical Observatory, National Astronomical Observatories, Chinese Academy of Sciences, Urumqi 830000, China;{\it guolianglv@xao.ac.cn} }

\author{Xi-zhen, Lu}
\affiliation{School of Physical Science and Technology, Xinjiang University, Urumqi 830046, People's Republic of China; {\it chunhuazhu@sina.cn}}

\author{Xiao-long, He}
\affiliation{Xinjiang Astronomical Observatory, National Astronomical Observatories, Chinese Academy of Sciences, Urumqi 830000, China;{\it guolianglv@xao.ac.cn} }

\renewcommand{\thefootnote}{\fnsymbol{footnote}} 
\footnotetext[1]{Corresponding authors: chunhuazhu@sina.cn, guolianglv@xao.ac.cn} 



\begin{abstract}
In order to explore how the ubiquitous stochastic low-frequency (SLF) variability of O-type stars is related to various stellar characteristics, we compiled a sample of 150 O-type stars observed via ground-based spectroscopic surveys, alongside photometric data obtained from the Transiting Exoplanet Survey Satellite (TESS). We analyzed 298 light curves obtained from TESS sectors 1-65 for the stars in our sample. Leveraging the spectroscopic parameters, we used Bonnsai to determine masses, radii, fractional main sequence ages, and mass-loss rates for stars of our sample. Subsequently, we identified possible correlations between the fitted parameters of SLF variability and stellar properties. Our analysis unveiled four significant correlations between the amplitude and stellar parameters, including mass, radius, fractional main sequence ages, and mass-loss rate. For stars with $M \gtrsim 30 M_{\odot}$, we observed a decrease in characteristic frequency and steepness with increasing radius. Finally, we compared various physical processes that may account for the SLF variability with our results. The observed SLF variability may arise from the combined effects of FeCZ and IGWs, with IGWs potentially more dominant in the early stages of stellar evolution, and the contribution of FeCZ becoming more significant as stars evolve. Meanwhile, our results indicate that the SLF variability of O-type stars bears certain signatures of the line-driven wind instability and granulation.

\end{abstract}

\keywords{stars: early-type --- stars: variables: general --- stars: statistics}


\section{Introduction}
Massive stars, with their substantial masses and immense luminosities, play a pivotal role in the Universe. 
Their presence significantly influences an array of astrophysical phenomena, contributing notably to the enrichment of stellar populations 
\citep[e.g.,][]{2005ApJ...626..864M,2017MNRAS.471.2151H}, the galactic chemical evolution \citep[e.g.,][]{2013ARA&A..51..457N, 2015MNRAS.449.1057G}, 
and the energetic dynamics of the cosmos \citep[e.g.,][]{2008IAUS..250....3M, 2012ARA&A..50..107L}.
Beyond their active role in the Universe, massive stars exhibit their radiance across vast spatial extents, owing to their remarkable luminosities.
The majority of massive stars undergo a cataclysmic demise through a supernova \citep[e.g.,][]{2012ARA&A..50..107L, 2019NatAs...3..717M}, 
leaving behind exotic compact remnants that serve as ideal laboratories for testing General Relativity and exploring the 
Universe through gravitational waves \citep[e.g.,][]{2016PhRvL.116f1102A, 2017cgrc.book..291A, 2020MNRAS.496..182L}.
Consequently, massive stars serve as invaluable instruments for scrutinizing the ambient conditions in their distant abodes and the characteristics of the 
intervening interstellar medium \citep[e.g.,][]{2008RMxAC..33..169P, 2010MNRAS.408..731C, 2010AN....331..459K, 2010MNRAS.408L..31D}. 
Therefore, it is of utmost significance to verify and refine models pertaining to the structure and evolution of massive stars.

Intrinsic variability permeates the entire life cycle from the main sequence to end state of massive stars. Two commonly types of coherent pulsation modes have been identified in massive stars: gravity modes ($g$-modes), where the restoring force is buoyancy, and pressure modes ($p$-modes), where the restoring force is pressure \citep{2010aste.book.....A}. Through examining these coherent pulsation frequencies or period spacing patterns enables us to directly infer the internal rotation and chemical stratification in stars that have a convective core \citep[e.g.,][]{2017A&A...598A..74P,2015ApJ...810...16T,2012MNRAS.424.2380H,2017MNRAS.464.2249H,2022MNRAS.511.1529S, 2023arXiv231100038B}. Furthermore, forward seismic modeling of these frequencies provides constraints on the masses and ages of the stars \citep[e.g.,][]{2018ApJS..237...15A, 2019A&A...624A..75A, 2019ApJ...873L...4H,2023NatAs...7..913B}. On the other hand, massive stars typically exhibit strong stellar winds, which are derived by radiation pressure \citep{2000ARA&A..38..613K}, able to modify the ionizing radiation of massive stars dramatically, resulting in high mass-loss rates, and cause the line profile variability and photospheric brightness variations \citep[e.g.,][]{2011AJ....141...17R, 2015A&A...581A..75K}. 
Prior to the era of space-based photometric missions, ground-based telescopes provided fragmented photometric time series due to inadequate duty cycles, complicating mode identification \citep[e.g.,][]{2005MNRAS.362..612H, 2009MNRAS.396.1460D, 2013MNRAS.431.3396D}. In recent decades, space-based photometric missions such as COnvection internal ROtation and Transiting planets \citep[CoRoT;][]{2009IAUS..253...71B}, Kepler \citep{2010Sci...327..977B} and Transiting Exoplanet Survey Satellite \citep[TESS;][]{2015JATIS...1a4003R}, provided high-duty cycles, long-term, near-continuous, and high-precision data sets. The asteroseismic potential of space-based photometric missions is evidenced by many studies \citep[e.g.,][]{2018A&A...619A..97D, 2023A&A...676A..81Z, 2023A&A...676A..28T, 2023A&A...674A.204B, 2023A&A...678A..24R}. In the imminent horizon, the PLAnetary Transits and Oscillations of stars (PLATO) mission is poised to substantially augment the existing dataset \citep{2024arXiv240605447R}.

Beyond coherent pulsations and binarity, stochastic low-frequency (SLF) variability has been discerned through high-precision photometric data acquired by space-based telescopes \citep[e.g.,][]{2011A&A...533A...4B,2015MNRAS.453...89B,2019A&A...621A.135B,2020A&A...640A..36B,2023ApJ...955..123S}. SLF variability is observed across a wide range in mass, age, and metallicity in massive stars. \cite{2019A&A...621A.135B} developed the MCMC numerical scheme within a Bayesian framework for the detection of SLF variability and applied it to a sample composed of O, B, A, and F stars. Gaussian process regression has also been used to characterise SLF variability in massive stars \citep{2022A&A...668A.134B}. \cite{2019NatAs...3..760B} utilized a sample of 167 OB-type stars within the Galaxy and the Large Magellanic Cloud (LMC) to demonstrate that SLF variability appears to be insensitive to the metallicity of the star. By analyzing the photometric data of TESS Sectors 1-13 for a sample composed of 98 OB stars, \cite{2020A&A...639A..81B} also reported that SLF variability is ubiquitous in upper-main sequence stars. Additionally, \cite{2020A&A...640A..36B} integrated high-precision photometry and high-resolution spectroscopy, revealing statistically correlated properties of SLF variability with stellar properties. Their investigation indicated that more luminous stars exhibit larger amplitudes but smaller characteristic frequencies in their stochastic photometric variability for stars with $M > 20 M_{\odot}$. They also concluded that the relationship between SLF variability and macroturbulence broadening provides strong evidence for internal gravity waves (IGWs) in massive stars. Using a sample composed of 118 magnetic hot stars, \cite{2023ApJ...955..123S} explored the relationships between SLF variability and stellar parameters. They found a significant negative correlation between the surface dipole magnetic fields and the characteristic frequency of SLF variability. Through analyzing the TESS photometric data with a sample consisting of 26 Wolf-Rayet stars and eight luminous blue variables (LBVs), \cite{2021MNRAS.502.5038N} reported that SLF variability is ubiquitous in evolved massive stars. They suggested that SLF variability in evolved massive stars is most probably linked to pulsational activity. Based on a sample of 50 Galactic Wolf-Rayet stars, \cite{2022ApJ...925...79L} investigated the possible correlations between the fitting parameters of SLF variability and stellar wind characteristics.

While SLF variability is prevalent in massive stars, its origin continues to be a subject of debate. It could arise from subsurface convection zones \citep[e.g.,][]{2009A&A...499..279C, 2019ApJ...886L..15L, 2021ApJ...915..112C, 2023NatAs...7.1228A} or from the excitation of IGWs \citep[e.g.,][]{2019ApJ...876....4E, 2019MNRAS.482.5500R, 2020MNRAS.497.4231R, 2023A&A...674A.134R}, while instabilities in the stellar wind could also contribute \citep{2021A&A...648A..79K}. \cite{2021ApJ...915..112C} suggested that the SLF variability may be attributed to subsurface convection zones, with their models reproducing both the typical values of the observed SLF variability and the variation trends of SLF variability in the spectroscopic Hertzsprung-Russell diagram (sHRD). Three-dimensional radiation hydrodynamical simulations of SLF variability in massive star envelopes conducted by \cite{2022ApJ...924L..11S, 2023ApJ...951L..42S} are directly supported by spectroscopic and photometric data, although further models are necessary for comparison with various observational results. To understand how the IGWs excite photometric variability and compare it with observational SLF variability, it is important to know what spectrum of waves in frequency and wavenumber space is excited by convection. Theoretical work for characterizing these spectra mostly focuses on two mechanisms of excitation: through the Reynolds stresses of convective eddies or through the penetration of plumes. The former excitation mechanism of IGWs shows a power-law dependence in frequency, i.e., proportional to $f^{-\alpha}$, with wave frequency $f$ and exponent $\alpha$ \citep{1952RSPSA.211..564L, 1990ApJ...363..694G, 1999ApJ...520..859K, 2013MNRAS.430.2363L}. In a more recent publication, the mechanism of penetration of plumes is investigated in massive stars \citep[e.g.,][]{2019ApJ...876....4E, 2020A&A...641A..18H, 2023ApJ...954..171V}. Multidimensional hydrodynamical simulations have demonstrated that IGWs can be generated at the interface between the convective core and the radiative envelope in massive stars \citep[e.g.,][]{2015ApJ...815L..30R, 2020MNRAS.497.4231R, 2023MNRAS.522.2835L, 2023MNRAS.525.1601H,2023MNRAS.522.1706B, 2024MNRAS.531.1316T}. These IGWs are expected to produce low-frequency variability and a significant tangential velocity field near the stellar surface \citep[e.g.,][]{2013ApJ...772...21R, 2014ApJ...795...92C, 2019ApJ...876....4E, 2020A&A...641A..18H}. Using 10\,$M_{\odot}$ star model with solar-metallicity, \cite{2019ApJ...886L..15L} studied the observational signatures of IGWs generated by core convection and developed a wave transfer function that connected the amplitude at the base of the radiative zone to variations in surface luminosity. However, this function was found to be inconsistent with the observed SLF spectra, as it predicted regularly spaced peaks around 1 day$^{-1}$ that were not present. Utilizing three-dimensional simulation code, \cite{2023NatAs...7.1228A} conducted simulations of a 15\,$M_{\odot}$ stellar model with LMC metallicity at the ZAMS. Their results suggested that the photometric variability generated by IGWs exhibits lower amplitude and characteristic frequency compared to the observed SLF variability in OB-type stars. This discrepancy implies that an alternative process may be responsible for generating the observed red noise. Additionally, the effects of stellar winds on SLF variability warrant further discussion \citep{2021A&A...648A..79K, 2022ApJ...925...79L}.

In this paper, we detected SLF variability in 150 Galactic O-type stars using high-precision TESS photometric data. Utilizing the spectroscopic parameters determined from high-resolution spectroscopy, we employed Bayesian analysis to infer the stellar parameters for the stars in our sample. Furthermore, we examined the correlations between SLF variability and the stellar properties of these massive stars. We detailed our sample selection criteria in Section 2, and the methodology was outlined in Section 3. 
We presented our results in Section 4. In Section 5, we compared our statistical results with theoretical predictions derived from the various potential excitation mechanisms of SLF variability. Conclusions drawn from our study were presented in Section 6.

\section{Sample Selection}
Recent advancements in high-resolution and high-cadence spectroscopic instruments have predominantly concentrated on delineating pivotal properties of massive stars, encompassing spectral type, luminosity, effective temperature, rotational velocity, and surface gravity \citep[e.g.,][]{2014A&A...570L...6S, 2016ApJS..224....4M, 2017A&A...597A..22S, 2018A&A...613A..65H}. 
The Galactic O-star spectroscopic survey (GOSSS) is an ambitious long-term project whose immediate objective is obtaining homogeneous, high 
signal-to-noise ratio ($SNR \geqslant 300$), intermediate resolution ($R\thickapprox\,2500$), blue-violet ($3900-5100\;\overset{\circ}{\rm A}$) 
spectroscopic database for $\sim$2500 visually observable ($B < 14\;$mag) Galactic O-type stars. Concurrently, other spectrometric surveys, such as the Southern Galactic O- and WN-type Stars (OWN Survey) and the IACOB spectroscopic database, complement the GOSSS project by providing superior-quality, high-resolution spectra of both northern and southern Galactic OB stars \citep{2010RMxAC..38...30B, 2011BSRSL..80..514S}. Furthermore, Northern Massive Dim Stars survey (NoMaDS) is an extension project that complements OWN and IACOB by observing northern massive stars down to $B = 14\;$mag using the High Resolution Spectrograph at the Hobby-Eberly Telescope (HET) \citep{2012AAS...21922403P, 2012ASPC..465..484M}. 

To investigate the SLF variability of the Galactic O-type stars in detail, we utilized existing compilations of the Galactic O-type stars provided by spectroscopic surveys from both the northern and southern hemispheres, such as GOSSS and IACOB. Our original sample consisted of 590 Galactic O-type stars from the catalogs compiled by \cite{2014ApJS..211...10S} and \cite{2020A&A...638A.157H}. As demonstrated by \cite{2019A&A...621A.135B, 2020A&A...640A..36B}, it is necessary to exclude eclipsing binary (EB) systems in order to conduct SLF variability analysis, as binary interaction dominates the evolution of massive stars in these systems \citep{2012Sci...337..444S, 2014ApJS..215...15S}. Additionally, the substantial light contributions from the secondary component in EB systems can cause modulation of the photometric variability detected in the primary star during the binary phase, thereby limiting the scientific inference of SLF variability. Therefore, we excluded binaries from our sample in SLF variability analysis. 
\cite{2014ApJS..211...10S} revised the GOSSS sample to include new results from the literature and re-observed spectroscopic binaries (SBs) near quadrature or visual binaries (VBs) with close companions. Additionally, they used the results of the OWN survey and a series of papers \citep[e.g.,][]{2003IBVS.5480....1O, 2006OEJV...45....1O, 2007OEJV...72....1O, 2004IBVS.5495....1O, 2005IBVS.5644....1O} to include information on binaries. In order to reduce the inclusion of binary stars from the catalog provided by \cite{2014ApJS..211...10S}, we excluded all stars classified as binary stars (those classified as SB1, SB1E, SB2, SB2E, and SBE, etc.) and their candidates (those classified as SB1?, SB2?, and SB3?, etc.) labeled by \cite{2014ApJS..211...10S}. 
The catalog of stars provided by \cite{2020A&A...638A.157H} has been meticulously analyzed for classification in terms of spectroscopic variability, which includes factors such as binarity, pulsations, wind variability, and other sources of stellar variability. This classification was determined using multi-epoch high-resolution spectra. We also excluded the stars classified as binary stars and their candidates published by \cite{2020A&A...638A.157H}. After implementing these exclusion criteria, our sample contains 344 O-type stars.

Beyond the influence of binary stars, we further considered the potential impact of blended and nearby stars on the detection of SLF variability. Following the contamination assessment as described in Section \ref{sec3.1}, our finalized sample includes 150 Galactic O-type stars. Among the stars in our sample, 130 O-type stars are observed by the high-resolution spectroscopic survey IACOB. The remaining 20 O-type stars are only observed by the low-resolution spectroscopic survey GOSSS. Thirty-three stars from our sample have previously been investigated for SLF variability \citep[e.g.,][]{2019NatAs...3..760B, 2020A&A...640A..36B, 2020A&A...639A..81B}. Our current dataset includes observations from TESS Sectors 1 to 65, providing the most recent observational data available for these stars. This extensive dataset is expected to yield additional insights into the SLF variability of these stars. The remaining 117 stars in our sample have yet to be explored for SLF variability. Specific details for each star in our sample, such as the name, spectral type, luminosity class, effective temperature, surface gravity, macroturbulent velocity, and rotational velocity, are provided in Table \ref{table:stellar_parameters}.

\section{Method}
\subsection{Contamination assessment}\label{sec3.1}
The TESS space telescope conducted observations of stars within a 24 by 96$^{\circ}$ field of view. Each sector, defined as a specific portion of the sky, underwent scrutiny for a period of 27.3 days before the reorientation of TESS to a new position. However, due to TESS large plate scale of $21^{\prime \prime}$ ${\rm pixel}^{-1}$ \citep{2015JATIS...1a4003R}, 
astronomical object analysis is commonly affected by potential contamination from blended or nearby stars, particularly in crowded stellar 
fields \citep[e.g.,][]{2018AJ....156...83Z, 2023AJ....165..239P}. 
Various approaches have been utilized to limit the potential presence of contamination within a specific aperture mask \citep[e.g.,][]{2011ApJ...727...24T, 2021AJ....161...24G,2023AJ....165..141H}, while the influence of potential contamination on SLF variability remains to be determined. 
Therefore, prior to conducting SLF variability analysis on the sample, the meticulous selection of an appropriate aperture mask is of paramount importance to mitigate contamination flux.
\begin{figure*}
    \gridline{\fig{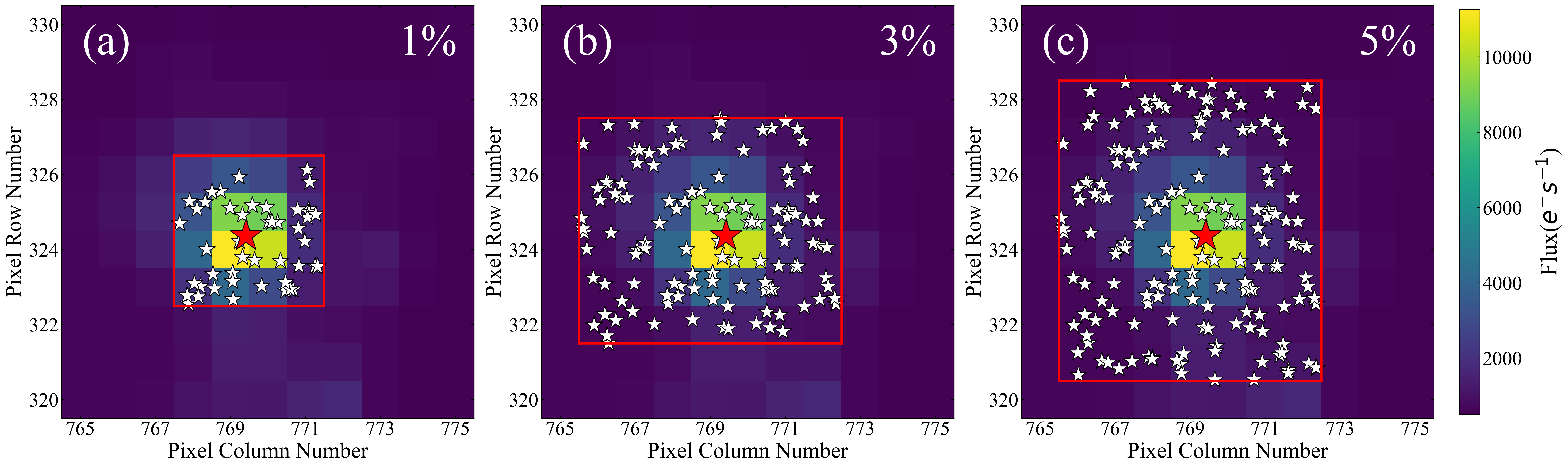}{1\textwidth}{}}
    \gridline{\fig{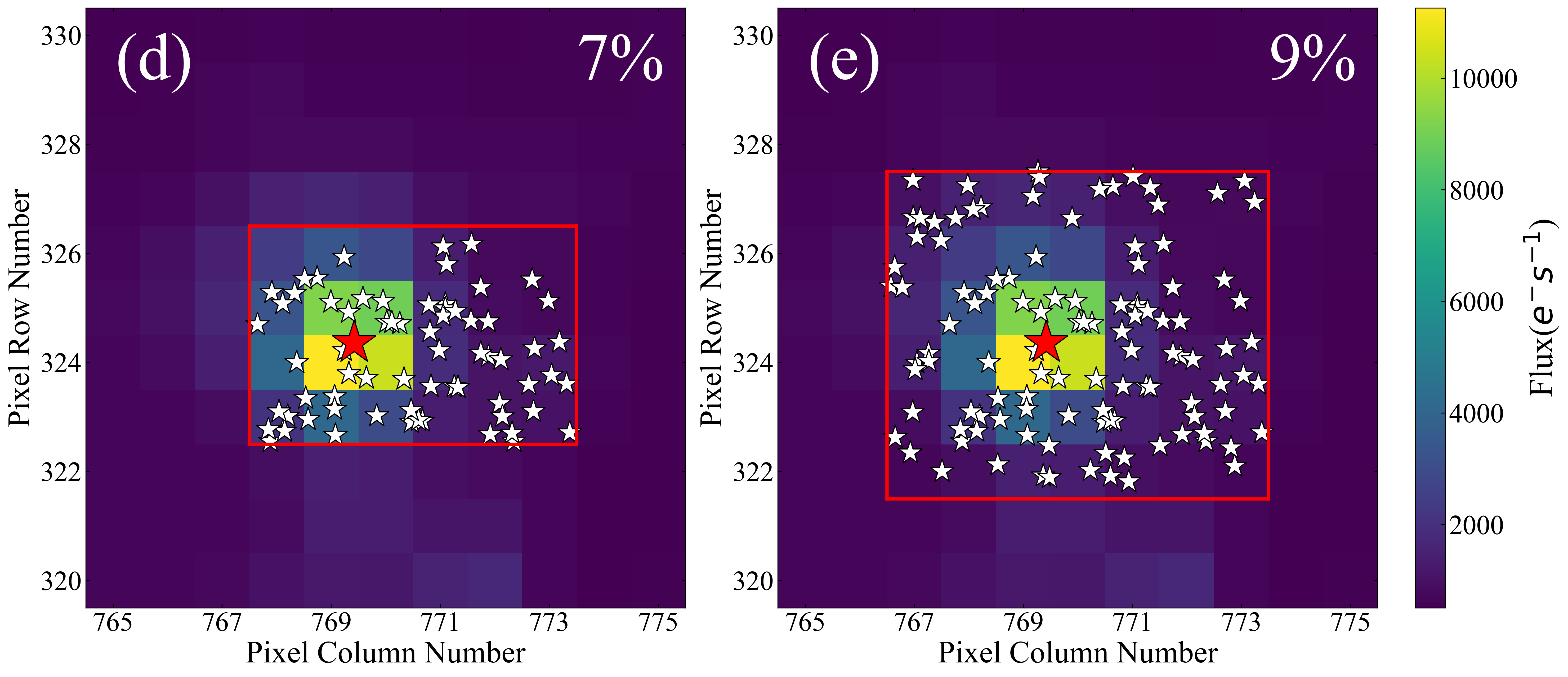}{0.7\textwidth}{}}
    \gridline{\fig{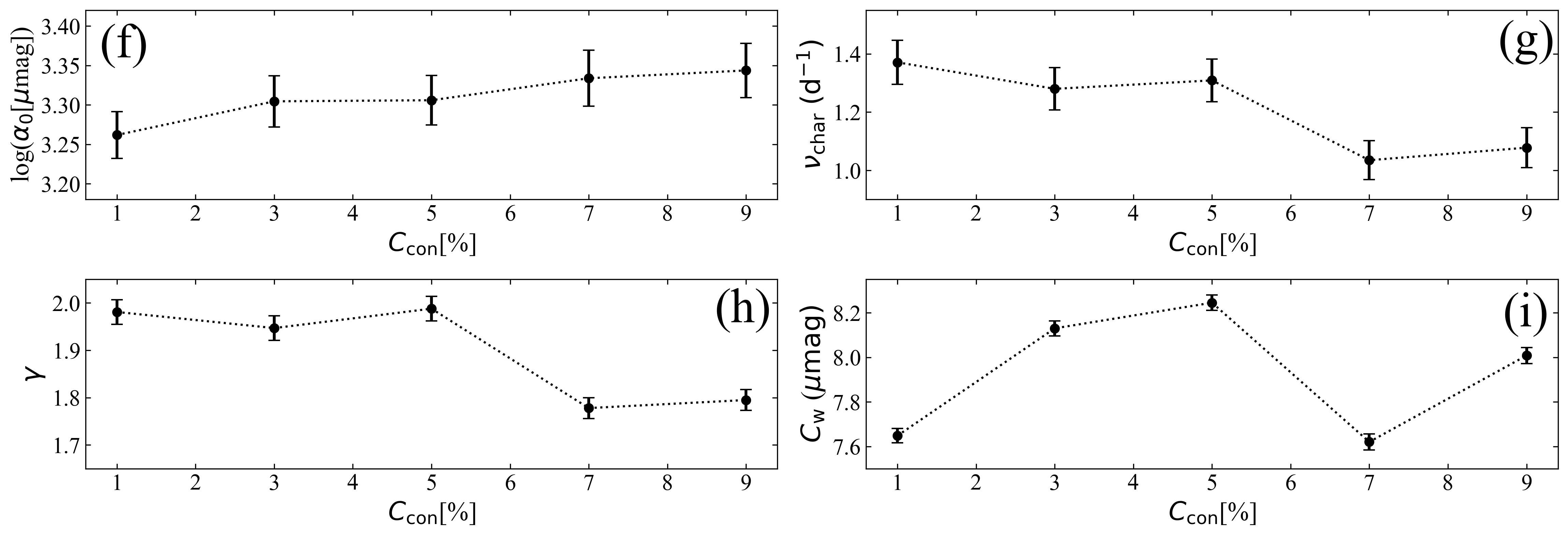}{1\textwidth}{}}
    \caption{The TPF of TESS sector 61 for HD\,69464, alongside aperture masks (red rectangles) corresponding to different $C_{\mathrm{con}}$ ranging from 1\% to 9\% with a step of 2\% are presented in panels (a) to (e). The position of HD\,69464 is denoted by a red pentagram, while nearby contaminating stars within the aperture masks are represented by white pentagrams. Panels (f) to (i) show the fitting parameters of SLF variability as functions of $C_{\mathrm{con}}$. Compared with the results at $C_{\mathrm{con}} = 1\%$, the value of $\log\alpha_{0}$ exhibits a slight increase with rising $C_{\mathrm{con}}$, and the relative error remains below 3\%, signifying a negligible impact from the $C_{\mathrm{con}}$. Regarding $\nu_{\mathrm{char}}$ and $\gamma$, the results are not significantly affected when $C_{\mathrm{con}} \leq 5\%$. However, when $C_{\mathrm{con}} > 5\%$, significant deviations in $\nu_{\mathrm{char}}$ and $\gamma$ are observed when compared to the results at $C_{\mathrm{con}} = 1\%$. \label{fig.1}}
\end{figure*}

Through the Mikulski Archive for Space Telescopes (MAST\footnote[1]{https://archive.stsci.edu/}), we acquired all available 2 min TESS Target Pixel Files (TPFs) spanning Sectors 1 to 65 for the Galactic O-type stars in our original sample. 
The aperture mask for each star is established by employed the thresholding method\footnote[2]{https://docs.lightkurve.org/reference/api/lightkurve.TessTargetPixelFile} using the Python package $Lightkurve$ \citep{2018ascl.soft12013L}. 
This methodology selects pixels with flux values surpassing a threshold, defined as a multiple of standard deviations above the median brightness, to constitute the photometric aperture mask. Therefore, the accuracy of the standard deviation for pixels with flux values directly affects the accuracy of the photometric apertures. The standard deviation is robustly estimated by the Median Absolute Deviation (MAD). In $Lightkurve$ python package\footnote[3]{http://docs.lightkurve.org/reference/api/lightkurve.KeplerTargetPixelFile.create$\_$threshold$\_$mask.html}, the default and unchangeable setting for determining the standard deviation is by multiplying the MAD by 1.4826. This approximation assumes that the TESS pixel response function (PRF) can be approximated as a Gaussian distribution \citep[refer to][for more details]{2010ApJ...713L..97B}.
The Gaia mission \citep{2016A&A...595A...1G}, with an angular resolution of $1^{\prime \prime}$, can be utilized as a supplementary tool for identifying potential sources of contamination within the photometric aperture mask \citep[e.g.,][]{2023RNAAS...7...18S}. 
To determine an appropriate aperture mask for each star, we initially generated the aperture masks using different threshold values $\sigma_{\mathrm{thr}} \in [10, 20, 30, 40, 50, 60, 70, 80, 90, 100]$.  
Employing the Python package $Astroquery$ \citep{2019AJ....157...98G}, we extracted Gaia $G,\,G_{\rm rp},$ and $\;G_{\rm bp}$ \citep{2022yCat.1355....0G} about stars within the corresponding sky regions defined by each aperture mask. We transformed the Gaia bands to TESS bands using a color-dependent 
transformation which published by \cite{2019AJ....158..138S}:
\begin{equation}\label{eq1}
    T = G - 0.00522555(G_{\mathrm{bp}} - G_{\mathrm{rp}})^3 + 0.0891337(G_{\mathrm{bp}} - G_{\mathrm{rp}})^2 -0.633923(G_{\mathrm{bp}} - G_{\mathrm{rp}}) + 0.0324473
\end{equation}
where $T$ is the magnitude in the TESS band. $G,\;G_{\rm rp},$ and $\;G_{\rm bp}$ are magnitudes of Gaia bands. When $G_{\rm rp}, \rm{and}\;G_{\rm bp}$ are missing, we adopted the transformation as follows:
\begin{equation}\label{eq2}
    T = G - 0.430,
\end{equation}
from the same work. After this step, we calculated the contamination ratios of different aperture masks for the star to determine the one with the minimum contamination ratio. 
Drawing inspiration from \cite{2021AJ....162..170H}, the contamination ratio, $C$, is defined by the following equation:
\begin{equation}\label{eq3}
    C = 100\% \times (1 - \frac{f_{\mathrm{target}}}{f_{\mathrm{total}}}) = 100\% \times [1 - 10^{0.4(T_{\mathrm{total}} - T_{\mathrm{target}})}],
\end{equation}
where $f_{\mathrm{target}}$ is the mean flux of target star, $f_{\mathrm{total}}$ refers to the total flux of all stars within the aperture, including target star. 
$T_{\mathrm{total}}$ and $T_{\mathrm{target}}$ represent the total magnitude of all stars within the aperture mask and the magnitude of the target star, respectively.
In this manner, the contamination ratio will tend to 100\% when the flux from blended or nearby stars within the aperture mask equals that of the target star.

Using HD\,69464 as an example, we illustrated the impact of varying contamination ratios ($C_{\rm con}$) on the fitting parameters of SLF variability. The TPF of TESS sector 61 for HD\,69464 is presented in panels (a) to (e) of Figure \ref{fig.1}. The contamination aperture mask is indicated by red rectangle in each panel with corresponding $C_{\rm{con}}$ (notated in the upper right corner). It should be noted that rectangular aperture is employed solely to establish an upper limit for contamination assessment within the aperture generated using the threshold method. During the photometry process, the aperture generated by the threshold method (which corresponds to the minimum $C_{\rm con}$) is employed for extracting the light curve. The position of HD\,69464 is denoted by a red pentagram, while nearby contaminating stars within the contamination aperture masks are represented by white pentagrams. The fitting parameters including $\log\,(\alpha_0\,[\mu\rm mag])$, $\nu_{\rm{char}}$, and $\gamma$, serve as crucial indicators for assessing SLF variability, particularly $\nu_{\rm{char}}$ and $\gamma$, which are pivotal in discerning the origin of SLF variability \citep[e.g.,][]{2019A&A...621A.135B, 2019NatAs...3..760B}. Therefore, it is imperative to evaluate the influence of contamination on these fitting parameters ($\log\,(\alpha_0\,[\mu\rm mag])$, $\nu_{\rm{char}}$, and $\gamma$). In the panels (f) to (i) of Figure \ref{fig.1}, we displayed the fitting parameters including amplitude ($\log\,(\alpha_0\,[\mu\rm mag])$), characteristic frequency ($\nu_{\rm{char}}$), steepness ($\gamma$) and white noise term ($C_{\rm{w}}$) as functions of $C_{\rm con}$. We adopted the fitting parameters derived from the aperture with $C_{\rm con}=1\%$ as our benchmark for evaluation. We observed a slight increase in the $\log\,(\alpha_0\,[\mu\rm mag])$ as the $C_{\rm con}$ rise. The relative errors of $\log\,(\alpha_0\,[\mu\rm mag])$, compared with the results at $C_{\rm con }= 1\%$, remain below 3\%, signifying a negligible impact from the $C_{\rm con}$. Regarding the $\nu_{\mathrm{char}}$ and $\gamma$, the results are insignificantly affected when the $C_{\rm con} \leq 5\%$. However, for $C_{\rm con} > 5\%$, significant deviations in $\nu_{\mathrm{char}}$ and $\gamma$ are observed compared to the results obtained at $C_{\rm con}=1\%$. When $C_{\mathrm{con}} > 5\%$,relative errors for $\nu_{\mathrm{char}}$ exceed 25\%, and relative errors for $\gamma$ surpass 10\% compared with the results at $C_{\mathrm{con}}=1\%$. Therefore, we optimized the aperture masks for all stars of our sample, ensuring that the $C_{\rm con}$ of each star's aperture mask is minimized. Stars with $C_{\rm con} > 5\%$ after optimization were excluded from the sample. Subsequently, the optimized aperture masks underwent visual inspection to ensure their accurate coverage of the target stars. Ultimately, our final sample comprises a total of 150 Galactic O-type stars.

\subsection{Iterative pre-whitening of coherent frequencies} 
Using the Python package $Lightkurve$ \citep{2018ascl.soft12013L}, we performed the aperture photometry and data reduction for the stars in our sample. During the photometry process, we adopted the optimized aperture masks as described in Section \ref{sec3.1}. 
To eliminate the background noise, we utilized the PLDCorrector\footnote[3]{http://docs.lightkurve.org/reference/api/lightkurve.correctors}. The background aperture masks were defined using the threshold method, specifically using pixels fainter than 1$\sigma$ above the median flux. 
Outliers were removed using the sigma-clipping method provided by $Lightkurve$. 
Additionally, we excluded photometric data points labeled as "Argabrightening" and "Cosmic ray in optimal aperture", based on the quality flags contained in each individual data cadence. 
Long-term instrumental trends, such as jumps and drifts, are removed through the application of a linear or low-order polynomial fit to the light curve.

We employed the Discrete Fourier Transform \citep[DFT;][]{1985MNRAS.213..773K} to compute the amplitude spectra of the light curves.
The Nyquist frequency refers to the highest frequency that can be accurately resolved by the sampling process, which can be calculated using equation $f_{\mathrm{Ny}} = 1/(2\Delta t)$, where $\Delta t$ represents the time interval between consecutive samples. 
Due to the TESS short cadence sampling interval of 120 seconds, we determined the Nyquist frequency to be $f_{\mathrm{Ny}} = 359.9\;\mathrm{day^{-1}}$.
Therefore, the significant coherent frequencies in the range of $0 < f < f_{\mathrm{Ny}}$ are removed during the iterative pre-whitening process. 
Using the PERIOD04 software \citep{2005CoAst.146...53L}, the significant coherent frequencies are identified from the amplitude spectra. We calculated and optimized the parameters of the frequencies obtained from the amplitude spectra using the following formula:
\begin{equation}\label{eq4}
    m = m_{0} +  \sum_{i}^{N} A_{i}\mathrm{sin}(f_{i}t + \varPhi_{i})
\end{equation}
where $N$ represents the number of coherent frequencies, while $m_{0}$ represents the zero-point. $A_{i}$, $f_{i}$, and $\varPhi_{i}$ denote the amplitude, frequency, and phase of the $i$th coherent pulsation, respectively. As the previous investigation conducted by \cite{2021A&A...656A.158B}, coherent frequencies with a signal-to-noise ratio (SNR) in the range from 3.5 to 4.0 from space photometry are indistinguishable from high-amplitude noise, while those falling within the range of $4.0 \leqslant  \mathrm{SNR} \leqslant  4.6$ have a modest chance of being significant frequencies. Therefore, the frequencies with SNR $\geqslant$ 4.6 are determined as the significant frequencies in this work. We employed a traditional pre-whitening technique to eliminate significant frequencies from the light curves. The significant peaks are identified and fitted using Equation (\ref{eq4}), following which the fitted functions were subtracted from the light curves. 

\subsection{Fitting parameters of SLF variability}
The SLF variability is commonly observed in the amplitude spectrum as the characteristic red noise component of a stochastic signal \citep[e.g.,][]{2019NatAs...3..760B, 2023ApJ...955..123S}. Based on the study by \cite{2021A&A...656A.158B}, following the elimination of frequencies with SNR $\geqslant$ 4.6, the light curves exclusively exhibit residual red noise (attributed to SLF variability) and white noise. Building on the studies of \cite{2002A&A...394..625S}, \cite{2011A&A...533A...4B}, and \cite{2019A&A...621A.135B,2019NatAs...3..760B}, the morphology of SLF variability in an amplitude spectrum for a massive star can be physically interpreted using a Lorentzian function:
\begin{equation}\label{eq5}
    \alpha(\nu) = \frac{\alpha_{\rm{0}}}{1 + (\frac{\nu}{\nu_{\rm{char}}})^{\gamma}} + C_{\rm{w}},
\end{equation}
where $\alpha_{\rm{0}}$ denotes the scaling factor, representing the amplitude at zero frequency, while $\nu_{\rm{char}}$ stands for the characteristic frequency.
Moreover, $\gamma$ signifies the logarithmic amplitude gradient, and $C_{\rm{w}}$ is a white noise term. We adopted a Bayesian Markov Chain Monte Carlo (MCMC) framework, employing the Python code $emcee$ \citep{2013PASP..125..306F}, to infer morphological characteristics from the residual amplitude spectra as delineated in quation \ref{eq5}. 

In the fitting of the residual amplitude spectra, flat priors and 128 parallel chains are employed. During the iterative process, each parallel chain used to construct a model is evaluated using the likelihood function published by \cite{1990ApJ...364..699A}:
\begin{equation}\label{eq6}
    \ln \mathfrak{L}  = - \sum_{i}^{N} \Bigg\{{\ln [M_{i}(\boldsymbol{\theta})]} + \frac{O_{i}}{[M_{i}(\boldsymbol{\theta})]}\Bigg\},
\end{equation}
where $N$ is the total number of data points within the residual amplitude spectrum, $O_{i}$ represents the $i$th point of the residual amplitude spectrum, and $M_{i}(\boldsymbol{\theta})$ represents the model with parameters $\boldsymbol{\theta}$ for the $i$th point. 
After burning the initial few hundred iterations, convergence is confirmed for the subsequent 2000 iterations through the utilization of the parameter variance criterion published by \cite{1992StaSc...7..457G}. 

The residual amplitude spectrum fit obtained through Equation (\ref{eq5}) yielded the amplitude ($\alpha_{0}$), characteristic frequency ($\nu_{\mathrm{char}}$), and steepness ($\gamma$) of the frequency spectrum that characterize the SLF variability present in the light curve of each massive star. Moreover, the parameters $\nu_{\mathrm{char}}$ and $\gamma$ are considered critical for distinguishing the source of the SLF variability of massive stars \citep[e.g.,][]{2019A&A...621A.135B}. 
At low frequencies, specifically those falling below 0.1 day$^{-1}$, the capability of TESS time series photometry to elucidate variability is constrained by the temporal duration of observations. 
Additionally, the existence of instrumental systematics in TESS data may significantly impact the amplitude spectrum at frequencies below 0.1 day$^{-1}$, given their association with variability manifesting periods of a similar order to the time series duration. 
Consequently, previous studies exclusively characterized variability above 0.1 day$^{-1}$ in light of this consideration \citep[e.g.,][]{2019NatAs...3..760B, 2019A&A...621A.135B, 2020A&A...640A..36B}. For all Galactic O-type stars included in our present TESS sample, the measured parameter $\nu_{\mathrm{char}}$ is significant at frequencies exceeding 0.1 day$^{-1}$.

\begin{figure}
    \gridline{\fig{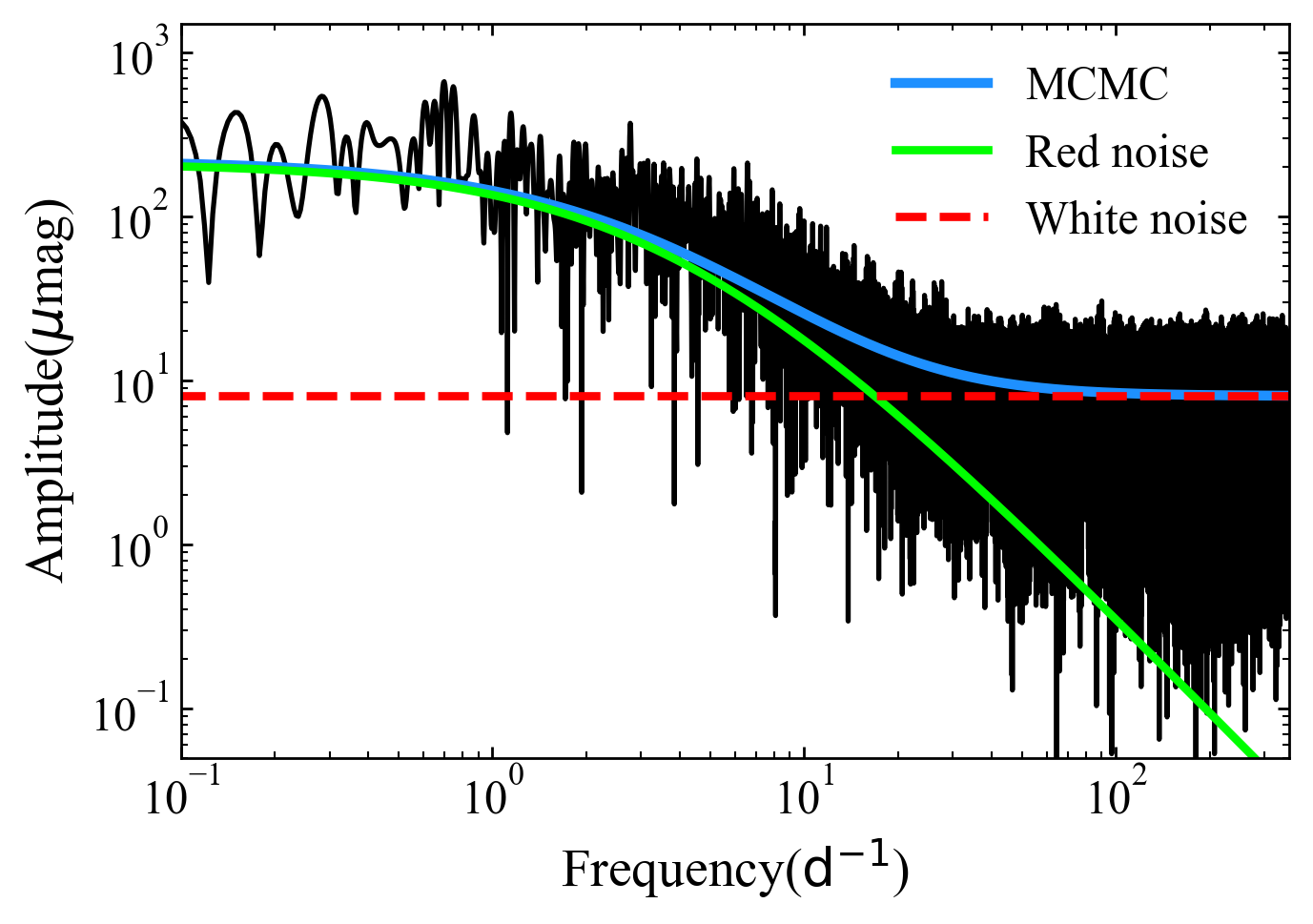}{0.4\textwidth}{}}
    \gridline{\fig{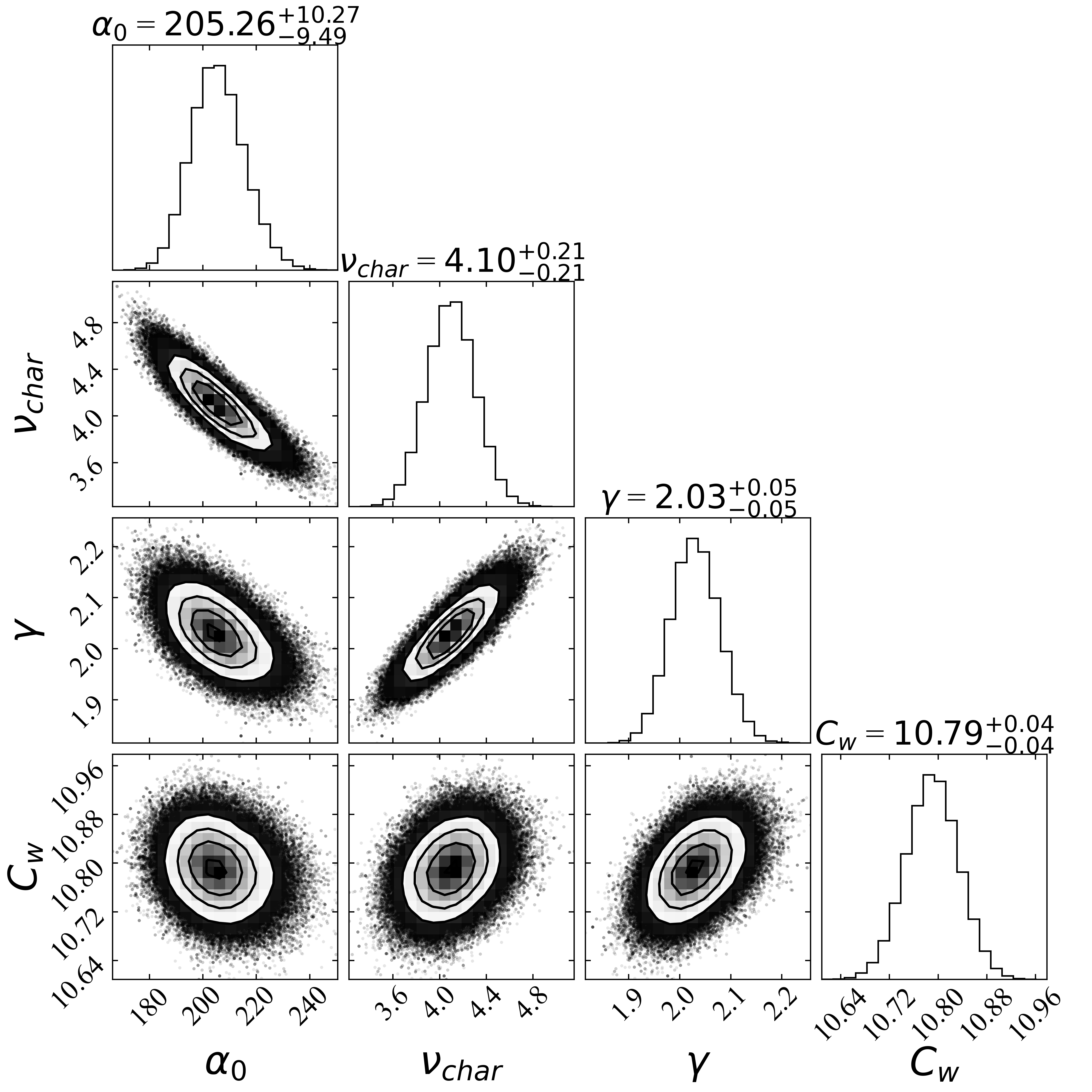}{0.45\textwidth}{}}
    \caption{Top panel: the logarithmic residual amplitude spectrum of the TESS short cadence photometric data for BD$\,+36\;4145$ is shown in black, with the best-fit model obtained using Equation (\ref{eq5}) depicted as the solid blue line. The red and white noise components are illustrated by the solid green line and red dashed lines, respectively. Bottom panel: posterior distributions from the Bayesian MCMC, with best-fit parameters.}
    \label{mcmc_temp}
 \end{figure}

\subsection{Astrophysical parameters}
Benefiting from the comprehensive observational efforts dedicated to spectroscopic surveys such as GOSSS, IACOB, and OWN over the past decade, and facilitated by the accessibility of stellar atmosphere codes, the stellar parameters of hundreds of O-type stars have been precisely determined \citep[e.g.,]{2012ASPC..465..484M, 2016AJ....152...31A, 2018A&A...613A..65H, 2020A&A...638A.157H}. 
These parameters include effective temperatures ($T_{\mathrm{eff}}$), surface gravity (log$\,g$), the projected rotational velocity ($v$sin$i$), macroturbulence velocity ($v_{\mathrm{macro}}$), and spectroscopic luminosity (log$\,(\mathcal{L/L_{\odot}})$) \citep[$\mathcal{L} \equiv T_{\mathrm{eff}}^4/g$;][]{2014A&A...564A..52L}. Among the stars in our sample, 130 O-type stars are observed by the high-resolution spectroscopic survey IACOB. \cite{2020A&A...638A.157H} performed quantitative spectroscopic analysis using two semi-automatized tools designed within the framework of the IACOB project, namely, IACOB-BROAD \citep{2014A&A...562A.135S} and IACOB-GBAT \citep{2011JPhCS.328a2021S}. These tools allow for determining the line-broadening and spectroscopic parameters, as well as the associated uncertainties, for large samples of O-type stars in a homogeneous, objective, and relatively fast manner. We obtained the stellar parameters from the results published by \cite{2018A&A...613A..65H} for the 130 stars in our sample. 
The absence of spectroscopic data for the remaining 20 stars forces us to derive the effective temperatures and luminosities from spectral type-effective temperature calibrations provided by \cite{1987A&A...177..217D}. 
Based on the most recent and comprehensive tabular datasets of effective temperature and spectral type, \cite{1987A&A...177..217D} established relationships in which log\,$T_{\mathrm{eff}}$ and log\,$L$ are expressed as Chebyshev polynomials based on two continuous analytical variables, denoted as $s$ (linked to spectral class) and $b$ (associated with luminosity class), resulting in log\,$T_{\mathrm{eff}}$\,($s, b$) and log\,$L$\,($s, b$). 
The 1$\sigma$ values for the determinations of the unit weight of log\,$T_{\mathrm{eff}}$ and log\,$L$ are 0.021 and 0.164, respectively \citep{1987A&A...177..217D}. These uncertainties will also be employed in this study. 

We utilized the Bayesian tool Bonnsai\footnote[4]{The BONNSAI web-service is available at www.astro.uni-bonn.de/stars/bonnsai.} \citep{2014A&A...570A..66S} to determine the comprehensive posterior probability distribution of stellar mass, radius, age, fractional main sequence age ($\tau_{\mathrm{age}}$), and wind mass-loss rate (log\,$\dot{M}$) for each individual star. In pursuit of this objective, we employed the rotating Milky Way stellar evolution models developed by \cite{2011A&A...530A.115B} to match the $T_{\mathrm{eff}}$, log\,$g$, $v$sin$i$, and log\,($\mathcal{L/L_{\odot}}$) values that were determined from the high-resolution spectra for each star. We adopted the Salpeter initial mass function \cite{1955ApJ...121..161S} as the prior for initial mass, and a flat prior for initial rotational velocity. For the stars lacking $v$sin$i$ values, we applied a Gaussian prior for initial rotational velocity, with a mean of 100 km s$^{-1}$ and a standard deviation of 125 km s$^{-1}$, following the approach outlined by \cite{2008A&A...479..541H}. We presented all stellar parameters and their uncertainties in Table \ref{table:stellar_parameters}.

\section{Result}\label{sec4}
\begin{figure*}
       \gridline{\fig{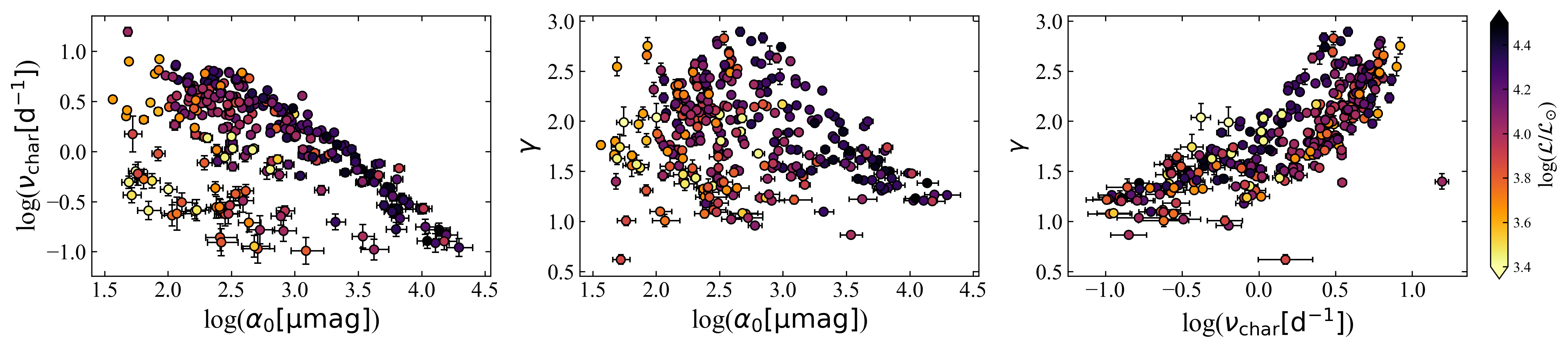}{1\textwidth}{}}
       \caption{The pairwise relationships between $\log\,(\alpha_0\,[\mu\rm mag])$, log\,$\nu_{\mathrm{char}}$, and $\gamma$ for the residual amplitude spectra of stars of our sample. Each star is marked as filled circle and color-coded based on spectroscopic luminosity.\label{slf_result}}
   \end{figure*}

We utilized Equation (\ref{eq6}) to derive the fitting parameters for the SLF variability in 298 residual amplitude spectra from 150 Galactic O-type stars in our sample.
The fitting parameters and their 1$\sigma$ statistical uncertainties are detailed in the Table \ref{table:slf_parameters}. The logarithmic residual amplitude spectra and corresponding fitting results can be found in Appendix B. Our analysis reveals that SLF variability is prevalent among Galactic O-type stars, which is consistent with previous studies \citep[e.g.,][]{2019NatAs...3..760B,2020A&A...640A..36B, 2023ApJ...955..123S}. 
Previous research indicated that $\log\,(\alpha_0\,[\mu\rm mag])$ ranges from 1 to 4, with the $\nu_{\mathrm{char}}$ distribution predominantly centering between 0.1 and 10 day$^{-1}$, and the log\,$\gamma$ distribution spans the interval from -0.5 to 1.3 \citep[e.g.,][]{2019NatAs...3..760B,2020A&A...640A..36B}. For magnetic hot stars, \cite{2023ApJ...955..123S} reported the distribution of $\log\,(\alpha_0\,[\mu\rm mag])$ is in the range of 0.5\textbf{\textendash}3, $\nu_{\mathrm{char}}$ primarily distributes in the range of 0\textbf{\textendash}7 day$^{-1}$, and $\gamma$ primarily falls between 0 and 5. In Figure \ref{slf_result}, we depicted the pairwise relationships between $\log\,(\alpha_0\,[\mu\rm mag])$, log\,$\nu_{\mathrm{char}}$, and $\gamma$ for the residual amplitude spectra of stars in our sample, using filled circles and color-coded based on spectroscopic luminosity.
In our study, $\log\,(\alpha_0\,[\mu\rm mag])$ ranges from 1.5 to 4.5, while $\nu_{\mathrm{char}}$ predominantly falls within the range of 0.1 to 8.3 day$^{-1}$. Additionally, the $\gamma$ distribution spans the interval from 0.6 to 2.9. Notably, in the left panel of Figure \ref{slf_result}, we observed a decrease in $\nu_{\mathrm{char}}$ with increasing $\log\,(\alpha_0\,[\mu\rm mag])$, indicating that stars with larger SLF variability amplitude tend to have smaller characteristic frequencies, consistent with findings reported by \cite{2020A&A...640A..36B}. Furthermore, within the same amplitude range, more luminous stars exhibit larger $\nu_{\mathrm{char}}$ values. In the middle panel of Figure \ref{slf_result}, a slight negative correlation between $\log\,(\alpha_0\,[\mu\rm mag])$ and $\gamma$ is evident. Lastly, the right panel of Figure \ref{slf_result} highlights a significant correlation between $\nu_{\mathrm{char}}$ and $\gamma$, indicating that stars with higher characteristic frequencies also tend to have steeper slopes.

    \begin{figure*}
      \centering
      \gridline{\fig{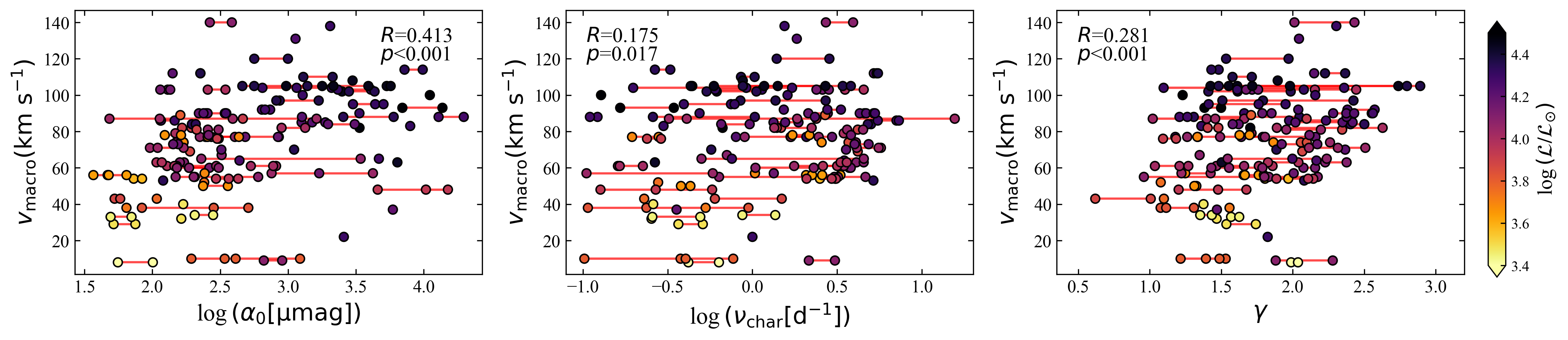}{1\textwidth}{}}
       \caption{Correlation between the fitting parameters of SLF and spectroscopic measured of $v_{\mathrm{macro}}$. Each star is depicted as a circle and color-coded based on spectroscopic luminosity. The data sets of individual stars observed in different sectors are connected by red lines. The Spearman's rank correlation coefficient (denoted as $R$) and the corresponding two-sided $p$-value (obtained from a t-test) are presented in each panel.\label{vmacro_slf}} 
   \end{figure*}

\begin{figure*}
      \centering
      \gridline{\fig{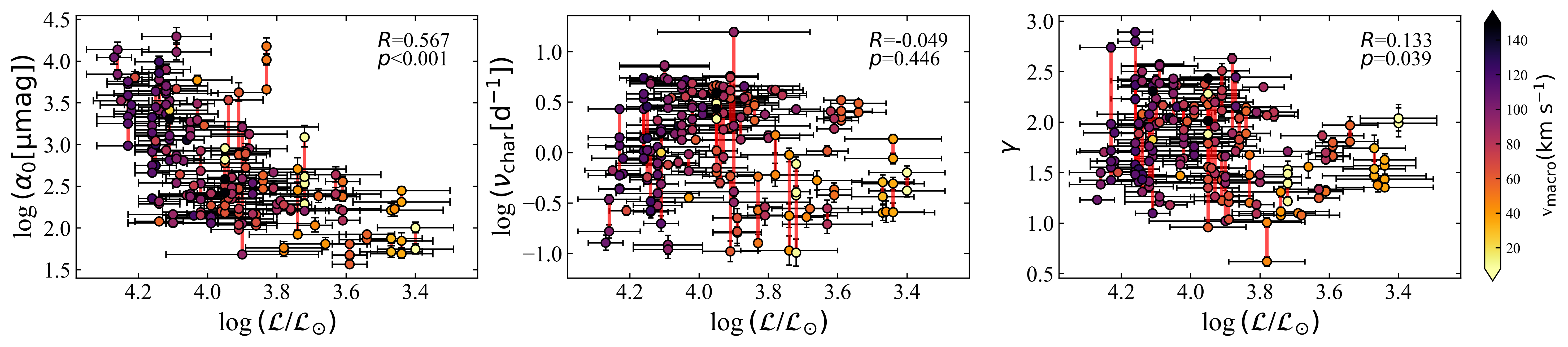}{1\textwidth}{}}
      \gridline{\fig{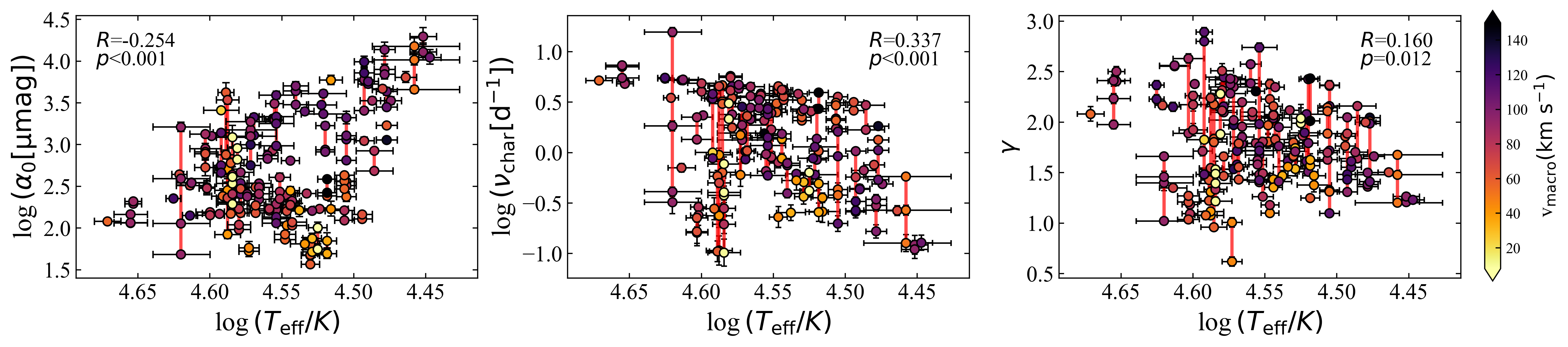}{1\textwidth}{}}
      \caption{Dependence of the fitting parameters of SLF variability on the log\,($\mathcal{L/L_{\odot}}$) (top panels) and $T_{\mathrm{eff}}$ (bottom panels). Each star is represented as a circle and color-coded by $v_{\mathrm{macro}}$ in km s$^{-1}$. The data sets of individual stars observed in different sectors are connected by red lines. \label{vmacro_relation}}
  \end{figure*}

   In their study, \cite{2020A&A...640A..36B} integrated TESS photometric data with high-resolution ground-based spectroscopy of 70 OB stars. They reported that $v_{\mathrm{macro}}$ shows a significant correlation with $\log\,(\alpha_0\,[\mu\rm mag])$ and $\nu_{\mathrm{char}}$, while exhibits no significant correlation with $\gamma$. In our sample of 150 O-type stars, 130 have been meticulously observed using the high-resolution IACOB survey. The stellar parameters for these stars are determined using two semi-automated tools developed within the IACOB project, ensuring a high level of precision and consistency. The remaining 20 stars in our sample were observed with the low-resolution GOSSS survey, with their stellar parameters inferred from spectral type-effective temperature calibrations. The inclusion of both high- and low-resolution observations in a sample may lead to heterogeneity. To mitigate this, we elected to focus solely on the 130 stars that were observed with high-resolution as the subsample for exploring the correlation between stellar properties and SLF variability in the subsequent analysis.
   In Figure \ref{vmacro_slf}, we presented the macroturbulence velocity, $v_{\mathrm{macro}}$, as the function of fitting parameters of SLF variability. Each star is represented by a circle in each panel of Figure \ref{vmacro_slf} and color-coded by spectroscopic luminosity. The data sets of individual stars observed in different sectors are connected by red lines.
   To assess the correlation between $v_{\mathrm{macro}}$ and the fitting parameters, we employed Spearman's rank correlation coefficient ($R$) and the corresponding two-sided $p$-value (obtained from a t-test). Our analysis reveals a correlation coefficient of 0.413 between $v_{\mathrm{macro}}$ and $\log\,(\alpha_0\,[\mu\rm mag])$, with a $p$-value much less than 0.001. This result suggests a significantly moderate correlation between $v_{\mathrm{macro}}$ and the amplitude of SLF variability. Additionally, our statistical analysis indicates a slight positive correlation between $\gamma$ and $v_{\mathrm{macro}}$ ($R = 0.281,\, p < 0.001$). We found only a marginal correlation between $\nu_{\mathrm{char}}$ and $v_{\mathrm{macro}}$ ($R = 0.175,\, p = 0.017$). In a separate study, \cite{2023ApJ...955..123S} reported that no significant correlations were found between the fitting parameters and the luminosities of the magnetic hot stars. In Figure \ref{vmacro_relation}, we illustrated the dependence of fitting parameters of SLF variability on log$\,(\mathcal{L/L_{\odot}})$ (top panels) and $T_{\mathrm{eff}}$ (bottom panels). Each star is depicted as a circle and color-coded by $v_{\mathrm{macro}}$ in km s$^{-1}$. The Spearman's rank correlation coefficients and the corresponding two-sided $p$-values are provided at the corners for the panels.
   We observed a significant correlation between $\log\,(\alpha_0\,[\mu\rm mag])$ and log$\,(\mathcal{L/L_{\odot}})$ ($R = 0.567,\, p < 0.001$), which is inconsistent with the results published by \cite{2023ApJ...955..123S}. No significant correlations are detected between $\nu_{\mathrm{char}}$ and log$\,(\mathcal{L/L_{\odot}})$, as well as between $\gamma$ and log$\,(\mathcal{L/L_{\odot}})$. In the bottom panels of Figure \ref{vmacro_relation}, we observed a significantly moderate correlation between $\log\,(\alpha_0\,[\mu\rm mag])$ and $T_{\mathrm{eff}}$ ($R = -0.254,\, p < 0.001$). $\nu_{\mathrm{char}}$ also shows a significantly moderate correlation with $T_{\mathrm{eff}}$ ($R = 0.337,\, p < 0.001$), while no relationship is detected between the $\gamma$ and $T_{\mathrm{eff}}$.

   \begin{figure*}
      \gridline{\fig{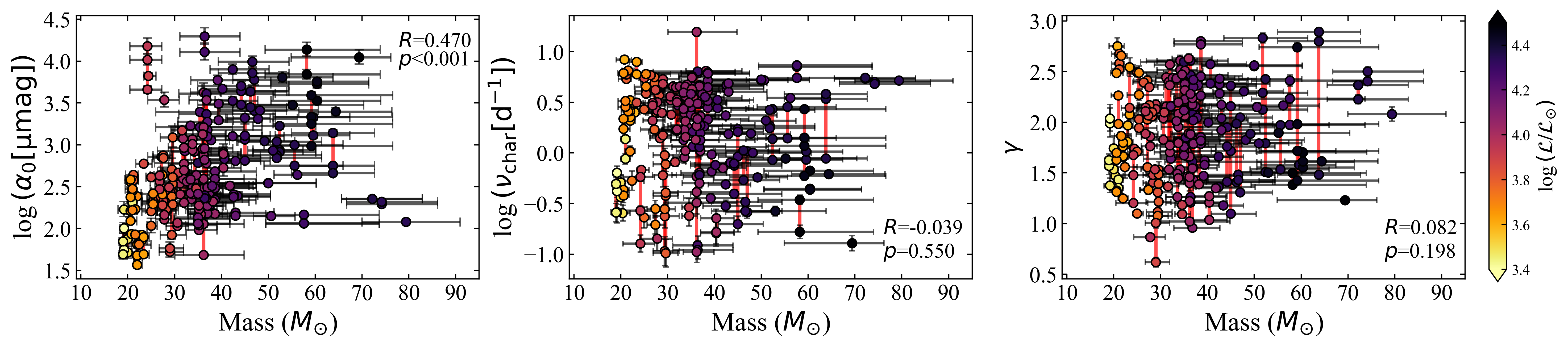}{1\textwidth}{}}
      \gridline{\fig{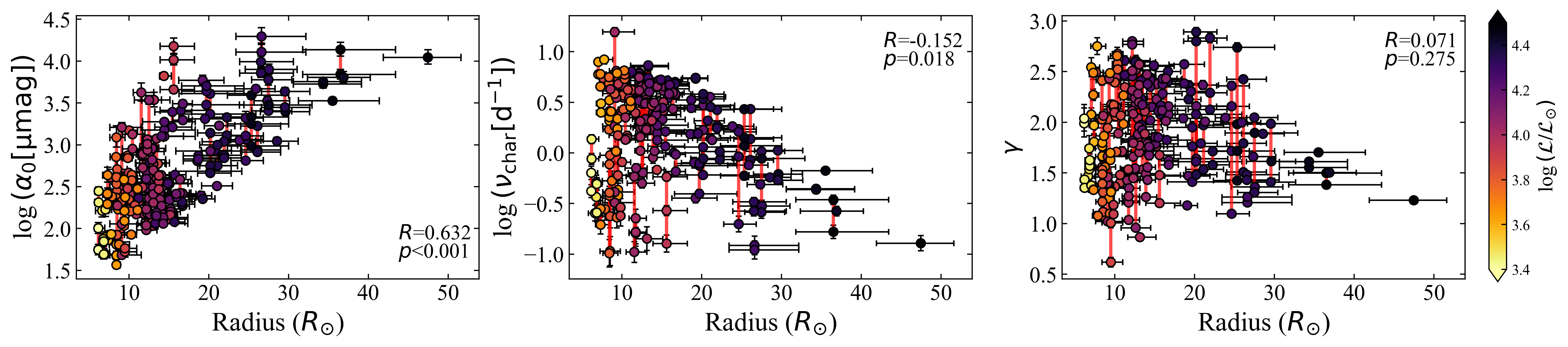}{1\textwidth}{}}
      \gridline{\fig{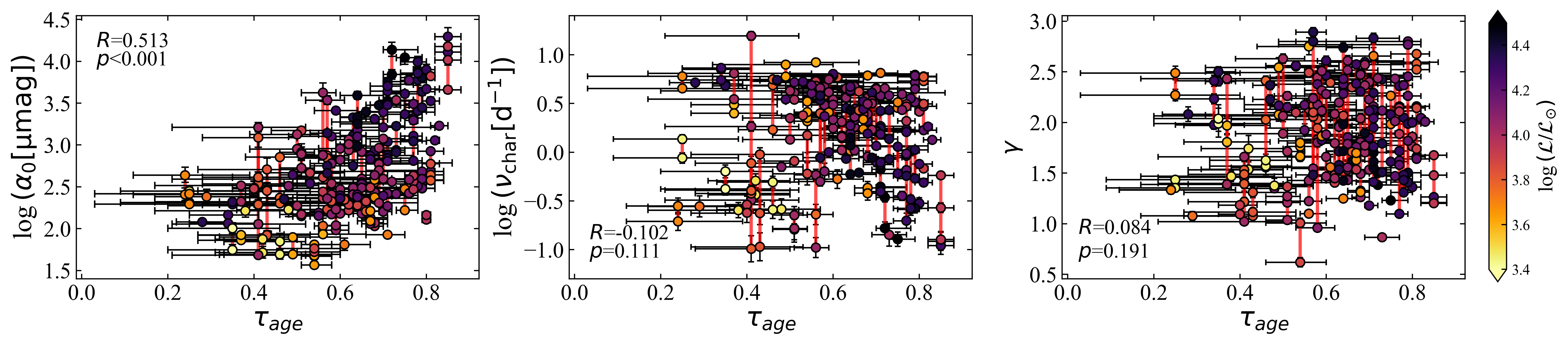}{1\textwidth}{}}
      \gridline{\fig{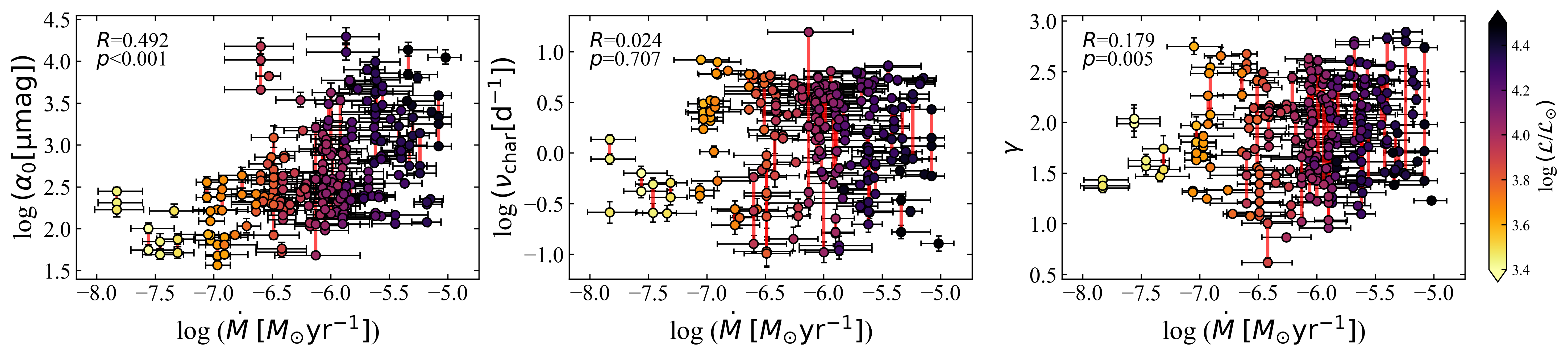}{1\textwidth}{}}
       \caption{The fitting parameters of SLF variability, $\alpha_{0}$, $\nu_{\mathrm{char}}$, and $\gamma$ as a function of various stellar parameters including mass, radius, fractional main sequence age ($\tau_{\mathrm{age}}$), and wind mass-loss rate (log\,$\dot{M}$). Each star is represented as a circle and color-coded by spectroscopic luminosity. The data sets of individual stars observed in different sectors are connected by red lines. \label{parameters_slf}}
    \end{figure*}

We then aimed to explore the correlations between the parameters characterizing the SLF variability component in the photometric fluctuations observed in light curves and various stellar parameters.
In Figure \ref{parameters_slf}, we presented the fitting parameters for $\log\,(\alpha_0\,[\mu\rm mag])$, log\,$\nu_{\mathrm{char}}$, and $\gamma$ as a function of the stellar mass ($M$), radius ($R$), $\tau_{\mathrm{age}}$, and log\,$\dot{M}$. The markers in the each panel have the same meaning as in Figure \ref{vmacro_slf}. 
In the first column of Figure \ref{parameters_slf}, we found that $\log\,(\alpha_0\,[\mu\rm mag])$ shows different significant correlations with stellar mass, stellar radius, $\tau_{\mathrm{age}}$, and log\,$\dot{M}$.
A linear relationship is revealed between $\log\,(\alpha_0\,[\mu\rm mag])$ and stellar mass with a significantly moderate correlation ($R = 0.470,\, p < 0.001$), indicating that more massive stars have larger amplitude of SLF variability. This result is consistent with the conclusion published by \cite{2020A&A...640A..36B}. It is noteworthy that some values of $\log\,(\alpha_0\,[\mu\rm mag])$ fall below 3 for the most massive stars ($M> 50 M_{\odot}$), which may be attributed to their specific properties such as chemical abundance, rotational velocity, and age. 
We also found that $\log\,(\alpha_0\,[\mu\rm mag])$ exhibits a significant linear relationship with stellar radius ($R = 0.632,\, p < 0.001$), suggesting that stars with larger radii exhibit higher values of $\log\,(\alpha_0\,[\mu\rm mag])$. The first column of Figure \ref{parameters_slf} reveals significant correlations between the $\log\,(\alpha_0\,[\mu\rm mag])$ and both $\tau_{\mathrm{age}}$ ($R = 0.513,\, p < 0.001$) and log\,$\dot{M}$ ($R = 0.492,\, p < 0.001$), suggesting that more evolved stars with higher mass-loss rates demonstrate greater amplitude of SLF variability. Furthermore, the correlation coefficient between $\log\,(\alpha_0\,[\mu\rm mag])$ and stellar radius is the highest, indicating that amplitude of SLF variability is more sensitive to radius compared to other stellar parameters.

\begin{figure}
    \centering
    \gridline{\fig{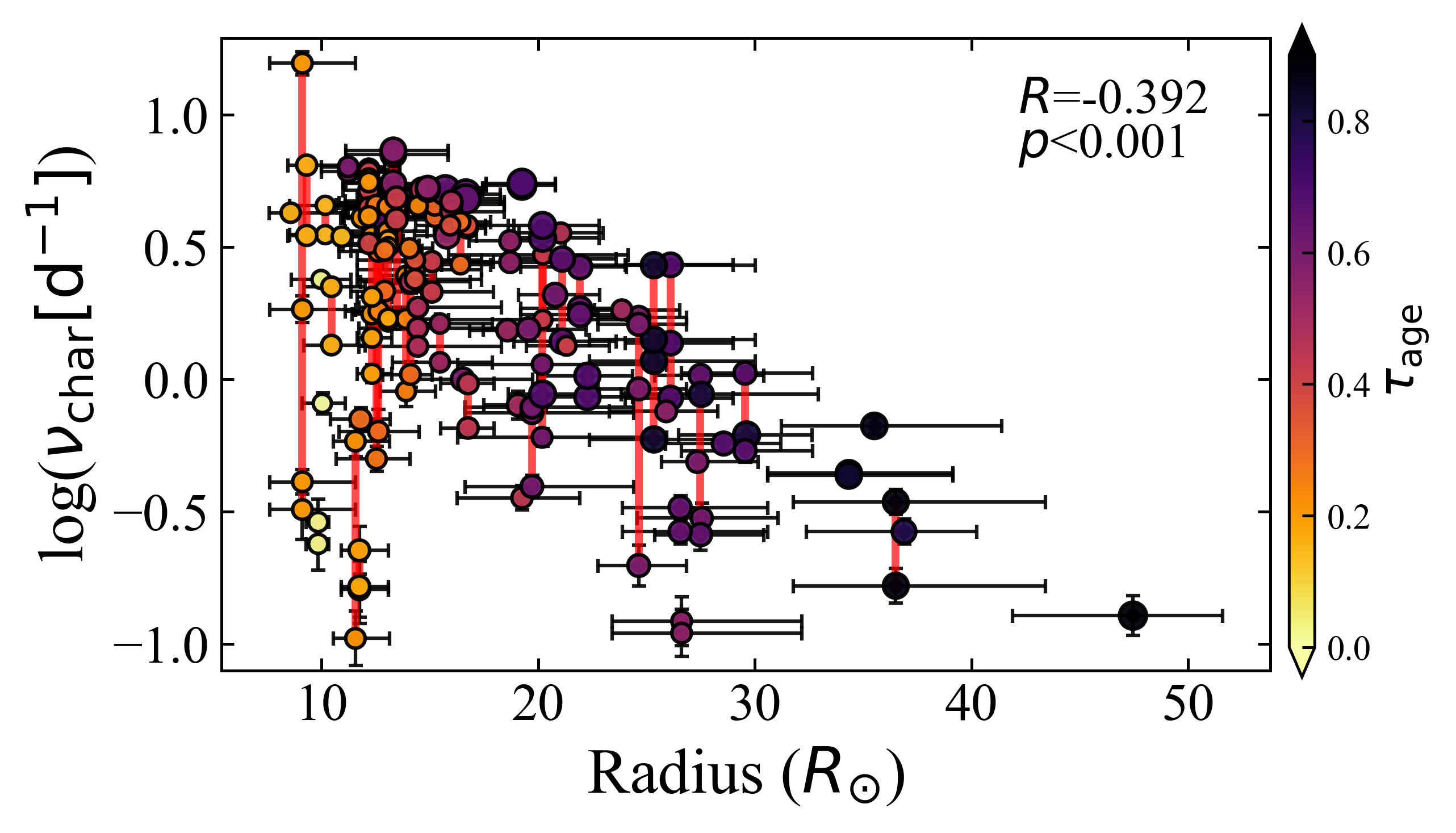}{0.5\textwidth}{}}
    \caption{The significant linear correlation between the $\nu_{\mathrm{char}}$ and stellar radius for stars with $M \geqslant 30\;M_{\odot}$. Each star is represented as a circle, color-coded according to its age $\tau_{\rm age}$. The data sets of individual stars observed in different sectors are connected by red lines. The circle size is directly proportional to the stellar mass. The Spearman's rank correlation coefficient (denoted as $R$) and the corresponding two-sided $p$-value (obtained from a t-test) are presented in upper right corner.\label{IGW}}
\end{figure}
    
In Figure \ref{parameters_slf}, the $\nu_{\mathrm{char}}$ values for stars with varying masses are uniformly distributed within the range of 0$\sim$10 day$^{-1}$, indicating no discernible correlation between stellar mass and the characteristic frequency of SLF variability. Although no significant correlation is detected between $\nu_{\mathrm{char}}$ and stellar radius for the stars of our sample, it should be noted that stars with radii larger than approximately 10\,$R_{\odot}$ (corresponding the stars with $M \geq 30\,M_{\odot}$) exhibit a negative linear relationship for $\nu_{\mathrm{char}}$. Our findings indicate that the characteristic frequency decreases as the radius increases for the more massive stars ($M \geq 30\,M_{\odot}$), while this correlation is absent for lower mass stars. In Figure~\ref{IGW}, we presented more details concerning the relationship between $\nu_{\rm char}$ and stellar radius for stars with $M \geq 30\,M_{\odot}$. Each star is depicted as a circle and color-coded based on $\tau_{\rm age}$. The size of each circle is directly proportional to the stellar mass. The data sets of individual stars observed in different sectors are connected by red lines. Figure~\ref{IGW} demonstrates a significantly moderate correlation between $\nu_{\rm char}$ and stellar radius for these massive stars ($R = -0.392,\, p < 0.001$). This correlation suggests that as the stellar radius increases, the characteristic frequency decreases for more massive stars. \citet{2020A&A...640A..36B} pointed out that more massive and more evolved stars not only exhibit larger amplitudes in their stochastic photometric variability but also smaller $\nu_{\mathrm{char}}$ values. Although no significant correlation between $\nu_{\mathrm{char}}$ and $\tau_{\mathrm{age}}$ is evident in our analysis, Figure~\ref{IGW} indicates that for the more massive stars, stellar evolution leads to a significant radius expansion, ultimately resulting in a smaller characteristic frequency. This result aligns with the findings of \citet{2020A&A...640A..36B}. On the other hand, \cite{2015ApJ...808L..31G} concluded that the turbulent pressure fraction of stellar surface is significantly correlated with the macroturbulent velocities derived from high-quality spectra of OB stars. Furthermore, the turbulent pressure fractions increase substantially for stars with $M \geq 30 M_{\odot}$. Therefore, the correlation between $\nu_{\mathrm{char}}$ and stars with radius $R \gtrsim 10\,R_{\odot}$ may suggest that SLF variability and macroturbulence broadening in spectral lines are excited by the same mechanism. Further details between the SLF variability and macroturbulence broadening will be discussed in Section 5.

The relationship between $\gamma$ and stellar radius exhibits similarity to the association observed between $\nu_{\mathrm{char}}$ and stellar radius, indicating that the steepness decreases as the radius increases for the more massive stars. Apart from this, no significant correlations between $\gamma$ and stellar parameters are found in our analysis. 
\cite{2019A&A...621A.135B} highlighted the significance of $\gamma$ and $\nu_{\mathrm{char}}$ as crucial parameters for discerning the origin of the SLF variability. 
The $\gamma$ values of O-type stars in our sample are distributed between 0.5 and 3. This distribution is consistent with hydrodynamical simulations, which predicted that an ensemble of hundreds IGWs generated by core convection produces an amplitude spectrum with $0.8 \leq \gamma \leq 3$ \citep[e.g.,][]{2013ApJ...772...21R, 2019ApJ...876....4E, 2020A&A...641A..18H}. 
Our statistical analysis has revealed complex relationships between the characteristic frequency and steepness of the SLF variability with stellar radius. Further exploration is needed to compare the extended samples within different mass ranges, particularly using $30\,M_{\odot}$ as the dividing line, which may reveal additional features in the fitting parameters of LSF variability and stellar parameters.

\section{Discussion}
The astrophysical origin of SLF variability in the light curves of massive stars has been firmly established \citep[e.g.,][]{2011A&A...533A...4B, 2015ApJ...806L..33A, 2019A&A...621A.135B, 2019NatAs...3..760B}. However, the question of its specific origin remains debated. Several concurrent physical mechanisms have been proposed as potential sources for exciting SLF variability: (i) subsurface convection \citep[e.g.,][]{2021ApJ...915..112C, 2022ApJ...924L..11S, 2023ApJ...951L..42S}; (ii) IGWs generated by convective cores of massive stars \citep[e.g.,][]{2019ApJ...886L..15L, 2019ApJ...876....4E, 2020A&A...641A..18H, 2021MNRAS.508..132L}; (iii) stellar winds or granulation \citep[e.g.,][]{2010ApJ...711L..35K, 2018MNRAS.476.1234A, 2018A&A...617A.121K, 2022ApJ...925...79L}. Our investigation presents a meticulously assembled sample of Galactic O-type stars for scrutinizing various theoretical predictions against observational results, focusing on the SLF variability of Galactic O-type stars. In this discussion section, we systematically compare the observational signatures of SLF variability from our sample with the theoretical predictions of the proposed excitation mechanisms.

\subsection{Subsurface convection}
Subsurface convection is prevalent in O-type stars, driven by narrow opacity peaks resulting from varying temperatures \citep{2022ApJ...926..221J}. These thin convective zones facilitate limited heat transport, and significantly impact the development of other stellar properties, such as magnetic fields and chemical mixing \citep[e.g.,][]{2009A&A...499..279C, 2017ApJ...843...68J, 2020ApJ...900..113J}. \cite{2009A&A...499..279C} substantiated three discernible trends for the iron convection zone (FeCZ) as a function of stellar parameters using empirical microturbulent velocities, suggesting a physical connection between sub-photospheric convective motions and small-scale stochastic velocities in the photosphere of OB-type stars. 
Utilizing non-rotating 1D stellar models with initial masses ranging from 5 to 120\,$M_{\odot}$, \cite{2021ApJ...915..112C} showed the relevant properties of FeCZ correlate very well with the timescale and amplitude of SLF variability, as well as with the amplitude of macroturbulence. \cite{2022ApJ...924L..11S} conducted multidimensional radiation hydrodynamics (RHD) simulations for two 35\,$M_{\odot}$ stars: one at ZAMS and the other halfway through its main sequence evolution. Their results indicate that surface turbulence produces SLF variability in model light curves with amplitudes and power-law slopes that are strikingly similar to those of observed stars. \cite{2023ApJ...951L..42S} employed 3D RHD simulations to investigate the evolution of the FeCZ region in a 13\,$M_{\odot}$ star, from ZAMS to the Hertzsprung Gap, while also reproducing SLF variability that aligns closely with observational data. By simulating 15\,$M_{\odot}$ stellar model with LMC metallicity at the ZAMS, \cite{2023NatAs...7.1228A} proposed that the photometric variability generated by IGWs exhibits lower amplitude and characteristic frequency than the observed SLF variability, implying that an alternative process generates the observed SLF variability. Furthermore, the multidimensional RHD simulations also indicate that the FeCZ should be responsible not only for the micro- and macroturbulence broadening in spectral lines, but also for the SLF variability observed in OB-type stars \citep[e.g.,][]{2021ApJ...915..112C, 2022ApJ...924L..11S,2023ApJ...951L..42S}. Using high-resolution spectra of 430 stars, \cite{2017A&A...597A..22S} exhaustively investigated the macroturbulence broadening in massive stars. Their results are more consistent with the turbulent motion caused by FeCZ as published by \cite{2015ApJ...808L..31G}. On the other hand, by integrating high-precision time-series photometry with high-resolution ground-based spectroscopy, \cite{2020A&A...640A..36B} elucidated the relationship between the properties of SLF variability and macroturbulent velocity broadening. 
   
\begin{figure*}
\
\gridline{\fig{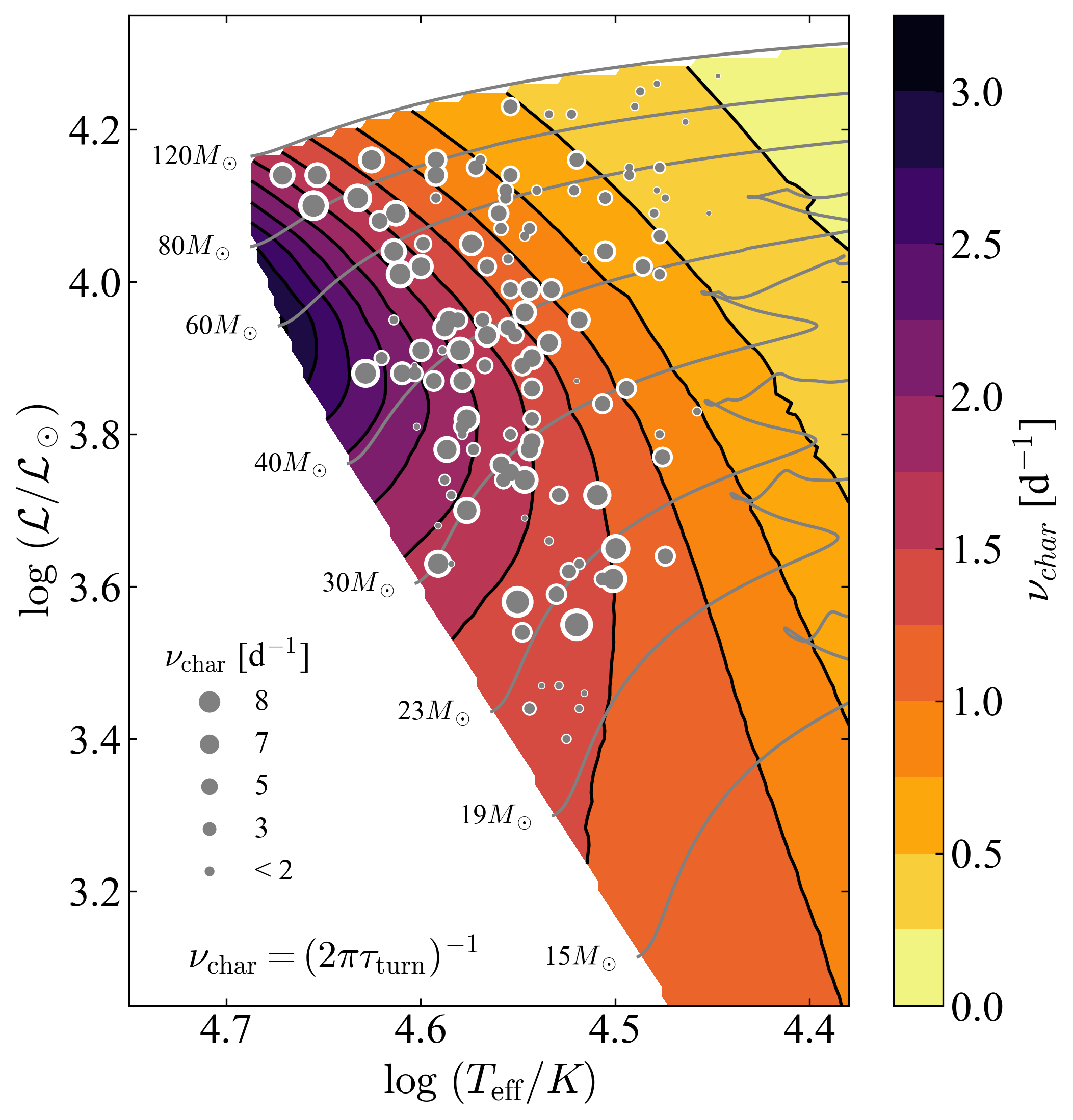}{0.45\textwidth}{}\fig{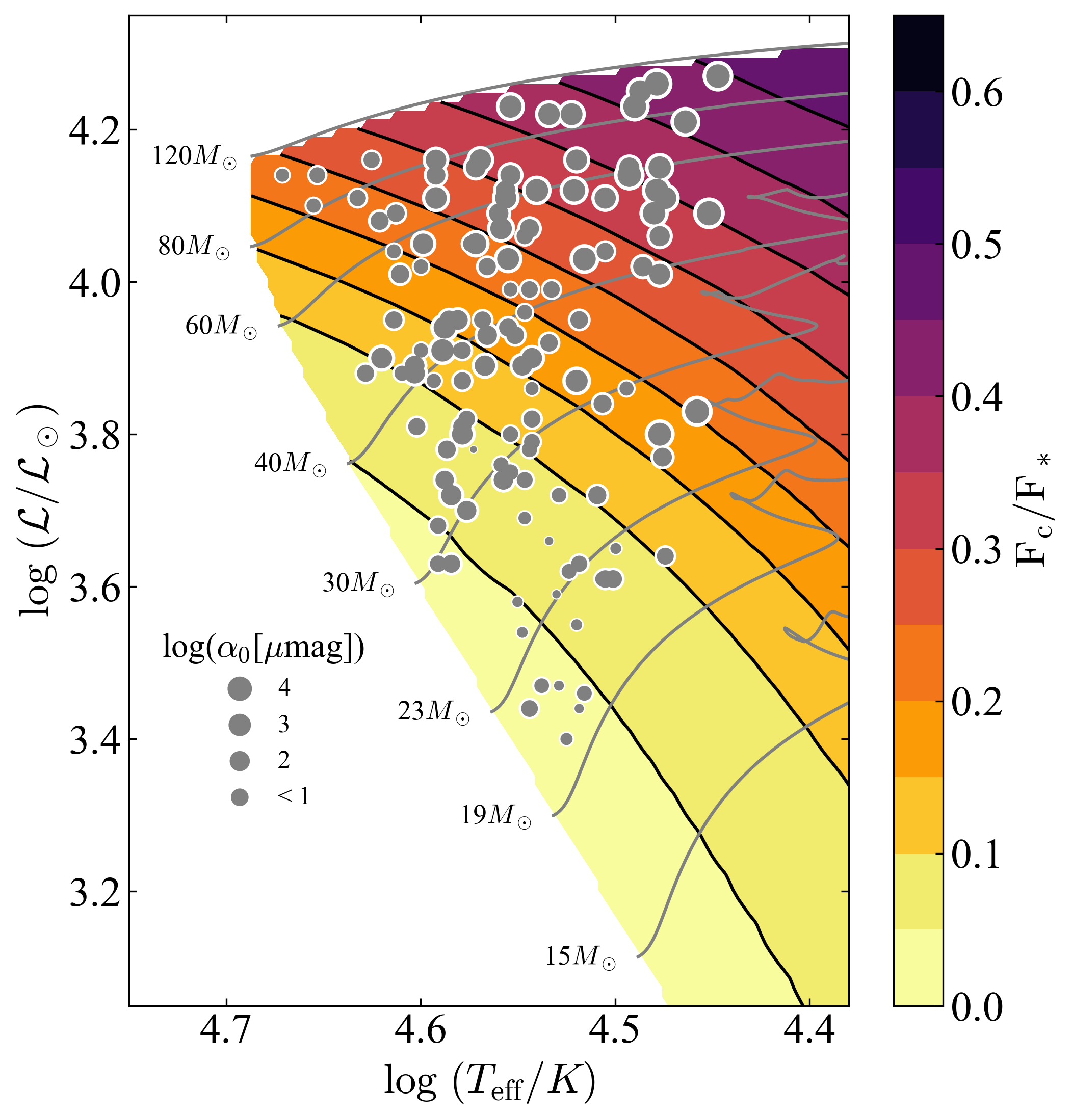}{0.45\textwidth}{}}
      \caption{The locations of stars in our sample are represented as gray circles in the spectroscopic Hertzsprung-Russell diagram. The size of each circle is directly proportional to $\nu_{\mathrm{char}}$ (left panel) and $\log\,(\alpha_0\,[\mu\rm mag])$ (right panel). Evolutionary tracks are depicted as gray solid lines. Additionally, the simulation results of $\nu_{\mathrm{char}}$ and $\log\,(\alpha_0\,[\mu\rm mag])$, computed by \cite{2021ApJ...915..112C}, are illustrated as black contour lines in the spectroscopic Hertzsprung-Russell diagrams.\label{sHRD}}
\end{figure*}

In Figure~\ref{sHRD}, we presented the positions of 130 O-type stars observed through the high-resolution IACOB survey. These stars are denoted by grey circles on the sHRD. The size of each circle is proportional to the characteristic frequency ($\nu_{\rm char}$) in the left panel, and to the logarithmic amplitude ($\log\,(\alpha_0\,[\mu\rm mag])$) in the right panel. To facilitate comparison with theoretical predictions, we also included simulation results from \cite{2021ApJ...915..112C} in Figure~\ref{sHRD}. In the left panel of Figure \ref{sHRD}, black contour lines represent theoretical $\nu_{\mathrm{char}}$ in the FeCZ as a function of stellar model locations. Similarly, in the right panel of Figure~\ref{sHRD}, black contour lines are included to represent the theoretical amplitude as a function of the stellar model locations. For more massive stars ($M \geqslant 40 M_{\odot}$), we observed that more luminous and more evolved stars have lower characteristic frequencies. However, this correlation is less pronounced for stars with masses within the range $30 M_{\odot}\leqslant M \leqslant 40 M_{\odot}$. For stars with lower masses ($M < 30 M_{\odot}$) the relationship becomes more difficult to discern. Our findings are consistent with \cite{2020A&A...640A..36B}, indicating a trend where more massive and more evolved stars have smaller $\nu_{\rm char}$ values. In the second row of Figure~\ref{parameters_slf}, our statistical analysis reveals that $\nu_{\rm char}$ exhibits a decrease with increasing stellar radius for stars with $R \geq 10 R_{\odot}$, which corresponds to stars with $M \geq 30 M_{\odot}$. Conversely, for stars with $R < 10 R_{\odot}$, the relationship between $\nu_{\rm char}$ and stellar radius is not evident. Additionally, Figure~\ref{IGW} indicates that the increase in stellar radius during evolution correlates with a significant reduction in characteristic frequency, an effect particularly evident in more massive stars. The variation trend of $\nu_{\mathrm{char}}$ across different stellar mass for O-type stars, as illustrated in the left panel of Figure~\ref{sHRD}, may correspond to the intricate correlations between $\nu_{\mathrm{char}}$ and stellar radius detailed in Section 4. This intricate trend of SLF variability observed across O-type stars of varying mass ranges potentially corresponds to the underlying evolution of the FeCZ.

\begin{figure*}
\centering
\gridline{\fig{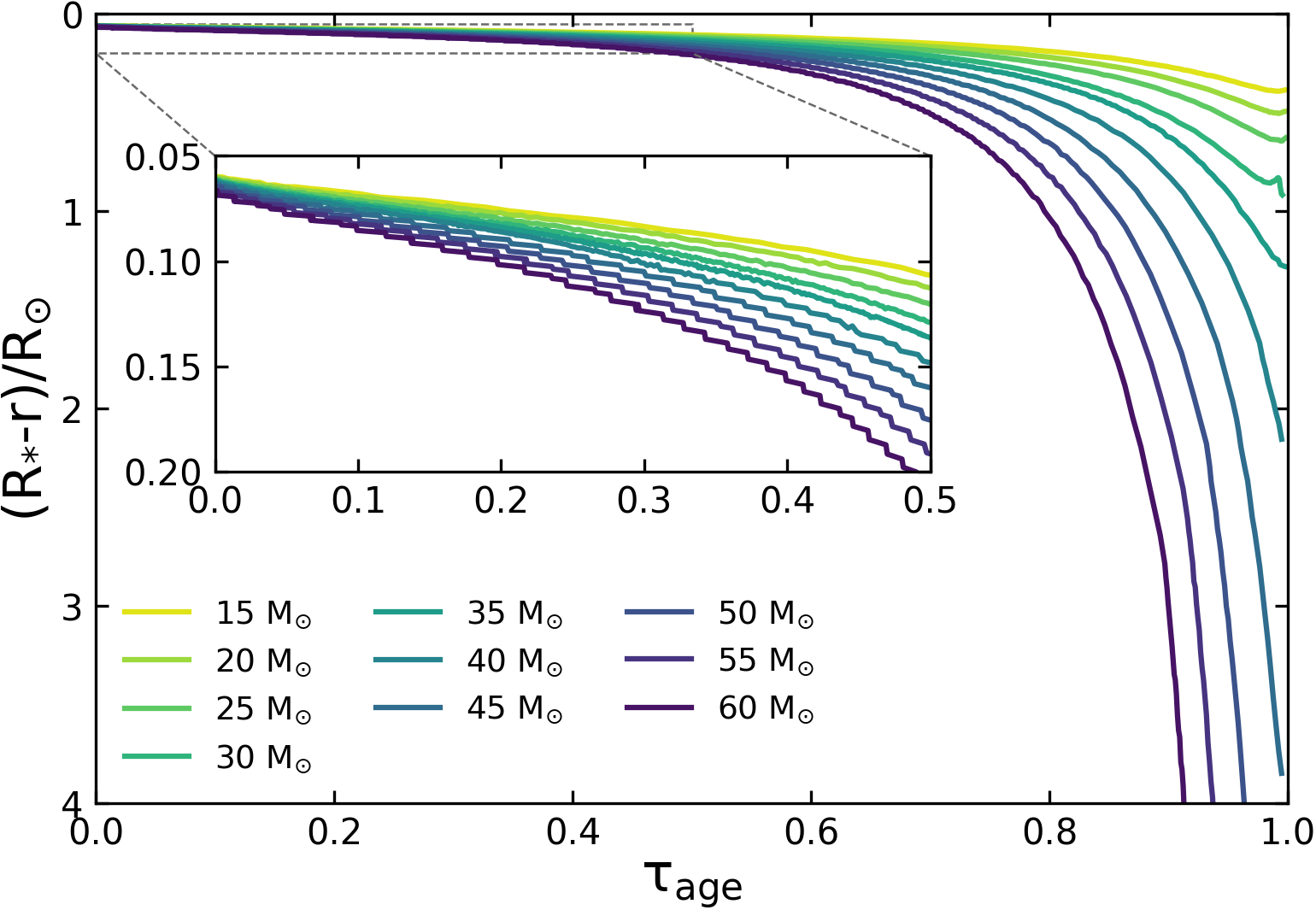}{0.4\textwidth}{}}
\caption{Evolution of the radial locations of the iron convective regions as function of time for stars with different masses. The inset illustrates a magnified view of the radial locations of the iron convective regions for stars transitioning from the zero-age main sequence to the mid-main sequence.\label{FeCZs}}
\end{figure*}

In the context of SLF variability excited by FeCZ, the theoretical characteristic frequency is determined by $(2\pi\tau_{\mathrm{c}})^{-1}$, where $\tau_{\mathrm{c}}$ is the average convective turnover time of FeCZ \citep{2021ApJ...915..112C}. During stellar evolution, the FeCZ moves towards a deeper interior, situated at a radius deeper from the surface of star. This movement results in a larger pressure scale height, higher velocities, and consequently, an increase in the $\tau_{\mathrm{c}}$ values \citep{2009A&A...499..279C}. Therefore, the characteristic frequency of SLF variability is observed to decrease as stellar age increases. We used version 23.05.1 of the Modules for Experiments in Stellar Astrophysics (MESA) code \citep[e.g.,][]{2011ApJS..192....3P, 2013ApJS..208....4P, 2015ApJS..220...15P, 2016ApJS..223...18P, 2018ApJS..234...34P, 2019ApJS..243...10P, 2023ApJS..265...15J} to calculate the evolution of the radial locations of the FeCZs as function of time for stellar evolution models with different masses at solar metallicity. In Figure \ref{FeCZs}, we demonstrated the evolution of radial locations of the FeCZs for stars with masses ranging from $15\,M_{\odot}$ to $60\,M_{\odot}$ in steps of $5\,M_{\odot}$. For the lower mass star models ($M < 30 M_{\odot}$), the FeCZ location moves $0.5 \sim 1 R_{\odot}$ towards the interior of the star from the zero-age main sequence (ZAMS) to terminal-age main sequence (TAMS). In the more massive stars ($M \geq 30 M_{\odot}$), the FeCZ location varies by more than $2 R_{\odot}$, moving significantly deeper into the interior of the star compared to stars with lower mass. As a result, the variation in the parameter $\tau_{\mathrm{c}}$ is pronounced. Therefore, we observed a significant correlation between the $\nu_{\mathrm{char}}$ and radius in stars with $M \geqslant 30 M_{\odot}$. Our analysis indicates the intricate variation trend of $\nu_{\mathrm{char}}$ on stellar radius across different mass O-type stars corresponds to the evolution of FeCZs.

In the right panel of Figure~\ref{sHRD}, the values of $F_{\mathrm{c}}/F_{\ast}$ demonstrate a trend of increasing with increasing log$\,(\mathcal{L/L_{\odot}})$ and decreasing log$\,T_{\mathrm{eff}}$. Based on the simulating results of non-rotating stellar models, theoretical predictions suggest that more massive and more evolved stars have a larger amplitude of SLF variability, which corresponds to the trend exhibited by the stars in sHRD \citep[e.g.,][]{2021ApJ...915..112C, 2022ApJ...924L..11S,2023ApJ...951L..42S}. The right panel of Figure~\ref{sHRD} reinforces a systematic association between SLF variability and FeCZ. On the other hand, our results evidenced significantly correlations between the amplitude of SLF variability and various stellar properties, including mass, radius, and fractional main sequence age, as detailed in Section~\ref{sec4}. These significantly positive relationships further provide evidence for the hypothesis that SLF variability is excited by FeCZ.

\cite{2021ApJ...915..112C} has proposed that the FeCZ not only drives the SLF variability but also induces macroturbulence broadening in massive stars. Based on a sample consisting of 70 stars observed by a high-resolution spectroscopic survey, \cite{2020A&A...640A..36B} has also identified the correlations between $v_{\rm marco}$ and fitting parameters of SLF variability. Our analysis (as demonstrated in Figure~\ref{vmacro_slf}), demonstrates significant positive correlations between both the amplitude and steepness of SLF variability and macroturbulence broadening, suggesting a potential link between these two phenomena. If macroturbulence broadening and SLF variability are indeed excited by the FeCZ, their variation trends with the same stellar parameter should align with the variation of FeCZ. \cite{2017A&A...597A..22S} proposed that macroturbulence broadening serves as the primary mechanism for line broadening in stars with  $M \geqslant 15 M_{\odot}$. Moreover, there exists a clear positive correlation between macroturbulence broadening and log$\,(\mathcal{L/L_{\odot}})$ for stars with log$\,(\mathcal{L/L_{\odot}})$ $\geqslant 3.5$. In Figure \ref{vmacro_relation}, we presented the dependence of fitting parameters of SLF variability on log$\,(\mathcal{L/L_{\odot}})$ (top panels) and $T_{\mathrm{eff}}$ (bottom panels). The top panels of Figure \ref{vmacro_relation} reveal a clear positive correlation between $\log\,(\alpha_0\,[\mu\rm mag])$ and log$\,(\mathcal{L/L_{\odot}})$ for the stars in our sample, consistent with the trend observed for macroturbulence broadening with log$\,(\mathcal{L/L_{\odot}})$ reported by \cite{2017A&A...597A..22S}. The significant positive correlations between log$\,(\mathcal{L/L_{\odot}})$ and both macroturbulence broadening and the amplitude of SLF variability are consistent with the theoretical predictions derived by \cite{2021ApJ...915..112C}. However, no significant relationships, analogous to those observed between macroturbulence broadening and log$\,(\mathcal{L/L_{\odot}})$, are found between the remaining fitting parameters ($\nu_{\mathrm{char}}, \gamma$) and log$\,(\mathcal{L/L_{\odot}})$. Our findings show significant correlations between log\,$T_{\mathrm{eff}}$ and both $\log\,(\alpha_0\,[\mu\rm mag])$ and log\,$\nu_{\mathrm{char}}$. Nevertheless, due to the limited temperature range of our sample, there are statistical constraints on exploring the relationships between the fitting parameters of SLF variability and log\,$T_{\mathrm{eff}}$.

It should be noted that although the simulations successfully replicate the associated trend of SLF variability in the sHRD, the $\nu_{\mathrm{char}}$ values derived from our sample appear to be approximately three times larger than those derived from the theoretical results. The study by \cite{2021ApJ...915..112C} utilized a simplified parameter of the Mixing Length Theory ($\alpha_{\mathrm{MLT}}$) to determine $\tau_{\rm c}$, potentially overlooking the intricate phenomenology of the FeCZ. Although this simplification effectively captures the correlations between SLF variability and stellar properties, the significant discrepancy in the absolute values between observations and simulations suggests that there may be additional factors influencing the SLF variability. These results prompt that SLF variability could be influenced by different dominant mechanisms at various stages of stellar evolution \citep{2023Ap&SS.368..107B}. On the other hand, current multidimensional numerical simulations are primarily limited to non-rotating early-type stars, however, OB stars typically exhibit rapid rotation \citep[e.g.,][]{2000ARA&A..38..143M, 2013A&A...550A.109D, 2013A&A...560A..29R}. When the rotational period becomes comparable to or shorter than the convective turnover time, the properties of convection can be significantly altered \citep{1979GApFD..12..139S, 2019ApJ...874...83A}. Notably, at the highest rotation rates, the latitudinal structure of the FeCZ zone is substantially modified \citep{2008A&A...479L..37M}, potentially affecting how these regions influence the stellar surface. Future research that incorporates samples with varying rotation speeds could enhance our understanding of the role of the FeCZ in SLF variability. Additionally, the significantly correlation between amplitude of SLF variability and mass-loss rate indicates that other surface phenomena, such as clump size of stellar winds, should also be taken into consideration \citep[e.g.,][]{2022ApJ...925...79L}.

\subsection{Internal gravity waves}

Internal gravity waves are thought to be primarily generated by Reynolds stresses in the form of eddies or plume penetration, propagating within stably stratified regions where buoyancy serves as the dominant restoring force \citep{2000A&A...354..943M, 2013ApJ...772...21R, 2015ApJ...815L..30R, 2019ApJ...876....4E, 2023MNRAS.525.1601H, 2024MNRAS.531.1316T}. Due to their unique properties, IGWs have been proposed as a potential solution for several longstanding issues, including chemical mixing \citep{2000A&A...354..943M, 2017ApJ...848L...1R, 2023ApJ...942...53V, 2024ApJ...970..104V} and angular momentum transport \citep[e.g.,][]{2000AIPC..537..256K, 2008A&A...482..597T,2013ApJ...772...21R, 2014ApJ...796...17F}. \cite{2010Ap&SS.328..253S} investigated the possibility that IGWs can be stochastically excited by turbulent convection in massive stars, utilizing stellar models of main sequence stars with masses $M = 10\,M_{\odot}, 15\,M_{\odot},$ and 20$\,M_{\odot}$. Utilizing one-dimensional (1D) non-rotating models with solar metallicity across a range of initial masses from 2 to 30\,$M_{\odot}$, \cite{2013MNRAS.430.1736S} predicted that the convective cores of stars with masses $M \geq 2\,M_{\odot}$ will induce observable surface brightness fluctuations while on the main sequence. Building on the simulations conducted by \cite{2013ApJ...772...21R}, \cite{2015ApJ...806L..33A} compared velocity spectra derived from two-dimensional hydrodynamical simulations of IGWs in a differentially rotating massive star with observed spectra. They suggested that signatures of IGWs may manifest in the residual light curves that have been prewhitened for the detected heat-driven mode frequencies associated with OB stars. By simulating five different stellar masses (3\,$M_{\odot}$, 5\,$M_{\odot}$, 7\,$M_{\odot}$, 10\,$M_{\odot}$, and 13\,$M_{\odot}$) at the middle of the main sequence, \cite{2023A&A...674A.134R} demonstrated that surface perturbations caused by IGWs are more pronounced in stars with higher masses. However, no discernible trends were observed in the frequency slopes across varying stellar masses. \cite{2023ApJ...954..171V} performed 3D RHD simulations, reaching up to 90\% of the total stellar radius, for three 7 $M_{\odot}$ stars of different ages to investigate how the propagation of IGWs significantly depends on the evolutionary stages of stars. They found that the steepness of the frequency spectrum at the surface increases from ZAMS to the older models, with older stars also showing more modes in their spectra. Moreover, the results of numerical simulations indicate that the steepness of SLF variability induced by IGWs excited through plume penetration differs significantly from that induced by IGWs excited through eddies \citep[e.g.,][]{2013ApJ...772...21R, 2013MNRAS.430.2363L, 2016A&A...588A.122P, 2019ApJ...876....4E, 2020A&A...641A..18H, 2023A&A...674A.134R}.

In Figure~\ref{vmacro_slf}, our analysis reveals significant correlations between $v_{\rm macro}$ with both the amplitude and steepness of SLF variability. As depicted in the top panels of Figure~\ref{vmacro_relation}, the amplitude of SLF variability exhibits a trend analogous to that of macroturbulence broadening for log$\,(\mathcal{L/L_{\odot}})$. These correlations suggest a potential connection between the SLF variability and $v_{\rm macro}$. Building upon the insights gained from 2D simulations of velocity spectra induced by IGWs, \textcite{2015ApJ...806L..33A} proposed that these IGWs may be responsible for detectable time-dependent line broadening in OB stars. Additionally, they suggested that IGWs could also account for the SLF variability observed in these stars. Drawing from non-rotating models across a range of initial masses from 2 to 30\,$M_{\odot}$ at various evolutionary stages, \cite{2013MNRAS.430.1736S} concluded that the characteristic frequency of SLF variability decreases with stellar evolution. Similarly, \cite{2023ApJ...954..171V}, through 3D RHD simulations involving 7 $M_{\odot}$ of stars at different ages, observed that the characteristic frequency decays with the stellar evolution. As illustrated in Figure~\ref{parameters_slf}, the $\nu_{\mathrm{char}}$ decreases as the stellar radius increases for the stars with $R \gtrsim 10\;R_{\odot}$ (corresponding the stars with $M \gtrsim 30\,M_{\odot}$), while no significant correlation is observed for the stars with $R \lesssim 10 R_{\odot}$. Figure~\ref{IGW} further elucidates that stellar evolution results in substantial radius expansion, culminating in a reduced characteristic frequency of SLF variability. The left panel of Figure~\ref{sHRD} corroborates this, showing that more massive and evolved stars exhibit lower $\nu_{\rm char}$ values. The observed trend in the variation of $\nu_{\rm char}$ within our sample is consistent with theoretical predictions from multidimensional simulations for IGWs \citep[e.g.,][]{2013MNRAS.430.1736S, 2023ApJ...954..171V}.

\begin{figure}
    \centering
    \gridline{\fig{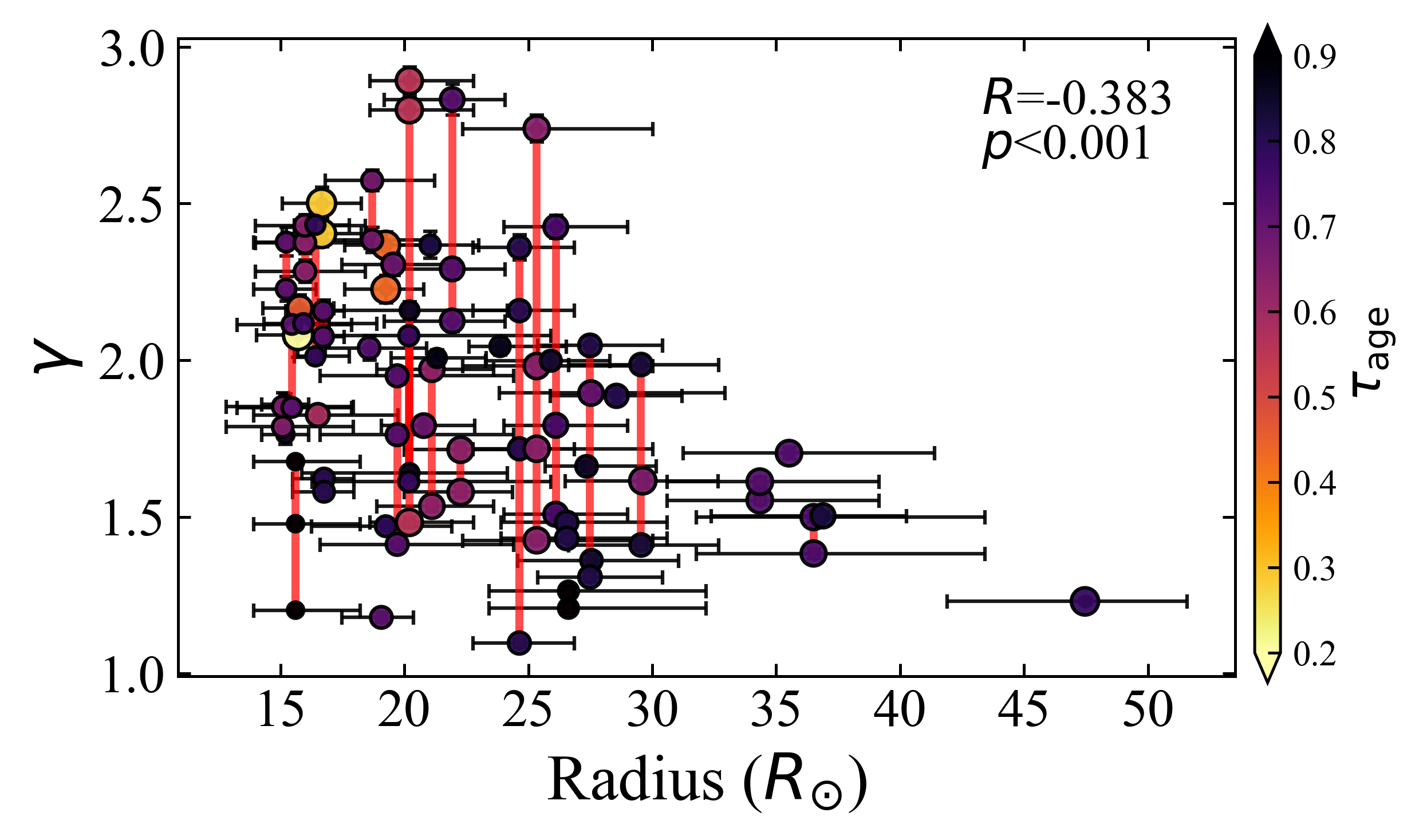}{0.5\textwidth}{}}
    \caption{The significant correlation between the $\gamma$ and stellar radius for stars with $R \gtrsim 15 R_{\odot}$. The markers have the same meaning as in Figure~\ref{IGW}. \label{IGW1}}
\end{figure}

In multidimensional simulations conducted by previously studies, the spectra (velocity, temperature, or kinetic energy spectra) of IGWs generated by various mechanisms and physical conditions exhibit differing steepness \citep[e.g.,][]{2013ApJ...772...21R, 2013MNRAS.430.2363L, 2016A&A...588A.122P, 2019ApJ...876....4E, 2020A&A...641A..18H, 2023A&A...674A.134R}. Using 1D stellar evolution codes, \cite{2016A&A...588A.122P} investigated the IGWs generated by penetrative plumes below the upper convective envelope and examined the influence of the steepness of the thermal adjustment layer on wave transmission into the radiative zone. \cite{2019ApJ...876....4E} conducted 3D simulations that spanned from 1\% to 90\% of the stellar radius for a 3\,$M_{\odot}$ star. Their findings indicated that IGWs excited by plumes produce spectra with $\gamma$ values ranging from 0 to 3 at low harmonic degrees. Furthermore, \cite{2023A&A...674A.134R} performed 3D simulations for five different stellar masses at the middle of the main sequence. Their results revealed that IGWs yield spectra with $\gamma$ values between 0 and 2, which align well with theoretical predictions reported in previous works \citep[e.g.,][]{2016A&A...588A.122P, 2019ApJ...876....4E, 2020A&A...641A..18H}.
Conversely, IGWs excited by eddies penetration would generate larger steepness in amplitude spectra \citep{1999ApJ...520..859K, 2013MNRAS.430.2363L}. As shown in Figure~\ref{slf_result}, the steepness derived from the amplitude spectra falls within the range of 0.5 to 3, aligning more closely with the hypothesis that SLF variability is driven by IGWs induced by plume penetration. \cite{2023A&A...674A.134R} conducted simulations of IGW generation and propagation across stars of varying masses and concluded that no discernible trends are observed in frequency slopes with respect to different stellar masses. Figure~\ref{parameters_slf} reveals a uniform distribution of steepness of SLF variability in our sample within the 0.5 to 3 range across stars of different masses, corroborating the simulation results provided by \cite{2023A&A...674A.134R}. In the second row of Figure~\ref{parameters_slf}, the spectral index $\gamma$ exhibits a trend similar to that of the characteristic frequency $\nu_{\rm char}$ with respect to stellar radius. In Figure~\ref{IGW1}, we presented the relationship between $\gamma$ and stellar radius for stars with $R \gtrsim 15 R_{\odot}$. The expansion in stellar radius due to stellar evolution leads to a decrease in $\gamma$, implying that the influence of IGWs on SLF variability may weaken as stars age, which is consistent with the simulating predictions published by \cite{2023ApJ...954..171V}.

Simulations of SLF variability driven by IGWs in stars of varying masses indicate that more massive stars tend to exhibit greater amplitudes of SLF variability \citep[e.g.,][]{2013MNRAS.430.1736S, 2015ApJ...806L..33A, 2023A&A...674A.134R}. In Figure \ref{parameters_slf}, our analysis reveals a significant positive correlation between the $\log\,(\alpha_0\,[\mu\rm mag])$ and stellar mass, which suggests the amplitude of SLF variability increase with the increase of mass of star. This finding aligns with theoretical predictions from prior studies \citep[e.g.,][]{2013MNRAS.430.1736S, 2015ApJ...806L..33A}. Based on the results of 2D simulations for intermediate-mass stars, \cite{2020MNRAS.497.4231R} concluded that the amplitude of SLF variability is decrease as the star ages. In 3D RHD simulations, the propagation of IGWs exhibits greater sensitivity to stellar radius. The depletion of IGWs in the upper region of the radiative zone in older stars is attributed to a combination of various separate effects, with the most significant being the evanescence of IGWs as they propagate towards the stellar surface \citep{2023ApJ...954..171V}. IGWs only exhibit wave-like characteristics within turning radius \citep{2019ApJ...876....4E, 2020MNRAS.497.4231R}. As the age of the star increases, the turning radius migrates towards the interior of the star, resulting in the increased depletion of IGWs within the radiative zone, and ultimately led to the amplitudes of IGWs decrease with stellar evolution \citep{2023ApJ...954..171V}. Contrary to these predictions, as illustrated in Figure~\ref{parameters_slf}, our statistical findings indicate a positive correlation between the amplitude of SLF variability, $\log\,(\alpha_0\,[\mu\rm mag])$, and $\tau_{\mathrm{age}}$. The right panel of Figure~\ref{sHRD} further corroborates that more massive and more evolved stars possess larger amplitudes of SLF variability. Our results imply that the role of IGWs in shaping SLF variability is more pronounced in the initial stages of stellar evolution, with their effect on SLF variability potentially decreasing as the star ages.

\subsection{Stellar winds and granulation}
The stellar winds of hot stars are primarily propelled by radiative acceleration in the spectral lines of elements such as C, O, Si, and Fe \citep{1970ApJ...159..879L, 1975ApJ...195..157C, 2008A&ARv..16..209P}. The perturbations stemming from the instability of the line-driven wind propagate into reverse shocks, furnishing an explanation for the pervasive X-ray emission observed in hot stars \citep{1988ApJ...335..914O, 1997A&A...322..878F}. Furthermore, the effect of wind blanketing results in a small fraction of the radiative flux emitted by hot stars being absorbed by their winds and redistributed towards longer wavelengths \citep{1985ApJ...294..286A, 2002ApJ...579..774C, 2021A&A...648A..79K}, providing another mechanism that can cause SLF variability in massive stars \citep{1997A&A...322..878F,2018A&A...617A.121K, 2021A&A...648A..79K}. \cite{2021A&A...648A..79K} applied two types of boundary perturbations to evaluate the photometric variations in light curve of massive stars, and pointed that the SLF variability of OB stars bears certain signatures of the line-driven wind instability. 

The simulations suggest that SLF variability becomes stronger with increasing mass-loss rates, indicating that larger values of $\log\,(\alpha_0\,[\mu\rm mag])$, $\nu_{\mathrm{char}}$, and $\gamma$ will be observed for stars as their mass-loss rates increase. In Figure \ref{parameters_slf}, the significant positive trend ($R = 0.463;\; p < 0.001$) between the amplitude of SLF variability and the mass-loss rate is consistent with the theoretical predictions published by \cite{2021A&A...648A..79K}. However, there is no discernible rise in the $\nu_{\mathrm{char}}$ values with the rising mass-loss rate (shown in Figure \ref{parameters_slf}). Furthermore, we only observed a marginal positive correlation between $\gamma$ and mass-loss rate ($R = 0.148;\; p = 0.01$). The study conducted by \cite{2022ApJ...925...79L} investigated the correlations between SLF variability and winds of Galactic Wolf-Rayet stars. It discerned significant associations between $\nu_{\mathrm{char}}$ and the terminal velocity, as well as the effective temperature for WN stars. In comparison with the investigation by \cite{2022ApJ...925...79L}, our findings solely confirmed the relationship between the amplitude and the mass-loss rate. The absence of a discernible trend in the associations between $\nu_{\mathrm{char}}$ and mass-loss rate in our statistical findings may suggest that the influence of stellar winds on the SLF variability of O-type stars could be relatively less significant than that of Wolf-Rayet stars.
   
   \begin{figure}
      \gridline{\fig{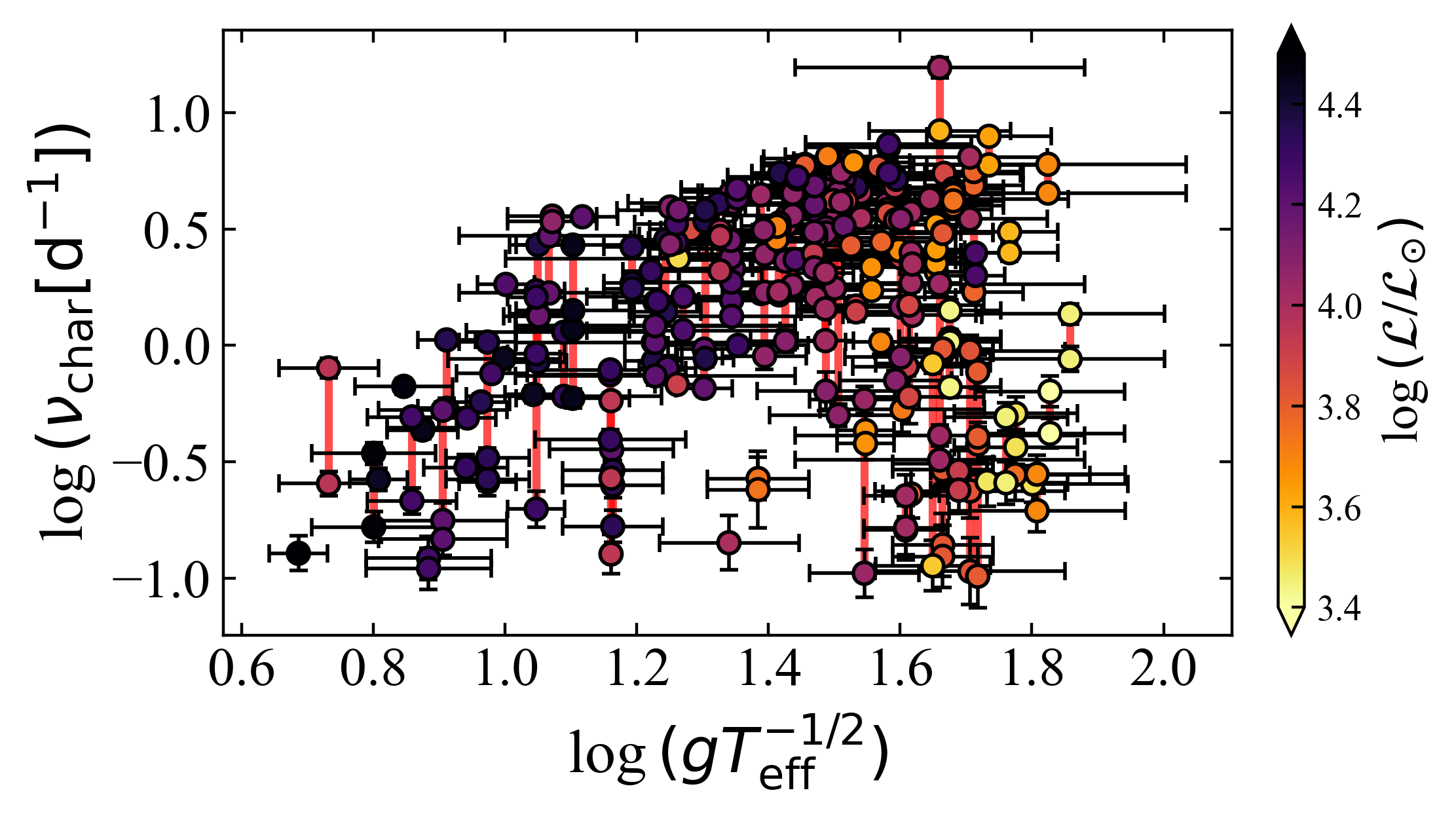}{0.45\textwidth}{}}
      \gridline{\fig{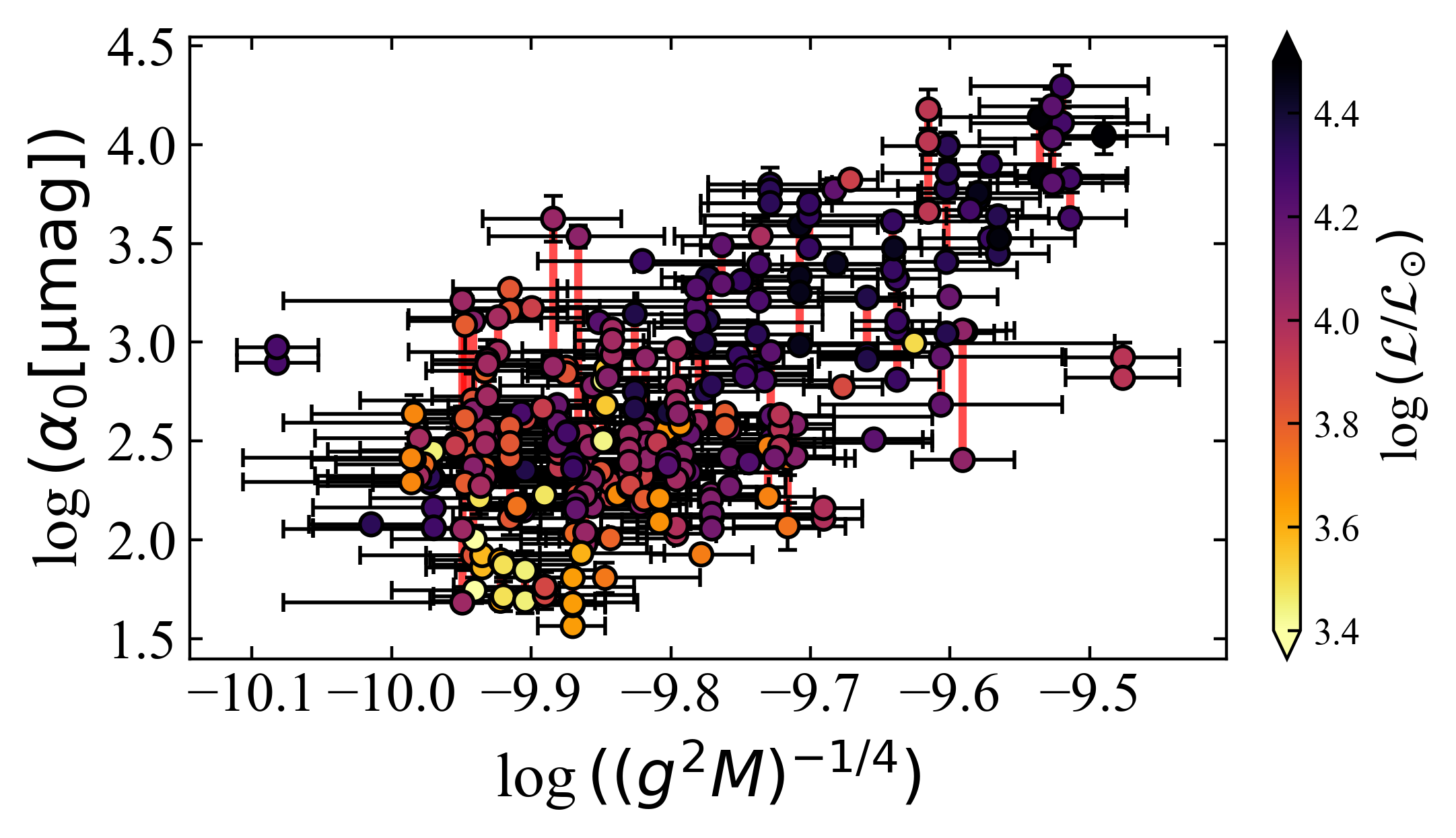}{0.45\textwidth}{}}
      \caption{The scaling relation between characteristic frequencies ($\nu_{\mathrm{char}}$) as a function of $g \cdot T_{\mathrm{eff}}^{-1/2}$ (top panel), and the scaling relation between the amplitude and $(g^2M)^{-1/4}$ (bottom panel) are illustrated. The markers in the each panel have the same meaning as in Figure \ref{parameters_slf}. \label{scaling_relation}}
  \end{figure}

As demonstrated by previous studies, a strong correlation exists between granulation background signal and the stellar parameters for thousands of pulsating solar-type and red giant stars \citep{2011ApJ...741..119M, 2014A&A...570A..41K}. This suggests that the physics of surface convection in the Sun can be appropriately scaled to stars on the red giant branch. Furthermore, \cite{2010ApJ...711L..35K} documented the SLF variability in A-type stars attributed to granulation. Using the CoRoT light curves for a sample of O, B, A, and F stars, \cite{2019A&A...621A.135B} parameterized the morphology of SLF variability and inferred the origin responsible for this variability in photometric observations. Additionally, they explored the scaling relationship between granulation and stellar parameters within their sample and concluded that no significant scaling relationship exists for SLF variability in OB stars. It is worth noting that this conclusion may be influenced by the limited sample size, as the study by \cite{2019A&A...621A.135B} includes only 15 OB stars in their sample. If the scaling relationship between granulation and stellar parameters holds valid for the subsurface convection zones in high-mass stars, one would reasonably expect a comparably systematic and robust correlation between the measured parameters of SLF variability and the stellar parameters. 

In order to determine whether the SLF variability bears certain signatures of the granulation background signal, and can scale with stellar parameters in high-mass stars, we applied the scaling relations previously identified in solar-like stars to measure the correlations between the SLF variability and stellar parameters:
   \begin{equation}\label{eq7}
       \nu_{\mathrm{gran}} \propto g \cdot T^{-1/2}_{\mathrm{eff}}
   \end{equation}
   \begin{equation}\label{eq8}
      A_{\mathrm{gran}} \propto (g^{2}M)^{-1/4} 
    \end{equation}
where $g$ is the surface gravity, $T_{\mathrm{eff}}$ is the effective temperature, and $M$ is the stellar mass. Moreover, $\nu_{\mathrm{gran}}$ and $A_{\mathrm{gran}}$ respectively denote the amplitude and characteristic frequency of granulation background signal. In the top panel of Figure \ref{scaling_relation}, we depicted $\nu_{\mathrm{char}}$ as a function of $g \cdot T_{\mathrm{eff}}^{-1/2}$. The stars are represented as circles and are color-coded based on their spectroscopic luminosities. We observed that $\nu_{\mathrm{char}}$ increases as $g \cdot T^{-1/2}_{\mathrm{eff}}$ rises for stars where $g \cdot T^{-1/2}_{\mathrm{eff}}$ is less than $10^{1.6}$. In the bottom panel of Figure \ref{scaling_relation}, we presented the scaling relation between $\alpha_{0}$ and $(g^2M)^{-1/4}$. A notable positive trend is evident between $\alpha_{0}$ and $(g^2M)^{-1/4}$, indicating that granulation may contribute to the amplitude of SLF variability. Our results suggest that SLF variability may indeed exhibit scaling relations with stellar parameters. However, it is worth noting that the specific form of the scaling relations and their inconsistency with the scaling relations between granulation background signals and stellar parameters need to be carefully considered.

\section{Conclusions}
In this paper, we compiled a sample of 150 O-type stars observed via ground-based spectroscopic surveys by the IACOB and GOSSS projects, supplemented with high-precision photometric data from the TESS mission. Employing the methodology developed by \cite{2019A&A...621A.135B}, we investigated the SLF variability of the stars in our sample by analyzing approximately 300 light curves derived from the short cadence photometric data of TESS sectors 1-65. This process initially entailed removing significant frequencies associated with rotational modulation and/or coherent pulsation modes through traditional iterative pre-whitening. Subsequently, we fitted the residual amplitude spectra using a Bayesian Markov Chain Monte Carlo framework to determine the fitting parameters of SLF variability, including amplitude, characteristic frequency, steepness, and white-noise term. Our sample of stars and their determined fitting parameters are provided in Table \ref{table:stellar_parameters}. Our analysis confirms the pervasive nature of SLF variability in massive stars, consistent with previous studies \citep{2011A&A...533A...4B,2015MNRAS.453...89B, 2019A&A...621A.135B, 2020A&A...640A..36B,2023ApJ...955..123S}.

To improve the homogeneity of our sample, we elected to focus solely on the 130 stars that were observed with high-resolution as the subsample for exploring the correlation between stellar properties and SLF variability. Drawing upon the spectroscopic parameters obtained from high-resolution spectroscopy, we employed the Bayesian tool Bonnsai to determine the stellar parameters, including mass, radius, fractional main sequence age, and mass-loss rate. The Spearman's rank correlation coefficient ($R$) and the corresponding two-sided $p$-value are adopted to evaluate the correlations between the fitting parameters of SLF variability and stellar parameters. In Figure \ref{vmacro_relation}, we observed a significant correlation between $\log\,(\alpha_0\,[\mu\rm mag])$ and log$\,(\mathcal{L/L_{\odot}})$. The amplitude of SLF variability exhibits a moderately significant correlation with $T_{\mathrm{eff}}$. $\nu_{\mathrm{char}}$ also shows a moderate correlation with $T_{\mathrm{eff}}$, while no relationship is detected between the $\gamma$ and $T_{\mathrm{eff}}$. In Figure \ref{parameters_slf}, our statistical results reveal significant correlations between the amplitude of SLF variability and various stellar parameters. The amplitude shows a positive correlation with stellar mass, with a $R$ value of 0.448 and a $p$-value less than 0.001. Furthermore, we observed a positive correlation between the amplitude and stellar radius, with a $R$ value of 0.6 and a $p$-value less than 0.001. Additionally, the amplitude exhibits positive correlations with the fractional main sequence age and mass-loss rate. We observed that the characteristic frequency, $\nu_{\mathrm{char}}$, decreases with increasing radius for stars with $R \gtrsim 10\;R_{\odot} $. Furthermore, no significant correlations were observed between $\nu_{\mathrm{char}}$ and other stellar parameters, including mass, fractional main sequence age, and mass-loss rate. Similarly, $\gamma$ only exhibits correlations with the stellar radius for stars with $R \gtrsim 15\;R_{\odot}$. The utilization of samples composed of massive stars from open clusters in future studies may aid in further developing the determination of the relationship between fitting parameters of SLF variability and stellar properties.

As the origin of SLF variability is currently under debate, we compared our statistical results with various potential excitation mechanisms. The stars of our sample are placed in Figure \ref{sHRD}. We found that the variations in SLF variability for stars with $M \geqslant 30\;M_{\odot}$, are consistent with the characteristics that excitation by FeCZ. We identified a clear positive correlation between $\log\,(\alpha_0\,[\mu\rm mag])$ and log$\,(\mathcal{L/L_{\odot}})$, consistent with the observed trend for macroturbulence broadening with the log$\,(\mathcal{L/L_{\odot}})$. No significant relationships, similar to those observed between macroturbulence broadening and the log$\,(\mathcal{L/L_{\odot}})$, are apparent between the remaining fitting parameters ($\nu_{\mathrm{char}}, \gamma$) and log$\,(\mathcal{L/L_{\odot}})$. The variation in the characteristic frequency of SLF variability for stars with $M \geq 30\,M_{\odot}$ also corroborates the variations of the FeCZ. Although the simulations successfully replicate the trend of SLF variability in the sHRD, the $\nu_{\mathrm{char}}$ values derived from our sample are approximately three times larger than those from theoretical results. We also compared our statistical results with predictions from IGWs simulations. Our results demonstrate a positive correlation between the amplitude of SLF variability and stellar mass, which is consistent with the characteristics of SLF variability excited by IGWs in multidimensional simulations. The variations in characteristic frequency ($\nu_{\mathrm{char}}$) and steepness ($\gamma$) are also consistent with theoretical predictions. However, the evolutionary trend of amplitude with age is inconsistent with the projections of 3D simulations. Our findings suggest that considering the impact of stellar winds on SLF variability is crucial, as we identified a significant correlation between amplitude and mass-loss rate. We observed a scaling correlation between the amplitude of SLF fluctuations and stellar parameters. This suggests that SLF variability, similar to granulation, exhibits scaling behavior with respect to stellar parameters. However, further exploration is required to determine the specific nature of the relationships between the fitting parameters of SLF variability and stellar parameters.

Finally, our findings indicate that the FeCZ hypothesis effectively captures the general trend of SLF variability, albeit with some discrepancies in the precise prediction of absolute values. Statistical correlations within our sample align with the predictions of the IGWs hypothesis. However, the predictions of the IGW hypothesis regarding the amplitude variations of SLF variability with respect to stellar age contradict our results. This discrepancy may imply that SLF variability in stars is a result of the combined excitation by both the FeCZ and IGWs mechanisms, with IGWs potentially dominating the variability in the initial phases of stellar evolution, while the influence of FeCZ becomes more pronounced as stars progress through their lifecycle. Further expansion of our sample is necessary to obtain more definitive evidence. It is highly plausible that each process may dominate at different evolutionary stages of stars, directly contributing to photometric variability or indirectly influencing it. On the other hand, our sample provided more stars for numerical simulation work similar to \cite{2022ApJ...924L..11S, 2023ApJ...951L..42S}. In future investigations, we will analyze the SLF variability of massive stars within the Galaxy, the Large Magellanic Cloud, and the Small Magellanic Cloud. This will enable us to unravel the distinctive features of SLF variability across stars of varying masses, metallicities, and evolutionary stages.


\textbf{Acknowledgments:} This work received the generous support of the National Natural Science Foundation of China under grants U2031204,
12163005, 12373038, and 12288102; the Natural Science Foundation of Xinjiang Nos. 2022D01D85 and 2022TSYCLJ0006; the science research grants from the China Manned Space Project with No. CMS-CSST-2021-A10. This paper includes data collected with the TESS mission, obtained from the MAST data archive at the Space Telescope Science Institute (STScI). All 2-minute cadence TessTargetPixelFiles utilized in the paper can be accessed via:\dataset[10.17909/t9-yk4w-zc73]{\doi{10.17909/t9-yk4w-zc73}}. Funding for the TESS mission is provided by the NASA Explorer Program. STScI is operated by the Association of Universities for Research in Astronomy, Inc., under NASA contract NAS 5-26555. This research has made use of the SIMBAD database, operated at CDS, Strasbourg, France.

$Facility:$ TESS

$Software:$ Python, Astropy \citep{2013A&A...558A..33A, 2018AJ....156..123A, 2022ApJ...935..167A}, astroquery \citep{2019AJ....157...98G}, Lightkurve \citep{2018ascl.soft12013L}, matplotlib \citep{2007CSE.....9...90H}, Numpy \citep{2020Natur.585..357H}, Scipy \citep{2020NatMe..17..261V}, Period04 \citep{2005CoAst.146...53L}, Bonnsai \citep{2014A&A...570A..66S}.

\appendices
\renewcommand\thefigure{\Alph{section}\arabic{figure}}
\section[]{Appendix A}
\startlongtable
\begin{longrotatetable}
\begin{deluxetable*}{lccccclccccc}
\tablecaption{\label{table:stellar_parameters}The stellar parameters of stars in our sample.}
    \tablehead{
    \colhead{Name} & \colhead{Sp.type} & \colhead{LumClass} & 
    \colhead{log\,$\mathcal{L/L_{\odot}}$} & \colhead{$T_{\rm eff}$} & \colhead{log\,$g$}& \colhead{Mass} & \colhead{Radius} & \colhead{$\tau_{\mathrm{age}}$} & \colhead{log\,$\dot{M}$} & \colhead{v$_{\mathrm macro}$} & \colhead{$v$sin$i$} \\
    \colhead{} & \colhead{} & \colhead{} & 
    \colhead{} & \colhead{(k$K$)} & \colhead{}& \colhead{($M_{\odot}$)} & \colhead{$(R_{\odot}$)} & \colhead{} & \colhead{($M_{\odot}$yr$^{-1}$)} & \colhead{(km s$^{-1}$)} & \colhead{(km s$^{-1}$)} 
    }
\startdata
$^1$BD +36 4145& O8.5& V (n)& $3.75 \pm 0.16$& $35.80 \pm 0.90$& $3.86 \pm 0.15$& $25.40^{+0.78}_{-2.36}$& $9.11^{+0.84}_{-1.51}$& $0.58^{+0.09}_{-0.12}$& $-6.58^{+0.32}_{-0.27}$& 203.0&   \\
$^1$BD +62 2078& O7& V ((f))z& $3.74 \pm 0.14$& $38.70 \pm 0.80$& $4.00 \pm 0.14$& $29.40^{+0.10}_{-2.44}$& $8.53^{+0.26}_{-1.25}$& $0.43^{+0.10}_{-0.16}$& $-6.49^{+0.22}_{-0.28}$& 38.0& 38.0 \\
$^1$CPD -47 2963& O5& I fc& $4.16 \pm 0.05$& $37.10 \pm 0.50$& $3.51 \pm 0.04$& $59.60^{+0.78}_{-8.13}$& $22.27^{+0.09}_{-2.28}$& $0.63^{+0.03}_{-0.03}$& $-5.30^{+0.14}_{-0.11}$& 67.0& 110.0 \\
$^1$HD 101190& O6& IV ((f))& $3.91 \pm 0.05$& $39.80 \pm 0.50$& $3.88 \pm 0.04$& $38.60^{+0.67}_{-2.06}$& $11.87^{+0.64}_{-0.83}$& $0.51^{+0.02}_{-0.03}$& $-5.96^{+0.11}_{-0.09}$& 49.0& 74.0 \\
$^1$HD 101223& O8& V& $3.96 \pm 0.06$& $35.20 \pm 0.50$& $3.62 \pm 0.05$& $35.20^{+0.27}_{-2.41}$& $15.23^{+0.22}_{-1.30}$& $0.70^{+0.02}_{-0.01}$& $-5.95^{+0.12}_{-0.14}$& 55.0& 63.0 \\
$^1$HD 135591& O8& IV ((f))& $3.99 \pm 0.05$& $35.00 \pm 0.50$& $3.58 \pm 0.04$& $37.40^{+0.04}_{-2.57}$& $16.73^{+0.84}_{-1.52}$& $0.71^{+0.01}_{-0.02}$& $-5.87^{+0.12}_{-0.10}$& 60.0& 60.0 \\
$^1$HD 14434& O5.5& IV nn(f)p& $3.78 \pm 0.12$& $38.60 \pm 1.10$& $3.96 \pm 0.11$& $25.00^{+0.29}_{-1.13}$& $7.65^{+0.88}_{-0.33}$& $0.56^{+0.10}_{-0.21}$& $-6.46^{+0.13}_{-0.17}$& 417.0&   \\
$^1$HD 149038& O9.7& Iab& $4.11 \pm 0.05$& $29.80 \pm 0.50$& $3.18 \pm 0.04$& $42.40^{+0.33}_{-3.72}$& $27.35^{+0.81}_{-1.66}$& $0.81^{+0.02}_{-0.02}$& $-5.69^{+0.07}_{-0.20}$& 52.0& 90.0 \\
\enddata
\tablecomments{The physical parameters of stars in our sample. The first column lists the names of the stars. The superscript $^1$ in the star names indicates stars observed by high-resolution spectroscopic survey IACOB, while the superscript $^2$ indicates stars observed by low-resolution. The second and third columns denote the spectral type and luminosity classification of each star, respectively. The spectroscopic luminosity is provided in the fourth column, while the effective temperature is presented in the fifth column. The sixth column indicates the surface gravity, followed by the seventh column representing the stellar mass, and the eighth column indicating the stellar radius. The ninth column displays the fractional main sequence age of each star. Additionally, the tenth column presents the mass-loss rate, while the eleventh and twelfth columns respectively detail the macroturbulent velocity and rotational projection velocity of the stars. }
\end{deluxetable*}
\end{longrotatetable}

\startlongtable
\begin{deluxetable*}{c c c c c c c }
\tablecaption{\label{table:slf_parameters}Results of SLF variability for stars of our sample}
    \tablehead{
        \colhead{Name} &  \colhead{Sector} &  \colhead{$\alpha_{0}$}    &  \colhead{$\nu_{\mathrm{char}}$} &  \colhead{$\gamma$} &  \colhead{C$_{\mathrm{w}}$}  \\
        \colhead{}     &  \colhead{}       &\colhead{($\mu\mathrm{mag}$)} &   \colhead{(day$^{-1}$)}&     \colhead{}      &  \colhead{($\mu\mathrm{mag}$)} 
    }
\startdata
BD +36 4145& 15& 205.47$^{+10.41}_{-9.74}$& 4.10$^{+0.21}_{-0.21}$& 2.03$^{+0.05}_{-0.05}$& 10.79$^{+0.04}_{-0.04}$ \\
BD +36 4145& 14& 212.29$^{+10.51}_{-9.82}$& 3.84$^{+0.19}_{-0.19}$& 1.94$^{+0.04}_{-0.04}$& 8.05$^{+0.03}_{-0.03}$ \\
BD +62 2078& 16& 509.37$^{+184.53}_{-117.71}$& 0.11$^{+0.04}_{-0.03}$& 1.08$^{+0.03}_{-0.03}$& 8.16$^{+0.04}_{-0.04}$ \\
BD +62 2078& 17& 283.13$^{+72.17}_{-50.67}$& 0.24$^{+0.07}_{-0.06}$& 1.11$^{+0.04}_{-0.04}$& 8.79$^{+0.04}_{-0.05}$ \\
BD +62 2078& 24& 83.68$^{+10.90}_{-8.87}$& 0.95$^{+0.16}_{-0.15}$& 1.31$^{+0.06}_{-0.05}$& 9.24$^{+0.04}_{-0.04}$ \\
CPD -47 2963 AB& 8& 2127.86$^{+206.18}_{-177.24}$& 0.86$^{+0.07}_{-0.07}$& 1.58$^{+0.02}_{-0.02}$& 7.80$^{+0.04}_{-0.04}$ \\
CPD -47 2963 AB& 9& 1294.56$^{+111.57}_{-96.76}$& 1.03$^{+0.07}_{-0.07}$& 1.71$^{+0.02}_{-0.02}$& 6.12$^{+0.03}_{-0.03}$ \\
HD 101190 AaAb& 38& 149.41$^{+6.96}_{-6.43}$& 4.08$^{+0.17}_{-0.17}$& 2.26$^{+0.04}_{-0.04}$& 3.75$^{+0.01}_{-0.01}$ \\
\enddata
\tablecomments{The fitting parameters of SLF variability for the stars in our sample. The first column lists the names of the stars included in our study, while the second column indicates the sector of TESS photometric data used for detecting SLF variability. Columns three through six present the results of fitting parameters of SLF variability.}
\end{deluxetable*}

\section[]{Appendix B}
\begin{figure}
    \centering
    \gridline{\fig{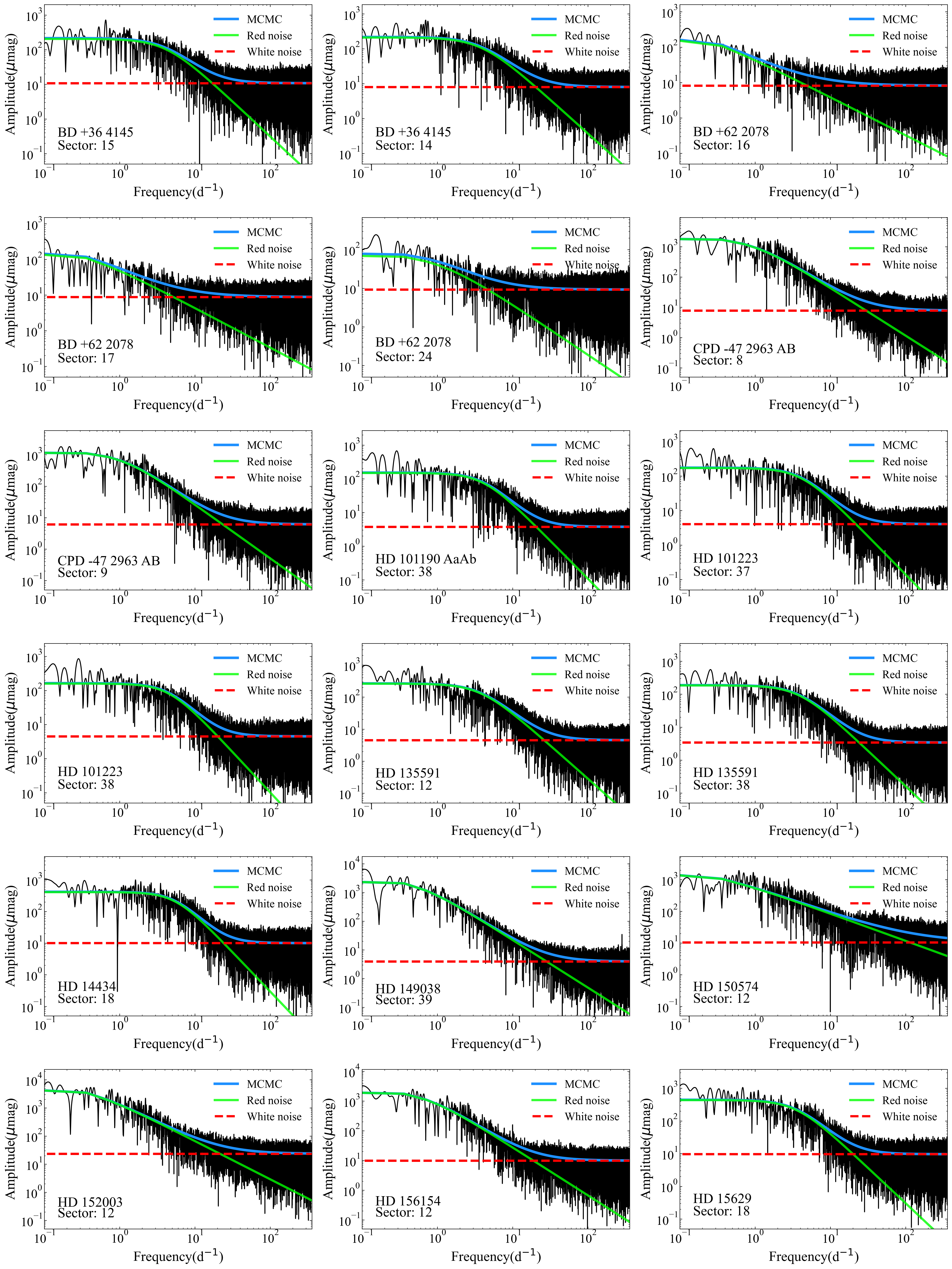}{0.93\textwidth}{}}
       \caption{Fitting results of the logarithmic residual amplitude spectra of stars in our sample. The markers in the each panel have the same meaning as in the top panel of Figure \ref{mcmc_temp}. The complete figure set (16 images) is available in the online journal.}
\end{figure}

%

\vspace{5mm}




\clearpage
\bibliography{sample631.bib}{}

\begin{thebibliography}{}
\expandafter\ifx\csname natexlab\endcsname\relax\def\natexlab#1{#1}\fi
\providecommand{\url}[1]{\href{#1}{#1}}
\providecommand{\dodoi}[1]{doi:~\href{http://doi.org/#1}{\nolinkurl{#1}}}
\providecommand{\doeprint}[1]{\href{http://ascl.net/#1}{\nolinkurl{http://ascl.net/#1}}}
\providecommand{\doarXiv}[1]{\href{https://arxiv.org/abs/#1}{\nolinkurl{https://arxiv.org/abs/#1}}}

\bibitem[{{Abbott}(2017)}]{2017cgrc.book..291A}
{Abbott}, B.~P. 2017, in Centennial of General Relativity: A Celebration, ed. C.~A. {Zen Vasconcellos}, 291--311, \dodoi{10.1142/9789814699662_0011}

\bibitem[{{Abbott} {et~al.}(2016){Abbott}, {Abbott}, {Abbott}, {Abernathy}, {Acernese}, {Ackley}, {Adams}, {Adams}, {Addesso}, {Adhikari}, {Adya}, {Affeldt}, {Agathos}, {Agatsuma}, {Aggarwal}, {Aguiar}, {Aiello}, {Ain}, {Ajith}, {Allen}, {Allocca}, {Altin}, {Anderson}, {Anderson}, {Arai}, {Arain}, {Araya}, {Arceneaux}, {Areeda}, {Arnaud}, {Arun}, {Ascenzi}, {Ashton}, {Ast}, {Aston}, {Astone}, {Aufmuth}, {Aulbert}, {Babak}, {Bacon}, {Bader}, {Baker}, {Baldaccini}, {Ballardin}, {Ballmer}, {Barayoga}, {Barclay}, {Barish}, {Barker}, {Barone}, {Barr}, {Barsotti}, {Barsuglia}, {Barta}, {Bartlett}, {Barton}, {Bartos}, {Bassiri}, {Basti}, {Batch}, {Baune}, {Bavigadda}, {Bazzan}, {Behnke}, {Bejger}, {Belczynski}, {Bell}, {Bell}, {Berger}, {Bergman}, {Bergmann}, {Berry}, {Bersanetti}, {Bertolini}, {Betzwieser}, {Bhagwat}, {Bhandare}, {Bilenko}, {Billingsley}, {Birch}, {Birney}, {Birnholtz}, {Biscans}, {Bisht}, {Bitossi}, {Biwer}, {Bizouard}, {Blackburn}, {Blair}, {Blair}, {Blair}, {Bloemen}, {Bock}, {Bodiya}, {Boer}, {Bogaert}, {Bogan}, {Bohe}, {Bojtos}, {Bond}, {Bondu}, {Bonnand}, {Boom}, {Bork}, {Boschi}, {Bose}, {Bouffanais}, {Bozzi}, {Bradaschia}, {Brady}, {Braginsky}, {Branchesi}, {Brau}, {Briant}, {Brillet}, {Brinkmann}, {Brisson}, {Brockill}, {Brooks}, {Brown}, {Brown}, {Brown}, {Buchanan}, {Buikema}, {Bulik}, {Bulten}, {Buonanno}, {Buskulic}, {Buy}, {Byer}, {Cabero}, {Cadonati}, {Cagnoli}, {Cahillane}, {Bustillo}, {Callister}, {Calloni}, {Camp}, {Cannon}, {Cao}, {Capano}, {Capocasa}, {Carbognani}, {Caride}, {Casanueva Diaz}, {Casentini}, {Caudill}, {Cavagli{\`a}}, {Cavalier}, {Cavalieri}, {Cella}, {Cepeda}, {Baiardi}, {Cerretani}, {Cesarini}, {Chakraborty}, {Chalermsongsak}, {Chamberlin}, {Chan}, {Chao}, {Charlton}, {Chassande-Mottin}, {Chen}, {Chen}, {Cheng}, {Chincarini}, {Chiummo}, {Cho}, {Cho}, {Chow}, {Christensen}, {Chu}, {Chua}, {Chung}, {Ciani}, {Clara}, {Clark}, {Cleva}, {Coccia}, {Cohadon}, {Colla}, {Collette}, {Cominsky}, {Constancio}, {Conte}, {Conti}, {Cook}, {Corbitt}, {Cornish}, {Corsi}, {Cortese}, {Costa}, {Coughlin}, {Coughlin}, {Coulon}, {Countryman}, {Couvares}, {Cowan}, {Coward}, {Cowart}, {Coyne}, {Coyne}, {Craig}, {Creighton}, {Creighton}, {Cripe}, {Crowder}, {Cruise}, {Cumming}, {Cunningham}, {Cuoco}, {Dal Canton}, {Danilishin}, {D'Antonio}, {Danzmann}, {Darman}, {Da Silva Costa}, {Dattilo}, {Dave}, {Daveloza}, {Davier}, {Davies}, {Daw}, {Day}, {De}, {DeBra}, {Debreczeni}, {Degallaix}, {De Laurentis}, {Del{\'e}glise}, {Del Pozzo}, {Denker}, {Dent}, {Dereli}, {Dergachev}, {DeRosa}, {De Rosa}, {DeSalvo}, {Dhurandhar}, {D{\'\i}az}, {Di Fiore}, {Di Giovanni}, {Di Lieto}, {Di Pace}, {Di Palma}, {Di Virgilio}, {Dojcinoski}, {Dolique}, {Donovan}, {Dooley}, {Doravari}, {Douglas}, {Downes}, {Drago}, {Drever}, {Driggers}, {Du}, {Ducrot}, {Dwyer}, {Edo}, {Edwards}, {Effler}, {Eggenstein}, {Ehrens}, {Eichholz}, {Eikenberry}, {Engels}, {Essick}, {Etzel}, {Evans}, {Evans}, {Everett}, {Factourovich}, {Fafone}, {Fair}, {Fairhurst}, {Fan}, {Fang}, {Farinon}, {Farr}, {Farr}, {Favata}, {Fays}, {Fehrmann}, {Fejer}, {Feldbaum}, {Ferrante}, {Ferreira}, {Ferrini}, {Fidecaro}, {Finn}, {Fiori}, {Fiorucci}, {Fisher}, {Flaminio}, {Fletcher}, {Fong}, {Fournier}, {Franco}, {Frasca}, {Frasconi}, {Frede}, {Frei}, {Freise}, {Frey}, {Frey}, {Fricke}, {Fritschel}, {Frolov}, {Fulda}, {Fyffe}, {Gabbard}, {Gair}, {Gammaitoni}, {Gaonkar}, {Garufi}, {Gatto}, {Gaur}, {Gehrels}, {Gemme}, {Gendre}, {Genin}, {Gennai}, {George}, {Gergely}, {Germain}, {Ghosh}, {Ghosh}, {Ghosh}, {Giaime}, {Giardina}, {Giazotto}, {Gill}, {Glaefke}, {Gleason}, {Goetz}, {Goetz}, {Gondan}, {Gonz{\'a}lez}, {Castro}, {Gopakumar}, {Gordon}, {Gorodetsky}, {Gossan}, {Gosselin}, {Gouaty}, {Graef}, {Graff}, {Granata}, {Grant}, {Gras}, {Gray}, {Greco}, {Green}, {Greenhalgh}, {Groot}, {Grote}, {Grunewald}, {Guidi}, {Guo}, {Gupta}, {Gupta}, {Gushwa}, {Gustafson}, {Gustafson}, {Hacker}, {Hall}, {Hall}, {Hammond}, {Haney}, {Hanke}, {Hanks}, {Hanna}, {Hannam}, {Hanson}, {Hardwick}, {Harms}, {Harry}, {Harry}, {Hart}, {Hartman}, {Haster}, {Haughian}, {Healy}, {Heefner}, {Heidmann}, {Heintze}, {Heinzel}, {Heitmann}, {Hello}, {Hemming}, {Hendry}, {Heng}, {Hennig}, {Heptonstall}, {Heurs}, {Hild}, {Hoak}, {Hodge}, {Hofman}, {Hollitt}, {Holt}, {Holz}, {Hopkins}, {Hosken}, {Hough}, {Houston}, {Howell}, {Hu}, {Huang}, {Huerta}, {Huet}, {Hughey}, {Husa}, {Huttner}, {Huynh-Dinh}, {Idrisy}, {Indik}, {Ingram}, {Inta}, {Isa}, {Isac}, {Isi}, {Islas}, {Isogai}, {Iyer}, {Izumi}, {Jacobson}, {Jacqmin}, {Jang}, {Jani}, {Jaranowski}, {Jawahar}, {Jim{\'e}nez-Forteza}, {Johnson}, {Johnson-McDaniel}, {Jones}, {Jones}, {Jonker}, {Ju}, {Haris}, {Kalaghatgi}, {Kalogera}, {Kandhasamy}, {Kang}, {Kanner}, {Karki}, {Kasprzack}, {Katsavounidis}, {Katzman}, {Kaufer}, {Kaur}, {Kawabe}, {Kawazoe}, {K{\'e}f{\'e}lian}, {Kehl}, {Keitel}, {Kelley}, {Kells}, {Kennedy}, {Keppel}, {Key}, {Khalaidovski}, {Khalili}, {Khan}, {Khan}, {Khan}, {Khazanov}, {Kijbunchoo}, {Kim}, {Kim}, {Kim}, {Kim}, {Kim}, {Kim}, {King}, {King}, {Kinzel}, {Kissel}, {Kleybolte}, {Klimenko}, {Koehlenbeck}, {Kokeyama}, {Koley}, {Kondrashov}, {Kontos}, {Koranda}, {Korobko}, {Korth}, {Kowalska}, {Kozak}, {Kringel}, {Krishnan}, {Kr{\'o}lak}, {Krueger}, {Kuehn}, {Kumar}, {Kumar}, {Kuo}, {Kutynia}, {Kwee}, {Lackey}, {Landry}, {Lange}, {Lantz}, {Lasky}, {Lazzarini}, {Lazzaro}, {Leaci}, {Leavey}, {Lebigot}, {Lee}, {Lee}, {Lee}, {Lee}, {Lenon}, {Leonardi}, {Leong}, {Leroy}, {Letendre}, {Levin}, {Levine}, {Li}, {Libson}, {Littenberg}, {Lockerbie}, {Logue}, {Lombardi}, {London}, {Lord}, {Lorenzini}, {Loriette}, {Lormand}, {Losurdo}, {Lough}, {Lousto}, {Lovelace}, {L{\"u}ck}, {Lundgren}, {Luo}, {Lynch}, {Ma}, {MacDonald}, {Machenschalk}, {MacInnis}, {Macleod}, {Maga{\~n}a-Sandoval}, {Magee}, {Mageswaran}, {Majorana}, {Maksimovic}, {Malvezzi}, {Man}, {Mandel}, {Mandic}, {Mangano}, {Mansell}, {Manske}, {Mantovani}, {Marchesoni}, {Marion}, {M{\'a}rka}, {M{\'a}rka}, {Markosyan}, {Maros}, {Martelli}, {Martellini}, {Martin}, {Martin}, {Martynov}, {Marx}, {Mason}, {Masserot}, {Massinger}, {Masso-Reid}, {Matichard}, {Matone}, {Mavalvala}, {Mazumder}, {Mazzolo}, {McCarthy}, {McClelland}, {McCormick}, {McGuire}, {McIntyre}, {McIver}, {McManus}, {McWilliams}, {Meacher}, {Meadors}, {Meidam}, {Melatos}, {Mendell}, {Mendoza-Gandara}, {Mercer}, {Merilh}, {Merzougui}, {Meshkov}, {Messenger}, {Messick}, {Meyers}, {Mezzani}, {Miao}, {Michel}, {Middleton}, {Mikhailov}, {Milano}, {Miller}, {Millhouse}, {Minenkov}, {Ming}, {Mirshekari}, {Mishra}, {Mitra}, {Mitrofanov}, {Mitselmakher}, {Mittleman}, {Moggi}, {Mohan}, {Mohapatra}, {Montani}, {Moore}, {Moore}, {Moraru}, {Moreno}, {Morriss}, {Mossavi}, {Mours}, {Mow-Lowry}, {Mueller}, {Mueller}, {Muir}, {Mukherjee}, {Mukherjee}, {Mukherjee}, {Mukund}, {Mullavey}, {Munch}, {Murphy}, {Murray}, {Mytidis}, {Nardecchia}, {Naticchioni}, {Nayak}, {Necula}, {Nedkova}, {Nelemans}, {Neri}, {Neunzert}, {Newton}, {Nguyen}, {Nielsen}, {Nissanke}, {Nitz}, {Nocera}, {Nolting}, {Normandin}, {Nuttall}, {Oberling}, {Ochsner}, {O'Dell}, {Oelker}, {Ogin}, {Oh}, {Oh}, {Ohme}, {Oliver}, {Oppermann}, {Oram}, {O'Reilly}, {O'Shaughnessy}, {Ott}, {Ottaway}, {Ottens}, {Overmier}, {Owen}, {Pai}, {Pai}, {Palamos}, {Palashov}, {Palomba}, {Pal-Singh}, {Pan}, {Pan}, {Pankow}, {Pannarale}, {Pant}, {Paoletti}, {Paoli}, {Papa}, {Paris}, {Parker}, {Pascucci}, {Pasqualetti}, {Passaquieti}, {Passuello}, {Patricelli}, {Patrick}, {Pearlstone}, {Pedraza}, {Pedurand}, {Pekowsky}, {Pele}, {Penn}, {Perreca}, {Pfeiffer}, {Phelps}, {Piccinni}, {Pichot}, {Pickenpack}, {Piergiovanni}, {Pierro}, {Pillant}, {Pinard}, {Pinto}, {Pitkin}, {Poeld}, {Poggiani}, {Popolizio}, {Post}, {Powell}, {Prasad}, {Predoi}, {Premachandra}, {Prestegard}, {Price}, {Prijatelj}, {Principe}, {Privitera}, {Prix}, {Prodi}, {Prokhorov}, {Puncken}, {Punturo}, {Puppo}, {P{\"u}rrer}, {Qi}, {Qin}, {Quetschke}, {Quintero}, {Quitzow-James}, {Raab}, {Rabeling}, {Radkins}, {Raffai}, {Raja}, {Rakhmanov}, {Ramet}, {Rapagnani}, {Raymond}, {Razzano}, {Re}, {Read}, {Reed}, {Regimbau}, {Rei}, {Reid}, {Reitze}, {Rew}, {Reyes}, {Ricci}, {Riles}, {Robertson}, {Robie}, {Robinet}, {Rocchi}, {Rolland}, {Rollins}, {Roma}, {Romano}, {Romano}, {Romanov}, {Romie}, {Rosi{\'n}ska}, {Rowan}, {R{\"u}diger}, {Ruggi}, {Ryan}, {Sachdev}, {Sadecki}, {Sadeghian}, {Salconi}, {Saleem}, {Salemi}, {Samajdar}, {Sammut}, {Sampson}, {Sanchez}, {Sandberg}, {Sandeen}, {Sanders}, {Sanders}, {Sassolas}, {Sathyaprakash}, {Saulson}, {Sauter}, {Savage}, {Sawadsky}, {Schale}, {Schilling}, {Schmidt}, {Schmidt}, {Schnabel}, {Schofield}, {Sch{\"o}nbeck}, {Schreiber}, {Schuette}, {Schutz}, {Scott}, {Scott}, {Sellers}, {Sengupta}, {Sentenac}, {Sequino}, {Sergeev}, {Serna}, {Setyawati}, {Sevigny}, {Shaddock}, {Shaffer}, {Shah}, {Shahriar}, {Shaltev}, {Shao}, {Shapiro}, {Shawhan}, {Sheperd}, {Shoemaker}, {Shoemaker}, {Siellez}, {Siemens}, {Sigg}, {Silva}, {Simakov}, {Singer}, {Singer}, {Singh}, {Singh}, {Singhal}, {Sintes}, {Slagmolen}, {Smith}, {Smith}, {Smith}, {Smith}, {Son}, {Sorazu}, {Sorrentino}, {Souradeep}, {Srivastava}, {Staley}, {Steinke}, {Steinlechner}, {Steinlechner}, {Steinmeyer}, {Stephens}, {Stevenson}, {Stone}, {Strain}, {Straniero}, {Stratta}, {Strauss}, {Strigin}, {Sturani}, {Stuver}, {Summerscales}, {Sun}, {Sutton}, {Swinkels}, {Szczepa{\'n}czyk}, {Tacca}, {Talukder}, {Tanner}, {T{\'a}pai}, {Tarabrin}, {Taracchini}, {Taylor}, {Theeg}, {Thirugnanasambandam}, {Thomas}, {Thomas}, {Thomas}, {Thorne}, {Thorne}, {Thrane}, {Tiwari}, {Tiwari}, {Tokmakov}, {Tomlinson}, {Tonelli}, {Torres}, {Torrie}, {T{\"o}yr{\"a}}, {Travasso}, {Traylor}, {Trifir{\`o}}, {Tringali}, {Trozzo}, {Tse}, {Turconi}, {Tuyenbayev}, {Ugolini}, {Unnikrishnan}, {Urban}, {Usman}, {Vahlbruch}, {Vajente}, {Valdes}, {Vallisneri}, {van Bakel}, {van Beuzekom}, {van den Brand}, {Van Den Broeck}, {Vander-Hyde}, {van der Schaaf}, {van Heijningen}, {van Veggel}, {Vardaro}, {Vass}, {Vas{\'u}th}, {Vaulin}, {Vecchio}, {Vedovato}, {Veitch}, {Veitch}, {Venkateswara}, {Verkindt},
  {Vetrano}, {Vicer{\'e}}, {Vinciguerra}, {Vine}, {Vinet}, {Vitale}, {Vo}, {Vocca}, {Vorvick}, {Voss}, {Vousden}, {Vyatchanin}, {Wade}, {Wade}, {Wade}, {Waldman}, {Walker}, {Wallace}, {Walsh}, {Wang}, {Wang}, {Wang}, {Wang}, {Wang}, {Ward}, {Ward}, {Warner}, {Was}, {Weaver}, {Wei}, {Weinert}, {Weinstein}, {Weiss}, {Welborn}, {Wen}, {We{\ss}els}, {Westphal}, {Wette}, {Whelan}, {Whitcomb}, {White}, {Whiting}, {Wiesner}, {Wilkinson}, {Willems}, {Williams}, {Williams}, {Williamson}, {Willis}, {Willke}, {Wimmer}, {Winkelmann}, {Winkler}, {Wipf}, {Wiseman}, {Wittel}, {Woan}, {Worden}, {Wright}, {Wu}, {Yablon}, {Yakushin}, {Yam}, {Yamamoto}, {Yancey}, {Yap}, {Yu}, {Yvert}, {Zadro{\.Z}ny}, {Zangrando}, {Zanolin}, {Zendri}, {Zevin}, {Zhang}, {Zhang}, {Zhang}, {Zhang}, {Zhao}, {Zhou}, {Zhou}, {Zhu}, {Zucker}, {Zuraw}, {Zweizig}, {LIGO Scientific Collaboration}, \& {Virgo Collaboration}}]{2016PhRvL.116f1102A}
{Abbott}, B.~P., {Abbott}, R., {Abbott}, T.~D., {et~al.} 2016, \prl, 116, 061102, \dodoi{10.1103/PhysRevLett.116.061102}

\bibitem[{{Abbott} \& {Hummer}(1985)}]{1985ApJ...294..286A}
{Abbott}, D.~C., \& {Hummer}, D.~G. 1985, \apj, 294, 286, \dodoi{10.1086/163297}

\bibitem[{{Aerts} {et~al.}(2010){Aerts}, {Christensen-Dalsgaard}, \& {Kurtz}}]{2010aste.book.....A}
{Aerts}, C., {Christensen-Dalsgaard}, J., \& {Kurtz}, D.~W. 2010, {Asteroseismology}, \dodoi{10.1007/978-1-4020-5803-5}

\bibitem[{{Aerts} \& {Rogers}(2015)}]{2015ApJ...806L..33A}
{Aerts}, C., \& {Rogers}, T.~M. 2015, \apjl, 806, L33, \dodoi{10.1088/2041-8205/806/2/L33}

\bibitem[{{Aerts} {et~al.}(2018{\natexlab{a}}){Aerts}, {Molenberghs}, {Michielsen}, {Pedersen}, {Bj{\"o}rklund}, {Johnston}, {Mombarg}, {Bowman}, {Buysschaert}, {P{\'a}pics}, {Sekaran}, {Sundqvist}, {Tkachenko}, {Truyaert}, {Van Reeth}, \& {Vermeyen}}]{2018ApJS..237...15A}
{Aerts}, C., {Molenberghs}, G., {Michielsen}, M., {et~al.} 2018{\natexlab{a}}, \apjs, 237, 15, \dodoi{10.3847/1538-4365/aaccfb}

\bibitem[{{Aerts} {et~al.}(2018{\natexlab{b}}){Aerts}, {Bowman}, {S{\'\i}mon-D{\'\i}az}, {Buysschaert}, {Johnston}, {Moravveji}, {Beck}, {De Cat}, {Triana}, {Aigrain}, {Castro}, {Huber}, \& {White}}]{2018MNRAS.476.1234A}
{Aerts}, C., {Bowman}, D.~M., {S{\'\i}mon-D{\'\i}az}, S., {et~al.} 2018{\natexlab{b}}, \mnras, 476, 1234, \dodoi{10.1093/mnras/sty308}

\bibitem[{{Aerts} {et~al.}(2019){Aerts}, {Pedersen}, {Vermeyen}, {Hendriks}, {Johnston}, {Tkachenko}, {P{\'a}pics}, {Debosscher}, {Briquet}, {Thoul}, {Rainer}, \& {Poretti}}]{2019A&A...624A..75A}
{Aerts}, C., {Pedersen}, M.~G., {Vermeyen}, E., {et~al.} 2019, \aap, 624, A75, \dodoi{10.1051/0004-6361/201834762}

\bibitem[{{Anders} {et~al.}(2023){Anders}, {Lecoanet}, {Cantiello}, {Burns}, {Hyatt}, {Kaufman}, {Townsend}, {Brown}, {Vasil}, {Oishi}, \& {Jermyn}}]{2023NatAs...7.1228A}
{Anders}, E.~H., {Lecoanet}, D., {Cantiello}, M., {et~al.} 2023, Nature Astronomy, 7, 1228, \dodoi{10.1038/s41550-023-02040-7}

\bibitem[{{Anderson} {et~al.}(1990){Anderson}, {Duvall}, \& {Jefferies}}]{1990ApJ...364..699A}
{Anderson}, E.~R., {Duvall}, Thomas~L., J., \& {Jefferies}, S.~M. 1990, \apj, 364, 699, \dodoi{10.1086/169452}

\bibitem[{{Arias} {et~al.}(2016){Arias}, {Walborn}, {Sim{\'o}n D{\'\i}az}, {Barb{\'a}}, {Ma{\'\i}z Apell{\'a}niz}, {Sab{\'\i}n-Sanjuli{\'a}n}, {Gamen}, {Morrell}, {Sota}, {Marco}, {Negueruela}, {Le{\~a}o}, {Herrero}, \& {Alfaro}}]{2016AJ....152...31A}
{Arias}, J.~I., {Walborn}, N.~R., {Sim{\'o}n D{\'\i}az}, S., {et~al.} 2016, \aj, 152, 31, \dodoi{10.3847/0004-6256/152/2/31}

\bibitem[{{Astropy Collaboration} {et~al.}(2013){Astropy Collaboration}, {Robitaille}, {Tollerud}, {Greenfield}, {Droettboom}, {Bray}, {Aldcroft}, {Davis}, {Ginsburg}, {Price-Whelan}, {Kerzendorf}, {Conley}, {Crighton}, {Barbary}, {Muna}, {Ferguson}, {Grollier}, {Parikh}, {Nair}, {Unther}, {Deil}, {Woillez}, {Conseil}, {Kramer}, {Turner}, {Singer}, {Fox}, {Weaver}, {Zabalza}, {Edwards}, {Azalee Bostroem}, {Burke}, {Casey}, {Crawford}, {Dencheva}, {Ely}, {Jenness}, {Labrie}, {Lim}, {Pierfederici}, {Pontzen}, {Ptak}, {Refsdal}, {Servillat}, \& {Streicher}}]{2013A&A...558A..33A}
{Astropy Collaboration}, {Robitaille}, T.~P., {Tollerud}, E.~J., {et~al.} 2013, \aap, 558, A33, \dodoi{10.1051/0004-6361/201322068}

\bibitem[{{Astropy Collaboration} {et~al.}(2018){Astropy Collaboration}, {Price-Whelan}, {Sip{\H{o}}cz}, {G{\"u}nther}, {Lim}, {Crawford}, {Conseil}, {Shupe}, {Craig}, {Dencheva}, {Ginsburg}, {VanderPlas}, {Bradley}, {P{\'e}rez-Su{\'a}rez}, {de Val-Borro}, {Aldcroft}, {Cruz}, {Robitaille}, {Tollerud}, {Ardelean}, {Babej}, {Bach}, {Bachetti}, {Bakanov}, {Bamford}, {Barentsen}, {Barmby}, {Baumbach}, {Berry}, {Biscani}, {Boquien}, {Bostroem}, {Bouma}, {Brammer}, {Bray}, {Breytenbach}, {Buddelmeijer}, {Burke}, {Calderone}, {Cano Rodr{\'\i}guez}, {Cara}, {Cardoso}, {Cheedella}, {Copin}, {Corrales}, {Crichton}, {D'Avella}, {Deil}, {Depagne}, {Dietrich}, {Donath}, {Droettboom}, {Earl}, {Erben}, {Fabbro}, {Ferreira}, {Finethy}, {Fox}, {Garrison}, {Gibbons}, {Goldstein}, {Gommers}, {Greco}, {Greenfield}, {Groener}, {Grollier}, {Hagen}, {Hirst}, {Homeier}, {Horton}, {Hosseinzadeh}, {Hu}, {Hunkeler}, {Ivezi{\'c}}, {Jain}, {Jenness}, {Kanarek}, {Kendrew}, {Kern}, {Kerzendorf}, {Khvalko}, {King}, {Kirkby}, {Kulkarni}, {Kumar}, {Lee}, {Lenz}, {Littlefair}, {Ma}, {Macleod}, {Mastropietro}, {McCully}, {Montagnac}, {Morris}, {Mueller}, {Mumford}, {Muna}, {Murphy}, {Nelson}, {Nguyen}, {Ninan}, {N{\"o}the}, {Ogaz}, {Oh}, {Parejko}, {Parley}, {Pascual}, {Patil}, {Patil}, {Plunkett}, {Prochaska}, {Rastogi}, {Reddy Janga}, {Sabater}, {Sakurikar}, {Seifert}, {Sherbert}, {Sherwood-Taylor}, {Shih}, {Sick}, {Silbiger}, {Singanamalla}, {Singer}, {Sladen}, {Sooley}, {Sornarajah}, {Streicher}, {Teuben}, {Thomas}, {Tremblay}, {Turner}, {Terr{\'o}n}, {van Kerkwijk}, {de la Vega}, {Watkins}, {Weaver}, {Whitmore}, {Woillez}, {Zabalza}, \& {Astropy Contributors}}]{2018AJ....156..123A}
{Astropy Collaboration}, {Price-Whelan}, A.~M., {Sip{\H{o}}cz}, B.~M., {et~al.} 2018, \aj, 156, 123, \dodoi{10.3847/1538-3881/aabc4f}

\bibitem[{{Astropy Collaboration} {et~al.}(2022){Astropy Collaboration}, {Price-Whelan}, {Lim}, {Earl}, {Starkman}, {Bradley}, {Shupe}, {Patil}, {Corrales}, {Brasseur}, {N{\"o}the}, {Donath}, {Tollerud}, {Morris}, {Ginsburg}, {Vaher}, {Weaver}, {Tocknell}, {Jamieson}, {van Kerkwijk}, {Robitaille}, {Merry}, {Bachetti}, {G{\"u}nther}, {Aldcroft}, {Alvarado-Montes}, {Archibald}, {B{\'o}di}, {Bapat}, {Barentsen}, {Baz{\'a}n}, {Biswas}, {Boquien}, {Burke}, {Cara}, {Cara}, {Conroy}, {Conseil}, {Craig}, {Cross}, {Cruz}, {D'Eugenio}, {Dencheva}, {Devillepoix}, {Dietrich}, {Eigenbrot}, {Erben}, {Ferreira}, {Foreman-Mackey}, {Fox}, {Freij}, {Garg}, {Geda}, {Glattly}, {Gondhalekar}, {Gordon}, {Grant}, {Greenfield}, {Groener}, {Guest}, {Gurovich}, {Handberg}, {Hart}, {Hatfield-Dodds}, {Homeier}, {Hosseinzadeh}, {Jenness}, {Jones}, {Joseph}, {Kalmbach}, {Karamehmetoglu}, {Ka{\l}uszy{\'n}ski}, {Kelley}, {Kern}, {Kerzendorf}, {Koch}, {Kulumani}, {Lee}, {Ly}, {Ma}, {MacBride}, {Maljaars}, {Muna}, {Murphy}, {Norman}, {O'Steen}, {Oman}, {Pacifici}, {Pascual}, {Pascual-Granado}, {Patil}, {Perren}, {Pickering}, {Rastogi}, {Roulston}, {Ryan}, {Rykoff}, {Sabater}, {Sakurikar}, {Salgado}, {Sanghi}, {Saunders}, {Savchenko}, {Schwardt}, {Seifert-Eckert}, {Shih}, {Jain}, {Shukla}, {Sick}, {Simpson}, {Singanamalla}, {Singer}, {Singhal}, {Sinha}, {Sip{\H{o}}cz}, {Spitler}, {Stansby}, {Streicher}, {{\v{S}}umak}, {Swinbank}, {Taranu}, {Tewary}, {Tremblay}, {de Val-Borro}, {Van Kooten}, {Vasovi{\'c}}, {Verma}, {de Miranda Cardoso}, {Williams}, {Wilson}, {Winkel}, {Wood-Vasey}, {Xue}, {Yoachim}, {Zhang}, {Zonca}, \& {Astropy Project Contributors}}]{2022ApJ...935..167A}
{Astropy Collaboration}, {Price-Whelan}, A.~M., {Lim}, P.~L., {et~al.} 2022, \apj, 935, 167, \dodoi{10.3847/1538-4357/ac7c74}

\bibitem[{{Augustson} \& {Mathis}(2019)}]{2019ApJ...874...83A}
{Augustson}, K.~C., \& {Mathis}, S. 2019, \apj, 874, 83, \dodoi{10.3847/1538-4357/ab0b3d}

\bibitem[{{Baglin} {et~al.}(2009){Baglin}, {Auvergne}, {Barge}, {Deleuil}, {Michel}, \& {CoRoT Exoplanet Science Team}}]{2009IAUS..253...71B}
{Baglin}, A., {Auvergne}, M., {Barge}, P., {et~al.} 2009, in Transiting Planets, ed. F.~{Pont}, D.~{Sasselov}, \& M.~J. {Holman}, Vol. 253, 71--81, \dodoi{10.1017/S1743921308026252}

\bibitem[{{Barb{\'a}} {et~al.}(2010){Barb{\'a}}, {Gamen}, {Arias}, {Morrell}, {Ma{\'\i}z Apell{\'a}niz}, {Alfaro}, {Walborn}, \& {Sota}}]{2010RMxAC..38...30B}
{Barb{\'a}}, R.~H., {Gamen}, R., {Arias}, J.~I., {et~al.} 2010, in Revista Mexicana de Astronomia y Astrofisica Conference Series, Vol.~38, Revista Mexicana de Astronomia y Astrofisica Conference Series, 30--32

\bibitem[{{Bellinger} {et~al.}(2023){Bellinger}, {de Mink}, {van Rossem}, \& {Justham}}]{2023arXiv231100038B}
{Bellinger}, E.~P., {de Mink}, S.~E., {van Rossem}, W.~E., \& {Justham}, S. 2023, arXiv e-prints, arXiv:2311.00038, \dodoi{10.48550/arXiv.2311.00038}

\bibitem[{{Blomme} {et~al.}(2011){Blomme}, {Mahy}, {Catala}, {Cuypers}, {Gosset}, {Godart}, {Montalban}, {Ventura}, {Rauw}, {Morel}, {Degroote}, {Aerts}, {Noels}, {Michel}, {Baudin}, {Baglin}, {Auvergne}, \& {Samadi}}]{2011A&A...533A...4B}
{Blomme}, R., {Mahy}, L., {Catala}, C., {et~al.} 2011, \aap, 533, A4, \dodoi{10.1051/0004-6361/201116949}

\bibitem[{{Blouin} {et~al.}(2023){Blouin}, {Mao}, {Herwig}, {Denissenkov}, {Woodward}, \& {Thompson}}]{2023MNRAS.522.1706B}
{Blouin}, S., {Mao}, H., {Herwig}, F., {et~al.} 2023, \mnras, 522, 1706, \dodoi{10.1093/mnras/stad1115}

\bibitem[{{Bogn{\'a}r} {et~al.}(2023){Bogn{\'a}r}, {S{\'o}dor}, {Clark}, \& {Kawaler}}]{2023A&A...674A.204B}
{Bogn{\'a}r}, Z., {S{\'o}dor}, {\'A}., {Clark}, I.~R., \& {Kawaler}, S.~D. 2023, \aap, 674, A204, \dodoi{10.1051/0004-6361/202245177}

\bibitem[{{Borucki} {et~al.}(2010){Borucki}, {Koch}, {Basri}, {Batalha}, {Brown}, {Caldwell}, {Caldwell}, {Christensen-Dalsgaard}, {Cochran}, {DeVore}, {Dunham}, {Dupree}, {Gautier}, {Geary}, {Gilliland}, {Gould}, {Howell}, {Jenkins}, {Kondo}, {Latham}, {Marcy}, {Meibom}, {Kjeldsen}, {Lissauer}, {Monet}, {Morrison}, {Sasselov}, {Tarter}, {Boss}, {Brownlee}, {Owen}, {Buzasi}, {Charbonneau}, {Doyle}, {Fortney}, {Ford}, {Holman}, {Seager}, {Steffen}, {Welsh}, {Rowe}, {Anderson}, {Buchhave}, {Ciardi}, {Walkowicz}, {Sherry}, {Horch}, {Isaacson}, {Everett}, {Fischer}, {Torres}, {Johnson}, {Endl}, {MacQueen}, {Bryson}, {Dotson}, {Haas}, {Kolodziejczak}, {Van Cleve}, {Chandrasekaran}, {Twicken}, {Quintana}, {Clarke}, {Allen}, {Li}, {Wu}, {Tenenbaum}, {Verner}, {Bruhweiler}, {Barnes}, \& {Prsa}}]{2010Sci...327..977B}
{Borucki}, W.~J., {Koch}, D., {Basri}, G., {et~al.} 2010, Science, 327, 977, \dodoi{10.1126/science.1185402}

\bibitem[{{Bowman}(2023)}]{2023Ap&SS.368..107B}
{Bowman}, D.~M. 2023, \apss, 368, 107, \dodoi{10.1007/s10509-023-04262-7}

\bibitem[{{Bowman} {et~al.}(2020){Bowman}, {Burssens}, {Sim{\'o}n-D{\'\i}az}, {Edelmann}, {Rogers}, {Horst}, {R{\"o}pke}, \& {Aerts}}]{2020A&A...640A..36B}
{Bowman}, D.~M., {Burssens}, S., {Sim{\'o}n-D{\'\i}az}, S., {et~al.} 2020, \aap, 640, A36, \dodoi{10.1051/0004-6361/202038224}

\bibitem[{{Bowman} \& {Dorn-Wallenstein}(2022)}]{2022A&A...668A.134B}
{Bowman}, D.~M., \& {Dorn-Wallenstein}, T.~Z. 2022, \aap, 668, A134, \dodoi{10.1051/0004-6361/202243545}

\bibitem[{{Bowman} \& {Michielsen}(2021)}]{2021A&A...656A.158B}
{Bowman}, D.~M., \& {Michielsen}, M. 2021, \aap, 656, A158, \dodoi{10.1051/0004-6361/202141726}

\bibitem[{{Bowman} {et~al.}(2019{\natexlab{a}}){Bowman}, {Aerts}, {Johnston}, {Pedersen}, {Rogers}, {Edelmann}, {Sim{\'o}n-D{\'\i}az}, {Van Reeth}, {Buysschaert}, {Tkachenko}, \& {Triana}}]{2019A&A...621A.135B}
{Bowman}, D.~M., {Aerts}, C., {Johnston}, C., {et~al.} 2019{\natexlab{a}}, \aap, 621, A135, \dodoi{10.1051/0004-6361/201833662}

\bibitem[{{Bowman} {et~al.}(2019{\natexlab{b}}){Bowman}, {Burssens}, {Pedersen}, {Johnston}, {Aerts}, {Buysschaert}, {Michielsen}, {Tkachenko}, {Rogers}, {Edelmann}, {Ratnasingam}, {Sim{\'o}n-D{\'\i}az}, {Castro}, {Moravveji}, {Pope}, {White}, \& {De Cat}}]{2019NatAs...3..760B}
{Bowman}, D.~M., {Burssens}, S., {Pedersen}, M.~G., {et~al.} 2019{\natexlab{b}}, Nature Astronomy, 3, 760, \dodoi{10.1038/s41550-019-0768-1}

\bibitem[{{Brott} {et~al.}(2011){Brott}, {de Mink}, {Cantiello}, {Langer}, {de Koter}, {Evans}, {Hunter}, {Trundle}, \& {Vink}}]{2011A&A...530A.115B}
{Brott}, I., {de Mink}, S.~E., {Cantiello}, M., {et~al.} 2011, \aap, 530, A115, \dodoi{10.1051/0004-6361/201016113}

\bibitem[{{Bryson} {et~al.}(2010){Bryson}, {Tenenbaum}, {Jenkins}, {Chandrasekaran}, {Klaus}, {Caldwell}, {Gilliland}, {Haas}, {Dotson}, {Koch}, \& {Borucki}}]{2010ApJ...713L..97B}
{Bryson}, S.~T., {Tenenbaum}, P., {Jenkins}, J.~M., {et~al.} 2010, \apjl, 713, L97, \dodoi{10.1088/2041-8205/713/2/L97}

\bibitem[{{Burssens} {et~al.}(2020){Burssens}, {Sim{\'o}n-D{\'\i}az}, {Bowman}, {Holgado}, {Michielsen}, {de Burgos}, {Castro}, {Barb{\'a}}, \& {Aerts}}]{2020A&A...639A..81B}
{Burssens}, S., {Sim{\'o}n-D{\'\i}az}, S., {Bowman}, D.~M., {et~al.} 2020, \aap, 639, A81, \dodoi{10.1051/0004-6361/202037700}

\bibitem[{{Burssens} {et~al.}(2023){Burssens}, {Bowman}, {Michielsen}, {Sim{\'o}n-D{\'\i}az}, {Aerts}, {Vanlaer}, {Banyard}, {Nardetto}, {Townsend}, {Handler}, {Mombarg}, {Vanderspek}, \& {Ricker}}]{2023NatAs...7..913B}
{Burssens}, S., {Bowman}, D.~M., {Michielsen}, M., {et~al.} 2023, Nature Astronomy, 7, 913, \dodoi{10.1038/s41550-023-01978-y}

\bibitem[{{Buysschaert} {et~al.}(2015){Buysschaert}, {Aerts}, {Bloemen}, {Debosscher}, {Neiner}, {Briquet}, {Vos}, {P{\'a}pics}, {Manick}, {Schmid}, {Van Winckel}, \& {Tkachenko}}]{2015MNRAS.453...89B}
{Buysschaert}, B., {Aerts}, C., {Bloemen}, S., {et~al.} 2015, \mnras, 453, 89, \dodoi{10.1093/mnras/stv1572}

\bibitem[{{Cantiello} {et~al.}(2021){Cantiello}, {Lecoanet}, {Jermyn}, \& {Grassitelli}}]{2021ApJ...915..112C}
{Cantiello}, M., {Lecoanet}, D., {Jermyn}, A.~S., \& {Grassitelli}, L. 2021, \apj, 915, 112, \dodoi{10.3847/1538-4357/ac03b0}

\bibitem[{{Cantiello} {et~al.}(2009){Cantiello}, {Langer}, {Brott}, {de Koter}, {Shore}, {Vink}, {Voegler}, {Lennon}, \& {Yoon}}]{2009A&A...499..279C}
{Cantiello}, M., {Langer}, N., {Brott}, I., {et~al.} 2009, \aap, 499, 279, \dodoi{10.1051/0004-6361/200911643}

\bibitem[{{Castor} {et~al.}(1975){Castor}, {Abbott}, \& {Klein}}]{1975ApJ...195..157C}
{Castor}, J.~I., {Abbott}, D.~C., \& {Klein}, R.~I. 1975, \apj, 195, 157, \dodoi{10.1086/153315}

\bibitem[{{Chatzopoulos} {et~al.}(2014){Chatzopoulos}, {Graziani}, \& {Couch}}]{2014ApJ...795...92C}
{Chatzopoulos}, E., {Graziani}, C., \& {Couch}, S.~M. 2014, \apj, 795, 92, \dodoi{10.1088/0004-637X/795/1/92}

\bibitem[{{Crowther} {et~al.}(2002){Crowther}, {Hillier}, {Evans}, {Fullerton}, {De Marco}, \& {Willis}}]{2002ApJ...579..774C}
{Crowther}, P.~A., {Hillier}, D.~J., {Evans}, C.~J., {et~al.} 2002, \apj, 579, 774, \dodoi{10.1086/342877}

\bibitem[{{Crowther} {et~al.}(2010){Crowther}, {Schnurr}, {Hirschi}, {Yusof}, {Parker}, {Goodwin}, \& {Kassim}}]{2010MNRAS.408..731C}
{Crowther}, P.~A., {Schnurr}, O., {Hirschi}, R., {et~al.} 2010, \mnras, 408, 731, \dodoi{10.1111/j.1365-2966.2010.17167.x}

\bibitem[{{Daszy{\'n}ska-Daszkiewicz} {et~al.}(2013){Daszy{\'n}ska-Daszkiewicz}, {Szewczuk}, \& {Walczak}}]{2013MNRAS.431.3396D}
{Daszy{\'n}ska-Daszkiewicz}, J., {Szewczuk}, W., \& {Walczak}, P. 2013, \mnras, 431, 3396, \dodoi{10.1093/mnras/stt418}

\bibitem[{{Davies} {et~al.}(2010){Davies}, {Bremer}, {Stanway}, {Birkinshaw}, \& {Lehnert}}]{2010MNRAS.408L..31D}
{Davies}, L.~J.~M., {Bremer}, M.~N., {Stanway}, E.~R., {Birkinshaw}, M., \& {Lehnert}, M.~D. 2010, \mnras, 408, L31, \dodoi{10.1111/j.1745-3933.2010.00922.x}

\bibitem[{{de Jager} \& {Nieuwenhuijzen}(1987)}]{1987A&A...177..217D}
{de Jager}, C., \& {Nieuwenhuijzen}, H. 1987, \aap, 177, 217

\bibitem[{{Deleuil} {et~al.}(2018){Deleuil}, {Aigrain}, {Moutou}, {Cabrera}, {Bouchy}, {Deeg}, {Almenara}, {H{\'e}brard}, {Santerne}, {Alonso}, {Bonomo}, {Bord{\'e}}, {Csizmadia}, {D{\`\i}az}, {Erikson}, {Fridlund}, {Gandolfi}, {Guenther}, {Guillot}, {Guterman}, {Grziwa}, {Hatzes}, {L{\'e}ger}, {Mazeh}, {Ofir}, {Ollivier}, {P{\"a}tzold}, {Parviainen}, {Rauer}, {Rouan}, {Schneider}, {Titz-Weider}, {Tingley}, \& {Weingrill}}]{2018A&A...619A..97D}
{Deleuil}, M., {Aigrain}, S., {Moutou}, C., {et~al.} 2018, \aap, 619, A97, \dodoi{10.1051/0004-6361/201731068}

\bibitem[{{Desmet} {et~al.}(2009){Desmet}, {Briquet}, {Thoul}, {Zima}, {De Cat}, {Handler}, {Ilyin}, {Kambe}, {Krzesinski}, {Lehmann}, {Masuda}, {Mathias}, {Mkrtichian}, {Telting}, {Uytterhoeven}, {Yang}, \& {Aerts}}]{2009MNRAS.396.1460D}
{Desmet}, M., {Briquet}, M., {Thoul}, A., {et~al.} 2009, \mnras, 396, 1460, \dodoi{10.1111/j.1365-2966.2009.14790.x}

\bibitem[{{Dufton} {et~al.}(2013){Dufton}, {Langer}, {Dunstall}, {Evans}, {Brott}, {de Mink}, {Howarth}, {Kennedy}, {McEvoy}, {Potter}, {Ram{\'\i}rez-Agudelo}, {Sana}, {Sim{\'o}n-D{\'\i}az}, {Taylor}, \& {Vink}}]{2013A&A...550A.109D}
{Dufton}, P.~L., {Langer}, N., {Dunstall}, P.~R., {et~al.} 2013, \aap, 550, A109, \dodoi{10.1051/0004-6361/201220273}

\bibitem[{{Edelmann} {et~al.}(2019){Edelmann}, {Ratnasingam}, {Pedersen}, {Bowman}, {Prat}, \& {Rogers}}]{2019ApJ...876....4E}
{Edelmann}, P.~V.~F., {Ratnasingam}, R.~P., {Pedersen}, M.~G., {et~al.} 2019, \apj, 876, 4, \dodoi{10.3847/1538-4357/ab12df}

\bibitem[{{Feldmeier} {et~al.}(1997){Feldmeier}, {Puls}, \& {Pauldrach}}]{1997A&A...322..878F}
{Feldmeier}, A., {Puls}, J., \& {Pauldrach}, A.~W.~A. 1997, \aap, 322, 878

\bibitem[{{Foreman-Mackey} {et~al.}(2013){Foreman-Mackey}, {Hogg}, {Lang}, \& {Goodman}}]{2013PASP..125..306F}
{Foreman-Mackey}, D., {Hogg}, D.~W., {Lang}, D., \& {Goodman}, J. 2013, \pasp, 125, 306, \dodoi{10.1086/670067}

\bibitem[{{Fuller} {et~al.}(2014){Fuller}, {Lecoanet}, {Cantiello}, \& {Brown}}]{2014ApJ...796...17F}
{Fuller}, J., {Lecoanet}, D., {Cantiello}, M., \& {Brown}, B. 2014, \apj, 796, 17, \dodoi{10.1088/0004-637X/796/1/17}

\bibitem[{{Gaia Collaboration}(2022)}]{2022yCat.1355....0G}
{Gaia Collaboration}. 2022, VizieR Online Data Catalog, I/355, \dodoi{10.26093/cds/vizier.1355}

\bibitem[{{Gaia Collaboration} {et~al.}(2016){Gaia Collaboration}, {Prusti}, {de Bruijne}, {Brown}, {Vallenari}, {Babusiaux}, {Bailer-Jones}, {Bastian}, {Biermann}, {Evans}, {Eyer}, {Jansen}, {Jordi}, {Klioner}, {Lammers}, {Lindegren}, {Luri}, {Mignard}, {Milligan}, {Panem}, {Poinsignon}, {Pourbaix}, {Randich}, {Sarri}, {Sartoretti}, {Siddiqui}, {Soubiran}, {Valette}, {van Leeuwen}, {Walton}, {Aerts}, {Arenou}, {Cropper}, {Drimmel}, {H{\o}g}, {Katz}, {Lattanzi}, {O'Mullane}, {Grebel}, {Holland}, {Huc}, {Passot}, {Bramante}, {Cacciari}, {Casta{\~n}eda}, {Chaoul}, {Cheek}, {De Angeli}, {Fabricius}, {Guerra}, {Hern{\'a}ndez}, {Jean-Antoine-Piccolo}, {Masana}, {Messineo}, {Mowlavi}, {Nienartowicz}, {Ord{\'o}{\~n}ez-Blanco}, {Panuzzo}, {Portell}, {Richards}, {Riello}, {Seabroke}, {Tanga}, {Th{\'e}venin}, {Torra}, {Els}, {Gracia-Abril}, {Comoretto}, {Garcia-Reinaldos}, {Lock}, {Mercier}, {Altmann}, {Andrae}, {Astraatmadja}, {Bellas-Velidis}, {Benson}, {Berthier}, {Blomme}, {Busso}, {Carry}, {Cellino}, {Clementini}, {Cowell}, {Creevey}, {Cuypers}, {Davidson}, {De Ridder}, {de Torres}, {Delchambre}, {Dell'Oro}, {Ducourant}, {Fr{\'e}mat}, {Garc{\'\i}a-Torres}, {Gosset}, {Halbwachs}, {Hambly}, {Harrison}, {Hauser}, {Hestroffer}, {Hodgkin}, {Huckle}, {Hutton}, {Jasniewicz}, {Jordan}, {Kontizas}, {Korn}, {Lanzafame}, {Manteiga}, {Moitinho}, {Muinonen}, {Osinde}, {Pancino}, {Pauwels}, {Petit}, {Recio-Blanco}, {Robin}, {Sarro}, {Siopis}, {Smith}, {Smith}, {Sozzetti}, {Thuillot}, {van Reeven}, {Viala}, {Abbas}, {Abreu Aramburu}, {Accart}, {Aguado}, {Allan}, {Allasia}, {Altavilla}, {{\'A}lvarez}, {Alves}, {Anderson}, {Andrei}, {Anglada Varela}, {Antiche}, {Antoja}, {Ant{\'o}n}, {Arcay}, {Atzei}, {Ayache}, {Bach}, {Baker}, {Balaguer-N{\'u}{\~n}ez}, {Barache}, {Barata}, {Barbier}, {Barblan}, {Baroni}, {Barrado y Navascu{\'e}s}, {Barros}, {Barstow}, {Becciani}, {Bellazzini}, {Bellei}, {Bello Garc{\'\i}a}, {Belokurov}, {Bendjoya}, {Berihuete}, {Bianchi}, {Bienaym{\'e}}, {Billebaud}, {Blagorodnova}, {Blanco-Cuaresma}, {Boch}, {Bombrun}, {Borrachero}, {Bouquillon}, {Bourda}, {Bouy}, {Bragaglia}, {Breddels}, {Brouillet}, {Br{\"u}semeister}, {Bucciarelli}, {Budnik}, {Burgess}, {Burgon}, {Burlacu}, {Busonero}, {Buzzi}, {Caffau}, {Cambras}, {Campbell}, {Cancelliere}, {Cantat-Gaudin}, {Carlucci}, {Carrasco}, {Castellani}, {Charlot}, {Charnas}, {Charvet}, {Chassat}, {Chiavassa}, {Clotet}, {Cocozza}, {Collins}, {Collins}, {Costigan}, {Crifo}, {Cross}, {Crosta}, {Crowley}, {Dafonte}, {Damerdji}, {Dapergolas}, {David}, {David}, {De Cat}, {de Felice}, {de Laverny}, {De Luise}, {De March}, {de Martino}, {de Souza}, {Debosscher}, {del Pozo}, {Delbo}, {Delgado}, {Delgado}, {di Marco}, {Di Matteo}, {Diakite}, {Distefano}, {Dolding}, {Dos Anjos}, {Drazinos}, {Dur{\'a}n}, {Dzigan}, {Ecale}, {Edvardsson}, {Enke}, {Erdmann}, {Escolar}, {Espina}, {Evans}, {Eynard Bontemps}, {Fabre}, {Fabrizio}, {Faigler}, {Falc{\~a}o}, {Farr{\`a}s Casas}, {Faye}, {Federici}, {Fedorets}, {Fern{\'a}ndez-Hern{\'a}ndez}, {Fernique}, {Fienga}, {Figueras}, {Filippi}, {Findeisen}, {Fonti}, {Fouesneau}, {Fraile}, {Fraser}, {Fuchs}, {Furnell}, {Gai}, {Galleti}, {Galluccio}, {Garabato}, {Garc{\'\i}a-Sedano}, {Gar{\'e}}, {Garofalo}, {Garralda}, {Gavras}, {Gerssen}, {Geyer}, {Gilmore}, {Girona}, {Giuffrida}, {Gomes}, {Gonz{\'a}lez-Marcos}, {Gonz{\'a}lez-N{\'u}{\~n}ez}, {Gonz{\'a}lez-Vidal}, {Granvik}, {Guerrier}, {Guillout}, {Guiraud}, {G{\'u}rpide}, {Guti{\'e}rrez-S{\'a}nchez}, {Guy}, {Haigron}, {Hatzidimitriou}, {Haywood}, {Heiter}, {Helmi}, {Hobbs}, {Hofmann}, {Holl}, {Holland}, {Hunt}, {Hypki}, {Icardi}, {Irwin}, {Jevardat de Fombelle}, {Jofr{\'e}}, {Jonker}, {Jorissen}, {Julbe}, {Karampelas}, {Kochoska}, {Kohley}, {Kolenberg}, {Kontizas}, {Koposov}, {Kordopatis}, {Koubsky}, {Kowalczyk}, {Krone-Martins}, {Kudryashova}, {Kull}, {Bachchan}, {Lacoste-Seris}, {Lanza}, {Lavigne}, {Le Poncin-Lafitte}, {Lebreton}, {Lebzelter}, {Leccia}, {Leclerc}, {Lecoeur-Taibi}, {Lemaitre}, {Lenhardt}, {Leroux}, {Liao}, {Licata}, {Lindstr{\o}m}, {Lister}, {Livanou}, {Lobel}, {L{\"o}ffler}, {L{\'o}pez}, {Lopez-Lozano}, {Lorenz}, {Loureiro}, {MacDonald}, {Magalh{\~a}es Fernandes}, {Managau}, {Mann}, {Mantelet}, {Marchal}, {Marchant}, {Marconi}, {Marie}, {Marinoni}, {Marrese}, {Marschalk{\'o}}, {Marshall}, {Mart{\'\i}n-Fleitas}, {Martino}, {Mary}, {Matijevi{\v{c}}}, {Mazeh}, {McMillan}, {Messina}, {Mestre}, {Michalik}, {Millar}, {Miranda}, {Molina}, {Molinaro}, {Molinaro}, {Moln{\'a}r}, {Moniez}, {Montegriffo}, {Monteiro}, {Mor}, {Mora}, {Morbidelli}, {Morel}, {Morgenthaler}, {Morley}, {Morris}, {Mulone}, {Muraveva}, {Musella}, {Narbonne}, {Nelemans}, {Nicastro}, {Noval}, {Ord{\'e}novic}, {Ordieres-Mer{\'e}}, {Osborne}, {Pagani}, {Pagano}, {Pailler}, {Palacin}, {Palaversa}, {Parsons}, {Paulsen}, {Pecoraro}, {Pedrosa}, {Pentik{\"a}inen}, {Pereira}, {Pichon}, {Piersimoni}, {Pineau}, {Plachy}, {Plum}, {Poujoulet}, {Pr{\v{s}}a}, {Pulone}, {Ragaini}, {Rago}, {Rambaux}, {Ramos-Lerate}, {Ranalli}, {Rauw}, {Read}, {Regibo}, {Renk}, {Reyl{\'e}}, {Ribeiro}, {Rimoldini}, {Ripepi}, {Riva}, {Rixon}, {Roelens}, {Romero-G{\'o}mez}, {Rowell}, {Royer}, {Rudolph}, {Ruiz-Dern}, {Sadowski}, {Sagrist{\`a} Sell{\'e}s}, {Sahlmann}, {Salgado}, {Salguero}, {Sarasso}, {Savietto}, {Schnorhk}, {Schultheis}, {Sciacca}, {Segol}, {Segovia}, {Segransan}, {Serpell}, {Shih}, {Smareglia}, {Smart}, {Smith}, {Solano}, {Solitro}, {Sordo}, {Soria Nieto}, {Souchay}, {Spagna}, {Spoto}, {Stampa}, {Steele}, {Steidelm{\"u}ller}, {Stephenson}, {Stoev}, {Suess}, {S{\"u}veges}, {Surdej}, {Szabados}, {Szegedi-Elek}, {Tapiador}, {Taris}, {Tauran}, {Taylor}, {Teixeira}, {Terrett}, {Tingley}, {Trager}, {Turon}, {Ulla}, {Utrilla}, {Valentini}, {van Elteren}, {Van Hemelryck}, {van Leeuwen}, {Varadi}, {Vecchiato}, {Veljanoski}, {Via}, {Vicente}, {Vogt}, {Voss}, {Votruba}, {Voutsinas}, {Walmsley}, {Weiler}, {Weingrill}, {Werner}, {Wevers}, {Whitehead}, {Wyrzykowski}, {Yoldas}, {{\v{Z}}erjal}, {Zucker}, {Zurbach}, {Zwitter}, {Alecu}, {Allen}, {Allende Prieto}, {Amorim}, {Anglada-Escud{\'e}}, {Arsenijevic}, {Azaz}, {Balm}, {Beck}, {Bernstein}, {Bigot}, {Bijaoui}, {Blasco}, {Bonfigli}, {Bono}, {Boudreault}, {Bressan}, {Brown}, {Brunet}, {Bunclark}, {Buonanno}, {Butkevich}, {Carret}, {Carrion}, {Chemin}, {Ch{\'e}reau}, {Corcione}, {Darmigny}, {de Boer}, {de Teodoro}, {de Zeeuw}, {Delle Luche}, {Domingues}, {Dubath}, {Fodor}, {Fr{\'e}zouls}, {Fries}, {Fustes}, {Fyfe}, {Gallardo}, {Gallegos}, {Gardiol}, {Gebran}, {Gomboc}, {G{\'o}mez}, {Grux}, {Gueguen}, {Heyrovsky}, {Hoar}, {Iannicola}, {Isasi Parache}, {Janotto}, {Joliet}, {Jonckheere}, {Keil}, {Kim}, {Klagyivik}, {Klar}, {Knude}, {Kochukhov}, {Kolka}, {Kos}, {Kutka}, {Lainey}, {LeBouquin}, {Liu}, {Loreggia}, {Makarov}, {Marseille}, {Martayan}, {Martinez-Rubi}, {Massart}, {Meynadier}, {Mignot}, {Munari}, {Nguyen}, {Nordlander}, {Ocvirk}, {O'Flaherty}, {Olias Sanz}, {Ortiz}, {Osorio}, {Oszkiewicz}, {Ouzounis}, {Palmer}, {Park}, {Pasquato}, {Peltzer}, {Peralta}, {P{\'e}turaud}, {Pieniluoma}, {Pigozzi}, {Poels}, {Prat}, {Prod'homme}, {Raison}, {Rebordao}, {Risquez}, {Rocca-Volmerange}, {Rosen}, {Ruiz-Fuertes}, {Russo}, {Sembay}, {Serraller Vizcaino}, {Short}, {Siebert}, {Silva}, {Sinachopoulos}, {Slezak}, {Soffel}, {Sosnowska}, {Strai{\v{z}}ys}, {ter Linden}, {Terrell}, {Theil}, {Tiede}, {Troisi}, {Tsalmantza}, {Tur}, {Vaccari}, {Vachier}, {Valles}, {Van Hamme}, {Veltz}, {Virtanen}, {Wallut}, {Wichmann}, {Wilkinson}, {Ziaeepour}, \& {Zschocke}}]{2016A&A...595A...1G}
{Gaia Collaboration}, {Prusti}, T., {de Bruijne}, J.~H.~J., {et~al.} 2016, \aap, 595, A1, \dodoi{10.1051/0004-6361/201629272}

\bibitem[{{Gatto} {et~al.}(2015){Gatto}, {Walch}, {Low}, {Naab}, {Girichidis}, {Glover}, {W{\"u}nsch}, {Klessen}, {Clark}, {Baczynski}, {Peters}, {Ostriker}, {Ib{\'a}{\~n}ez-Mej{\'\i}a}, \& {Haid}}]{2015MNRAS.449.1057G}
{Gatto}, A., {Walch}, S., {Low}, M. M.~M., {et~al.} 2015, \mnras, 449, 1057, \dodoi{10.1093/mnras/stv324}

\bibitem[{{Gelman} \& {Rubin}(1992)}]{1992StaSc...7..457G}
{Gelman}, A., \& {Rubin}, D.~B. 1992, Statistical Science, 7, 457, \dodoi{10.1214/ss/1177011136}

\bibitem[{{Giacalone} {et~al.}(2021){Giacalone}, {Dressing}, {Jensen}, {Collins}, {Ricker}, {Vanderspek}, {Seager}, {Winn}, {Jenkins}, {Barclay}, {Barkaoui}, {Cadieux}, {Charbonneau}, {Collins}, {Conti}, {Doyon}, {Evans}, {Ghachoui}, {Gillon}, {Guerrero}, {Hart}, {Jehin}, {Kielkopf}, {McLean}, {Murgas}, {Palle}, {Parviainen}, {Pozuelos}, {Relles}, {Shporer}, {Socia}, {Stockdale}, {Tan}, {Torres}, {Twicken}, {Waalkes}, \& {Waite}}]{2021AJ....161...24G}
{Giacalone}, S., {Dressing}, C.~D., {Jensen}, E. L.~N., {et~al.} 2021, \aj, 161, 24, \dodoi{10.3847/1538-3881/abc6af}

\bibitem[{{Ginsburg} {et~al.}(2019){Ginsburg}, {Sip{\H{o}}cz}, {Brasseur}, {Cowperthwaite}, {Craig}, {Deil}, {Guillochon}, {Guzman}, {Liedtke}, {Lian Lim}, {Lockhart}, {Mommert}, {Morris}, {Norman}, {Parikh}, {Persson}, {Robitaille}, {Segovia}, {Singer}, {Tollerud}, {de Val-Borro}, {Valtchanov}, {Woillez}, {Astroquery Collaboration}, \& {a subset of astropy Collaboration}}]{2019AJ....157...98G}
{Ginsburg}, A., {Sip{\H{o}}cz}, B.~M., {Brasseur}, C.~E., {et~al.} 2019, \aj, 157, 98, \dodoi{10.3847/1538-3881/aafc33}

\bibitem[{{Goldreich} \& {Kumar}(1990)}]{1990ApJ...363..694G}
{Goldreich}, P., \& {Kumar}, P. 1990, \apj, 363, 694, \dodoi{10.1086/169376}

\bibitem[{{Grassitelli} {et~al.}(2015){Grassitelli}, {Fossati}, {Sim{\'o}n-Di{\'a}z}, {Langer}, {Castro}, \& {Sanyal}}]{2015ApJ...808L..31G}
{Grassitelli}, L., {Fossati}, L., {Sim{\'o}n-Di{\'a}z}, S., {et~al.} 2015, \apjl, 808, L31, \dodoi{10.1088/2041-8205/808/1/L31}

\bibitem[{{Handberg} {et~al.}(2021){Handberg}, {Lund}, {White}, {Hall}, {Buzasi}, {Pope}, {Hansen}, {von Essen}, {Carboneau}, {Huber}, {Vanderspek}, {Fausnaugh}, {Tenenbaum}, {Jenkins}, \& {T'DA Collaboration}}]{2021AJ....162..170H}
{Handberg}, R., {Lund}, M.~N., {White}, T.~R., {et~al.} 2021, \aj, 162, 170, \dodoi{10.3847/1538-3881/ac09f1}

\bibitem[{{Handler} {et~al.}(2005){Handler}, {Shobbrook}, \& {Mokgwetsi}}]{2005MNRAS.362..612H}
{Handler}, G., {Shobbrook}, R.~R., \& {Mokgwetsi}, T. 2005, \mnras, 362, 612, \dodoi{10.1111/j.1365-2966.2005.09341.x}

\bibitem[{{Handler} {et~al.}(2012){Handler}, {Shobbrook}, {Uytterhoeven}, {Briquet}, {Neiner}, {Tshenye}, {Ngwato}, {van Winckel}, {Guggenberger}, {Raskin}, {Rodr{\'\i}guez}, {Mazumdar}, {Barban}, {Lorenz}, {Vandenbussche}, {{\c{S}}ahin}, {Medupe}, \& {Aerts}}]{2012MNRAS.424.2380H}
{Handler}, G., {Shobbrook}, R.~R., {Uytterhoeven}, K., {et~al.} 2012, \mnras, 424, 2380, \dodoi{10.1111/j.1365-2966.2012.21414.x}

\bibitem[{{Handler} {et~al.}(2017){Handler}, {Rybicka}, {Popowicz}, {Pigulski}, {Kuschnig}, {Zoc{\l}o{\'n}ska}, {Moffat}, {Weiss}, {Grant}, {Pablo}, {Whittaker}, {Ruci{\'n}ski}, {Ramiaramanantsoa}, {Zwintz}, \& {Wade}}]{2017MNRAS.464.2249H}
{Handler}, G., {Rybicka}, M., {Popowicz}, A., {et~al.} 2017, \mnras, 464, 2249, \dodoi{10.1093/mnras/stw2518}

\bibitem[{{Handler} {et~al.}(2019){Handler}, {Pigulski}, {Daszy{\'n}ska-Daszkiewicz}, {Irrgang}, {Kilkenny}, {Guo}, {Przybilla}, {Kahraman Ali{\c{c}}avu{\c{s}}}, {Kallinger}, {Pascual-Granado}, {Niemczura}, {R{\'o}{\.z}a{\'n}ski}, {Chowdhury}, {Buzasi}, {Mirouh}, {Bowman}, {Johnston}, {Pedersen}, {Sim{\'o}n-D{\'\i}az}, {Moravveji}, {Gazeas}, {De Cat}, {Vanderspek}, \& {Ricker}}]{2019ApJ...873L...4H}
{Handler}, G., {Pigulski}, A., {Daszy{\'n}ska-Daszkiewicz}, J., {et~al.} 2019, \apjl, 873, L4, \dodoi{10.3847/2041-8213/ab095f}

\bibitem[{{Harris} {et~al.}(2020){Harris}, {Millman}, {van der Walt}, {Gommers}, {Virtanen}, {Cournapeau}, {Wieser}, {Taylor}, {Berg}, {Smith}, {Kern}, {Picus}, {Hoyer}, {van Kerkwijk}, {Brett}, {Haldane}, {del R{\'\i}o}, {Wiebe}, {Peterson}, {G{\'e}rard-Marchant}, {Sheppard}, {Reddy}, {Weckesser}, {Abbasi}, {Gohlke}, \& {Oliphant}}]{2020Natur.585..357H}
{Harris}, C.~R., {Millman}, K.~J., {van der Walt}, S.~J., {et~al.} 2020, \nat, 585, 357, \dodoi{10.1038/s41586-020-2649-2}

\bibitem[{{Herwig} {et~al.}(2023){Herwig}, {Woodward}, {Mao}, {Thompson}, {Denissenkov}, {Lau}, {Blouin}, {Andrassy}, \& {Paul}}]{2023MNRAS.525.1601H}
{Herwig}, F., {Woodward}, P.~R., {Mao}, H., {et~al.} 2023, \mnras, 525, 1601, \dodoi{10.1093/mnras/stad2157}

\bibitem[{{Higgins} \& {Bell}(2023)}]{2023AJ....165..141H}
{Higgins}, M.~E., \& {Bell}, K.~J. 2023, \aj, 165, 141, \dodoi{10.3847/1538-3881/acb20c}

\bibitem[{{Holgado} {et~al.}(2018){Holgado}, {Sim{\'o}n-D{\'\i}az}, {Barb{\'a}}, {Puls}, {Herrero}, {Castro}, {Garcia}, {Ma{\'\i}z Apell{\'a}niz}, {Negueruela}, \& {Sab{\'\i}n-Sanjuli{\'a}n}}]{2018A&A...613A..65H}
{Holgado}, G., {Sim{\'o}n-D{\'\i}az}, S., {Barb{\'a}}, R.~H., {et~al.} 2018, \aap, 613, A65, \dodoi{10.1051/0004-6361/201731543}

\bibitem[{{Holgado} {et~al.}(2020){Holgado}, {Sim{\'o}n-D{\'\i}az}, {Haemmerl{\'e}}, {Lennon}, {Barb{\'a}}, {Cervi{\~n}o}, {Castro}, {Herrero}, {Meynet}, \& {Arias}}]{2020A&A...638A.157H}
{Holgado}, G., {Sim{\'o}n-D{\'\i}az}, S., {Haemmerl{\'e}}, L., {et~al.} 2020, \aap, 638, A157, \dodoi{10.1051/0004-6361/202037699}

\bibitem[{{Horst} {et~al.}(2020){Horst}, {Edelmann}, {Andr{\'a}ssy}, {R{\"o}pke}, {Bowman}, {Aerts}, \& {Ratnasingam}}]{2020A&A...641A..18H}
{Horst}, L., {Edelmann}, P.~V.~F., {Andr{\'a}ssy}, R., {et~al.} 2020, \aap, 641, A18, \dodoi{10.1051/0004-6361/202037531}

\bibitem[{{Hu} {et~al.}(2017){Hu}, {Naab}, {Glover}, {Walch}, \& {Clark}}]{2017MNRAS.471.2151H}
{Hu}, C.-Y., {Naab}, T., {Glover}, S. C.~O., {Walch}, S., \& {Clark}, P.~C. 2017, \mnras, 471, 2151, \dodoi{10.1093/mnras/stx1773}

\bibitem[{{Hunter} {et~al.}(2008){Hunter}, {Lennon}, {Dufton}, {Trundle}, {Sim{\'o}n-D{\'\i}az}, {Smartt}, {Ryans}, \& {Evans}}]{2008A&A...479..541H}
{Hunter}, I., {Lennon}, D.~J., {Dufton}, P.~L., {et~al.} 2008, \aap, 479, 541, \dodoi{10.1051/0004-6361:20078511}

\bibitem[{{Hunter}(2007)}]{2007CSE.....9...90H}
{Hunter}, J.~D. 2007, Computing in Science and Engineering, 9, 90, \dodoi{10.1109/MCSE.2007.55}

\bibitem[{{Jermyn} {et~al.}(2022){Jermyn}, {Anders}, \& {Cantiello}}]{2022ApJ...926..221J}
{Jermyn}, A.~S., {Anders}, E.~H., \& {Cantiello}, M. 2022, \apj, 926, 221, \dodoi{10.3847/1538-4357/ac4e89}

\bibitem[{{Jermyn} \& {Cantiello}(2020)}]{2020ApJ...900..113J}
{Jermyn}, A.~S., \& {Cantiello}, M. 2020, \apj, 900, 113, \dodoi{10.3847/1538-4357/ab9e70}

\bibitem[{{Jermyn} {et~al.}(2023){Jermyn}, {Bauer}, {Schwab}, {Farmer}, {Ball}, {Bellinger}, {Dotter}, {Joyce}, {Marchant}, {Mombarg}, {Wolf}, {Sunny Wong}, {Cinquegrana}, {Farrell}, {Smolec}, {Thoul}, {Cantiello}, {Herwig}, {Toloza}, {Bildsten}, {Townsend}, \& {Timmes}}]{2023ApJS..265...15J}
{Jermyn}, A.~S., {Bauer}, E.~B., {Schwab}, J., {et~al.} 2023, \apjs, 265, 15, \dodoi{10.3847/1538-4365/acae8d}

\bibitem[{{Jiang} {et~al.}(2017){Jiang}, {Cantiello}, {Bildsten}, {Quataert}, \& {Blaes}}]{2017ApJ...843...68J}
{Jiang}, Y.-F., {Cantiello}, M., {Bildsten}, L., {Quataert}, E., \& {Blaes}, O. 2017, \apj, 843, 68, \dodoi{10.3847/1538-4357/aa77b0}

\bibitem[{{Kallinger} \& {Matthews}(2010)}]{2010ApJ...711L..35K}
{Kallinger}, T., \& {Matthews}, J.~M. 2010, \apjl, 711, L35, \dodoi{10.1088/2041-8205/711/1/L35}

\bibitem[{{Kallinger} {et~al.}(2014){Kallinger}, {De Ridder}, {Hekker}, {Mathur}, {Mosser}, {Gruberbauer}, {Garc{\'\i}a}, {Karoff}, \& {Ballot}}]{2014A&A...570A..41K}
{Kallinger}, T., {De Ridder}, J., {Hekker}, S., {et~al.} 2014, \aap, 570, A41, \dodoi{10.1051/0004-6361/201424313}

\bibitem[{{Kim} \& {MacGregor}(2000)}]{2000AIPC..537..256K}
{Kim}, E.-J., \& {MacGregor}, K.~B. 2000, in American Institute of Physics Conference Series, Vol. 537, Waves in Dusty, Solar, and Space Plasmas, ed. F.~{Verheest}, M.~{Goossens}, M.~A. {Hellberg}, \& R.~{Bharuthram} (AIP), 256--263, \dodoi{10.1063/1.1324948}

\bibitem[{{Kraus} {et~al.}(2015){Kraus}, {Haucke}, {Cidale}, {Venero}, {Nickeler}, {N{\'e}meth}, {Niemczura}, {Tomi{\'c}}, {Aret}, {Kub{\'a}t}, {Kub{\'a}tov{\'a}}, {Oksala}, {Cur{\'e}}, {Kami{\'n}ski}, {Dimitrov}, {Fagas}, \& {Poli{\'n}ska}}]{2015A&A...581A..75K}
{Kraus}, M., {Haucke}, M., {Cidale}, L.~S., {et~al.} 2015, \aap, 581, A75, \dodoi{10.1051/0004-6361/201425383}

\bibitem[{{Krti{\v{c}}ka} \& {Feldmeier}(2018)}]{2018A&A...617A.121K}
{Krti{\v{c}}ka}, J., \& {Feldmeier}, A. 2018, \aap, 617, A121, \dodoi{10.1051/0004-6361/201731614}

\bibitem[{{Krti{\v{c}}ka} \& {Feldmeier}(2021)}]{2021A&A...648A..79K}
---. 2021, \aap, 648, A79, \dodoi{10.1051/0004-6361/202040148}

\bibitem[{{Kudritzki}(2010)}]{2010AN....331..459K}
{Kudritzki}, R.~P. 2010, Astronomische Nachrichten, 331, 459, \dodoi{10.1002/asna.200911342}

\bibitem[{{Kudritzki} \& {Puls}(2000)}]{2000ARA&A..38..613K}
{Kudritzki}, R.-P., \& {Puls}, J. 2000, \araa, 38, 613, \dodoi{10.1146/annurev.astro.38.1.613}

\bibitem[{{Kumar} {et~al.}(1999){Kumar}, {Talon}, \& {Zahn}}]{1999ApJ...520..859K}
{Kumar}, P., {Talon}, S., \& {Zahn}, J.-P. 1999, \apj, 520, 859, \dodoi{10.1086/307464}

\bibitem[{{Kurtz}(1985)}]{1985MNRAS.213..773K}
{Kurtz}, D.~W. 1985, \mnras, 213, 773, \dodoi{10.1093/mnras/213.4.773}

\bibitem[{{Langer}(2012)}]{2012ARA&A..50..107L}
{Langer}, N. 2012, \araa, 50, 107, \dodoi{10.1146/annurev-astro-081811-125534}

\bibitem[{{Langer} \& {Kudritzki}(2014)}]{2014A&A...564A..52L}
{Langer}, N., \& {Kudritzki}, R.~P. 2014, \aap, 564, A52, \dodoi{10.1051/0004-6361/201423374}

\bibitem[{{Le Saux} {et~al.}(2023){Le Saux}, {Baraffe}, {Guillet}, {Vlaykov}, {Morison}, {Pratt}, {Constantino}, \& {Goffrey}}]{2023MNRAS.522.2835L}
{Le Saux}, A., {Baraffe}, I., {Guillet}, T., {et~al.} 2023, \mnras, 522, 2835, \dodoi{10.1093/mnras/stad1067}

\bibitem[{{Lecoanet} {et~al.}(2021){Lecoanet}, {Cantiello}, {Anders}, {Quataert}, {Couston}, {Bouffard}, {Favier}, \& {Le Bars}}]{2021MNRAS.508..132L}
{Lecoanet}, D., {Cantiello}, M., {Anders}, E.~H., {et~al.} 2021, \mnras, 508, 132, \dodoi{10.1093/mnras/stab2524}

\bibitem[{{Lecoanet} \& {Quataert}(2013)}]{2013MNRAS.430.2363L}
{Lecoanet}, D., \& {Quataert}, E. 2013, \mnras, 430, 2363, \dodoi{10.1093/mnras/stt055}

\bibitem[{{Lecoanet} {et~al.}(2019){Lecoanet}, {Cantiello}, {Quataert}, {Couston}, {Burns}, {Pope}, {Jermyn}, {Favier}, \& {Le Bars}}]{2019ApJ...886L..15L}
{Lecoanet}, D., {Cantiello}, M., {Quataert}, E., {et~al.} 2019, \apjl, 886, L15, \dodoi{10.3847/2041-8213/ab5446}

\bibitem[{{Lenoir-Craig} {et~al.}(2022){Lenoir-Craig}, {St-Louis}, {Moffat}, {Pablo}, {Handler}, {Kuschnig}, {Popowicz}, {Wade}, \& {Zwintz}}]{2022ApJ...925...79L}
{Lenoir-Craig}, G., {St-Louis}, N., {Moffat}, A. F.~J., {et~al.} 2022, \apj, 925, 79, \dodoi{10.3847/1538-4357/ac397d}

\bibitem[{{Lenz} \& {Breger}(2005)}]{2005CoAst.146...53L}
{Lenz}, P., \& {Breger}, M. 2005, Communications in Asteroseismology, 146, 53, \dodoi{10.1553/cia146s53}

\bibitem[{{Lighthill}(1952)}]{1952RSPSA.211..564L}
{Lighthill}, M.~J. 1952, Proceedings of the Royal Society of London Series A, 211, 564, \dodoi{10.1098/rspa.1952.0060}

\bibitem[{{Lightkurve Collaboration} {et~al.}(2018){Lightkurve Collaboration}, {Cardoso}, {Hedges}, {Gully-Santiago}, {Saunders}, {Cody}, {Barclay}, {Hall}, {Sagear}, {Turtelboom}, {Zhang}, {Tzanidakis}, {Mighell}, {Coughlin}, {Bell}, {Berta-Thompson}, {Williams}, {Dotson}, \& {Barentsen}}]{2018ascl.soft12013L}
{Lightkurve Collaboration}, {Cardoso}, J. V. d.~M., {Hedges}, C., {et~al.} 2018, {Lightkurve: Kepler and TESS time series analysis in Python}, Astrophysics Source Code Library, record ascl:1812.013.
\newblock \doeprint{1812.013}

\bibitem[{{Liu} {et~al.}(2020){Liu}, {Shao}, {Zhao}, \& {Gao}}]{2020MNRAS.496..182L}
{Liu}, C., {Shao}, L., {Zhao}, J., \& {Gao}, Y. 2020, \mnras, 496, 182, \dodoi{10.1093/mnras/staa1512}

\bibitem[{{Lucy} \& {Solomon}(1970)}]{1970ApJ...159..879L}
{Lucy}, L.~B., \& {Solomon}, P.~M. 1970, \apj, 159, 879, \dodoi{10.1086/150365}

\bibitem[{{Mac Low} {et~al.}(2005){Mac Low}, {Balsara}, {Kim}, \& {de Avillez}}]{2005ApJ...626..864M}
{Mac Low}, M.-M., {Balsara}, D.~S., {Kim}, J., \& {de Avillez}, M.~A. 2005, \apj, 626, 864, \dodoi{10.1086/430122}

\bibitem[{{Maeder} {et~al.}(2008{\natexlab{a}}){Maeder}, {Georgy}, \& {Meynet}}]{2008A&A...479L..37M}
{Maeder}, A., {Georgy}, C., \& {Meynet}, G. 2008{\natexlab{a}}, \aap, 479, L37, \dodoi{10.1051/0004-6361:20079007}

\bibitem[{{Maeder} \& {Meynet}(2000)}]{2000ARA&A..38..143M}
{Maeder}, A., \& {Meynet}, G. 2000, \araa, 38, 143, \dodoi{10.1146/annurev.astro.38.1.143}

\bibitem[{{Maeder} {et~al.}(2008{\natexlab{b}}){Maeder}, {Meynet}, {Ekstr{\"o}m}, {Hirschi}, \& {Georgy}}]{2008IAUS..250....3M}
{Maeder}, A., {Meynet}, G., {Ekstr{\"o}m}, S., {Hirschi}, R., \& {Georgy}, C. 2008{\natexlab{b}}, in Massive Stars as Cosmic Engines, ed. F.~{Bresolin}, P.~A. {Crowther}, \& J.~{Puls}, Vol. 250, 3--16, \dodoi{10.1017/S1743921308020292}

\bibitem[{{Ma{\'\i}z Apell{\'a}niz} {et~al.}(2012){Ma{\'\i}z Apell{\'a}niz}, {Pellerin}, {Barb{\'a}}, {Sim{\'o}n-D{\'\i}az}, {Alfaro}, {Morrell}, {Sota}, {Penad{\'e}s Ordaz}, \& {Gallego Calvente}}]{2012ASPC..465..484M}
{Ma{\'\i}z Apell{\'a}niz}, J., {Pellerin}, A., {Barb{\'a}}, R.~H., {et~al.} 2012, in Astronomical Society of the Pacific Conference Series, Vol. 465, Proceedings of a Scientific Meeting in Honor of Anthony F. J. Moffat, ed. L.~{Drissen}, C.~{Robert}, N.~{St-Louis}, \& A.~F.~J. {Moffat}, 484, \dodoi{10.48550/arXiv.1109.1492}

\bibitem[{{Ma{\'\i}z Apell{\'a}niz} {et~al.}(2016){Ma{\'\i}z Apell{\'a}niz}, {Sota}, {Arias}, {Barb{\'a}}, {Walborn}, {Sim{\'o}n-D{\'\i}az}, {Negueruela}, {Marco}, {Le{\~a}o}, {Herrero}, {Gamen}, \& {Alfaro}}]{2016ApJS..224....4M}
{Ma{\'\i}z Apell{\'a}niz}, J., {Sota}, A., {Arias}, J.~I., {et~al.} 2016, \apjs, 224, 4, \dodoi{10.3847/0067-0049/224/1/4}

\bibitem[{{Mathur} {et~al.}(2011){Mathur}, {Hekker}, {Trampedach}, {Ballot}, {Kallinger}, {Buzasi}, {Garc{\'\i}a}, {Huber}, {Jim{\'e}nez}, {Mosser}, {Bedding}, {Elsworth}, {R{\'e}gulo}, {Stello}, {Chaplin}, {De Ridder}, {Hale}, {Kinemuchi}, {Kjeldsen}, {Mullally}, \& {Thompson}}]{2011ApJ...741..119M}
{Mathur}, S., {Hekker}, S., {Trampedach}, R., {et~al.} 2011, \apj, 741, 119, \dodoi{10.1088/0004-637X/741/2/119}

\bibitem[{{Modjaz} {et~al.}(2019){Modjaz}, {Guti{\'e}rrez}, \& {Arcavi}}]{2019NatAs...3..717M}
{Modjaz}, M., {Guti{\'e}rrez}, C.~P., \& {Arcavi}, I. 2019, Nature Astronomy, 3, 717, \dodoi{10.1038/s41550-019-0856-2}

\bibitem[{{Montalb{\'a}n} \& {Schatzman}(2000)}]{2000A&A...354..943M}
{Montalb{\'a}n}, J., \& {Schatzman}, E. 2000, \aap, 354, 943

\bibitem[{{Naz{\'e}} {et~al.}(2021){Naz{\'e}}, {Rauw}, \& {Gosset}}]{2021MNRAS.502.5038N}
{Naz{\'e}}, Y., {Rauw}, G., \& {Gosset}, E. 2021, \mnras, 502, 5038, \dodoi{10.1093/mnras/stab133}

\bibitem[{{Nomoto} {et~al.}(2013){Nomoto}, {Kobayashi}, \& {Tominaga}}]{2013ARA&A..51..457N}
{Nomoto}, K., {Kobayashi}, C., \& {Tominaga}, N. 2013, \araa, 51, 457, \dodoi{10.1146/annurev-astro-082812-140956}

\bibitem[{{Otero}(2003)}]{2003IBVS.5480....1O}
{Otero}, S.~A. 2003, Information Bulletin on Variable Stars, 5480, 1

\bibitem[{{Otero}(2006)}]{2006OEJV...45....1O}
---. 2006, Open European Journal on Variable Stars, 45, 1

\bibitem[{{Otero}(2007)}]{2007OEJV...72....1O}
---. 2007, Open European Journal on Variable Stars, 0072, 1

\bibitem[{{Otero} \& {Claus}(2004)}]{2004IBVS.5495....1O}
{Otero}, S.~A., \& {Claus}, F. 2004, Information Bulletin on Variable Stars, 5495, 1

\bibitem[{{Otero} \& {Wils}(2005)}]{2005IBVS.5644....1O}
{Otero}, S.~A., \& {Wils}, P. 2005, Information Bulletin on Variable Stars, 5644, 1

\bibitem[{{Owocki} {et~al.}(1988){Owocki}, {Castor}, \& {Rybicki}}]{1988ApJ...335..914O}
{Owocki}, S.~P., {Castor}, J.~I., \& {Rybicki}, G.~B. 1988, \apj, 335, 914, \dodoi{10.1086/166977}

\bibitem[{{P{\'a}pics} {et~al.}(2017){P{\'a}pics}, {Tkachenko}, {Van Reeth}, {Aerts}, {Moravveji}, {Van de Sande}, {De Smedt}, {Bloemen}, {Southworth}, {Debosscher}, {Niemczura}, \& {Gameiro}}]{2017A&A...598A..74P}
{P{\'a}pics}, P.~I., {Tkachenko}, A., {Van Reeth}, T., {et~al.} 2017, \aap, 598, A74, \dodoi{10.1051/0004-6361/201629814}

\bibitem[{{Paxton} {et~al.}(2011){Paxton}, {Bildsten}, {Dotter}, {Herwig}, {Lesaffre}, \& {Timmes}}]{2011ApJS..192....3P}
{Paxton}, B., {Bildsten}, L., {Dotter}, A., {et~al.} 2011, \apjs, 192, 3, \dodoi{10.1088/0067-0049/192/1/3}

\bibitem[{{Paxton} {et~al.}(2013){Paxton}, {Cantiello}, {Arras}, {Bildsten}, {Brown}, {Dotter}, {Mankovich}, {Montgomery}, {Stello}, {Timmes}, \& {Townsend}}]{2013ApJS..208....4P}
{Paxton}, B., {Cantiello}, M., {Arras}, P., {et~al.} 2013, \apjs, 208, 4, \dodoi{10.1088/0067-0049/208/1/4}

\bibitem[{{Paxton} {et~al.}(2015){Paxton}, {Marchant}, {Schwab}, {Bauer}, {Bildsten}, {Cantiello}, {Dessart}, {Farmer}, {Hu}, {Langer}, {Townsend}, {Townsley}, \& {Timmes}}]{2015ApJS..220...15P}
{Paxton}, B., {Marchant}, P., {Schwab}, J., {et~al.} 2015, \apjs, 220, 15, \dodoi{10.1088/0067-0049/220/1/15}

\bibitem[{{Paxton} {et~al.}(2016){Paxton}, {Marchant}, {Schwab}, {Bauer}, {Bildsten}, {Cantiello}, {Dessart}, {Farmer}, {Hu}, {Langer}, {Townsend}, {Townsley}, \& {Timmes}}]{2016ApJS..223...18P}
---. 2016, \apjs, 223, 18, \dodoi{10.3847/0067-0049/223/1/18}

\bibitem[{{Paxton} {et~al.}(2018){Paxton}, {Schwab}, {Bauer}, {Bildsten}, {Blinnikov}, {Duffell}, {Farmer}, {Goldberg}, {Marchant}, {Sorokina}, {Thoul}, {Townsend}, \& {Timmes}}]{2018ApJS..234...34P}
{Paxton}, B., {Schwab}, J., {Bauer}, E.~B., {et~al.} 2018, \apjs, 234, 34, \dodoi{10.3847/1538-4365/aaa5a8}

\bibitem[{{Paxton} {et~al.}(2019){Paxton}, {Smolec}, {Schwab}, {Gautschy}, {Bildsten}, {Cantiello}, {Dotter}, {Farmer}, {Goldberg}, {Jermyn}, {Kanbur}, {Marchant}, {Thoul}, {Townsend}, {Wolf}, {Zhang}, \& {Timmes}}]{2019ApJS..243...10P}
{Paxton}, B., {Smolec}, R., {Schwab}, J., {et~al.} 2019, \apjs, 243, 10, \dodoi{10.3847/1538-4365/ab2241}

\bibitem[{{Pedersen} \& {Bell}(2023)}]{2023AJ....165..239P}
{Pedersen}, M.~G., \& {Bell}, K.~J. 2023, \aj, 165, 239, \dodoi{10.3847/1538-3881/accc31}

\bibitem[{{Pellerin} {et~al.}(2012){Pellerin}, {Ma{\'\i}z Apell{\'a}niz}, {Sim{\'o}n-D{\'\i}az}, \& {Barb{\'a}}}]{2012AAS...21922403P}
{Pellerin}, A., {Ma{\'\i}z Apell{\'a}niz}, J., {Sim{\'o}n-D{\'\i}az}, S., \& {Barb{\'a}}, R.~H. 2012, in American Astronomical Society Meeting Abstracts, Vol. 219, American Astronomical Society Meeting Abstracts \#219, 224.03

\bibitem[{{Pin{\c{c}}on} {et~al.}(2016){Pin{\c{c}}on}, {Belkacem}, \& {Goupil}}]{2016A&A...588A.122P}
{Pin{\c{c}}on}, C., {Belkacem}, K., \& {Goupil}, M.~J. 2016, \aap, 588, A122, \dodoi{10.1051/0004-6361/201527663}

\bibitem[{{Przybilla} {et~al.}(2008){Przybilla}, {Bresolin}, {Butler}, {Kudritzki}, {Urbaneja}, \& {Venn}}]{2008RMxAC..33..169P}
{Przybilla}, N., {Bresolin}, F., {Butler}, K., {et~al.} 2008, in Revista Mexicana de Astronomia y Astrofisica Conference Series, Vol.~33, Revista Mexicana de Astronomia y Astrofisica Conference Series, 169--170, \dodoi{10.48550/arXiv.astro-ph/0703361}

\bibitem[{{Puls} {et~al.}(2008){Puls}, {Vink}, \& {Najarro}}]{2008A&ARv..16..209P}
{Puls}, J., {Vink}, J.~S., \& {Najarro}, F. 2008, \aapr, 16, 209, \dodoi{10.1007/s00159-008-0015-8}

\bibitem[{{Ram{\'\i}rez-Agudelo} {et~al.}(2013){Ram{\'\i}rez-Agudelo}, {Sim{\'o}n-D{\'\i}az}, {Sana}, {de Koter}, {Sab{\'\i}n-Sanjul{\'\i}an}, {de Mink}, {Dufton}, {Gr{\"a}fener}, {Evans}, {Herrero}, {Langer}, {Lennon}, {Ma{\'\i}z Apell{\'a}niz}, {Markova}, {Najarro}, {Puls}, {Taylor}, \& {Vink}}]{2013A&A...560A..29R}
{Ram{\'\i}rez-Agudelo}, O.~H., {Sim{\'o}n-D{\'\i}az}, S., {Sana}, H., {et~al.} 2013, \aap, 560, A29, \dodoi{10.1051/0004-6361/201321986}

\bibitem[{{Ratnasingam} {et~al.}(2019){Ratnasingam}, {Edelmann}, \& {Rogers}}]{2019MNRAS.482.5500R}
{Ratnasingam}, R.~P., {Edelmann}, P.~V.~F., \& {Rogers}, T.~M. 2019, \mnras, 482, 5500, \dodoi{10.1093/mnras/sty3086}

\bibitem[{{Ratnasingam} {et~al.}(2020){Ratnasingam}, {Edelmann}, \& {Rogers}}]{2020MNRAS.497.4231R}
---. 2020, \mnras, 497, 4231, \dodoi{10.1093/mnras/staa2296}

\bibitem[{{Ratnasingam} {et~al.}(2023){Ratnasingam}, {Rogers}, {Chowdhury}, {Handler}, {Vanon}, {Varghese}, \& {Edelmann}}]{2023A&A...674A.134R}
{Ratnasingam}, R.~P., {Rogers}, T.~M., {Chowdhury}, S., {et~al.} 2023, \aap, 674, A134, \dodoi{10.1051/0004-6361/202245727}

\bibitem[{{Rauer} {et~al.}(2024){Rauer}, {Aerts}, {Cabrera}, {Deleuil}, {Erikson}, {Gizon}, {Goupil}, {Heras}, {Lorenzo-Alvarez}, {Marliani}, {Martin-Garcia}, {Mas-Hesse}, {O'Rourke}, {Osborn}, {Pagano}, {Piotto}, {Pollacco}, {Ragazzoni}, {Ramsay}, {Udry}, {Appourchaux}, {Benz}, {Brandeker}, {G{\"u}del}, {Janot-Pacheco}, {Kabath}, {Kjeldsen}, {Min}, {Santos}, {Smith}, {Suarez}, {Werner}, {Aboudan}, {Abreu}, {Acu{\~n}a}, {Adams}, {Adibekyan}, {Affer}, {Agneray}, {Agnor}, {Aguirre B{\o}rsen-Koch}, {Ahmed}, {Aigrain}, {Al-Bahlawan}, {Alcacera Gil}, {Alei}, {Alencar}, {Alexander}, {Alfonso-Garz{\'o}n}, {Alibert}, {Allende Prieto}, {Almeida}, {Alonso Sobrino}, {Altavilla}, {Althaus}, {Alonso Alvarez Trujillo}, {Amarsi}, {Ammler-von Eiff}, {Am{\^o}res}, {Andrade}, {Antoniadis-Karnavas}, {Ant{\'o}nio}, {Aparicio del Moral}, {Appolloni}, {Arena}, {Armstrong}, {Aroca Aliaga}, {Asplund}, {Audenaert}, {Auricchio}, {Avelino}, {Baeke}, {Bailli{\'e}}, {Balado}, {Balestra}, {Ball}, {Ballans}, {Ballot}, {Barban}, {Barbary}, {Barbieri}, {Barcel{\'o} Forteza}, {Barker}, {Barklem}, {Barnes}, {Barrado Navascues}, {Barragan}, {Baruteau}, {Basu}, {Baudin}, {Baumeister}, {Bayliss}, {Bazot}, {Beck}, {Bedding}, {Belkacem}, {Bellinger}, {Benatti}, {Benomar}, {B{\'e}rard}, {Bergemann}, {Bergomi}, {Bernardo}, {Biazzo}, {Bignamini}, {Bigot}, {Billot}, {Binet}, {Biondi}, {Biondi}, {Birch}, {Bitsch}, {Bluhm Ceballos}, {B{\'o}di}, {Bogn{\'a}r}, {Boisse}, {Bolmont}, {Bonanno}, {Bonavita}, {Bonfanti}, {Bonfils}, {Bonito}, {Bonomo}, {B{\"o}rner}, {Boro Saikia}, {Borreguero Mart{\'\i}n}, {Borsa}, {Borsato}, {Bossini}, {Bouchy}, {Bou{\'e}}, {Boufleur}, {Boumier}, {Bourrier}, {Bowman}, {Bozzo}, {Bradley}, {Bray}, {Bressan}, {Breton}, {Brienza}, {Brito}, {Brogi}, {Brown}, {Brown}, {Brun}, {Bruno}, {Bruns}, {Buchhave}, {Bugnet}, {Buldgen}, {Burgess}, {Busatta}, {Busso}, {Buzasi}, {Caballero}, {Cabral}, {Calderone}, {Cameron}, {Cameron}, {Campante}, {Canto Martins}, {Cara}, {Carone}, {Carrasco}, {Casagrande}, {Casewell}, {Cassisi}, {Castellani}, {Castro}, {Catala}, {Catal{\'a}n Fern{\'a}ndez}, {Catelan}, {Cegla}, {Cerruti}, {Cessa}, {Chadid}, {Chaplin}, {Charpinet}, {Chiappini}, {Chiarucci}, {Chiavassa}, {Chinellato}, {Chirulli}, {Christensen-Dalsgaard}, {Church}, {Claret}, {Clarke}, {Claudi}, {Clermont}, {Coelho}, {Coelho}, {Cogato}, {Colom{\'e}}, {Condamin}, {Conseil}, {Corbard}, {Correia}, {Corsaro}, {Cosentino}, {Costes}, {Cottinelli}, {Covone}, {Creevey}, {Crida}, {Csizmadia}, {Cunha}, {Curry}, {da Costa}, {da Silva}, {Dalal}, {Damasso}, {Damiani}, {Damiani}, {Liduina das Chagas}, {Davies}, {Davies}, {Davies}, {Davison}, {de Almeida}, {de Angeli}, {Cabral de Barros}, {de Castro Le{\~a}o}, {Brito de Freitas}, {de Freitas}, {De Martino}, {Renan de Medeiros}, {de Paula}, {de Plaa}, {De Ridder}, {Deal}, {Decin}, {Deeg}, {Degl'Innocenti}, {Deheuvels}, {del Burgo}, {Del Sordo}, {Delgado-Mena}, {Demangeon}, {Denk}, {Derekas}, {Desidera}, {Dexet}, {Di Criscienzo}, {Di Giorgio}, {Di Mauro}, {Diaz Rial}, {D{\'\i}az-Garc{\'\i}a}, {Dima}, {Dinuzzi}, {Dionatos}, {Distefano}, {do Nascimento}, {Domingo}, {D'Orazi}, {Dorn}, {Doyle}, {Duarte}, {Ducellier}, {Dumaye}, {Dumusque}, {Dupret}, {Eggenberger}, {Ehrenreich}, {Eigm{\"u}ller}, {Eising}, {Emilio}, {Eriksson}, {Ermocida}, {Isidoro Escate Giribaldi}, {Eschen}, {Estrela}, {Evans}, {Fabbian}, {Fabrizio}, {Faria}, {Farina}, {Farinato}, {Feliz}, {Feltzing}, {Fenouillet}, {Ferrari}, {Ferraz-Mello}, {Fialho}, {Fienga}, {Figueira}, {Fiori}, {Flaccomio}, {Focardi}, {Foley}, {Fontignie}, {Ford}, {Fornazier}, {Forveille}, {Fossati}, {de Marca Franca}, {da Silva}, {Frasca}, {Fridlund}, {Furlan}, {Gabler}, {Gaido}, {Gallagher}, {Galli}, {Garcia}, {Garc{\'\i}a Hern{\'a}ndez}, {Garcia Munoz}, {Garc{\'\i}a-V{\'a}zquez}, {Garrido Haba}, {Gaulme}, {Gauthier}, {Gehan}, {Gent}, {Georgieva}, {Ghigo}, {Giana}, {Gill}, {Girardi}, {Giuliatti Winter}, {Giusi}, {Gomes da Silva}, {G{\'o}mez Zazo}, {Gomez-Lopez}, {Isai Gonz{\'a}lez Hern{\'a}ndez}, {Gonzalez Murillo}, {Gorius}, {Gouel}, {Goulty}, {Granata}, {Grenfell}, {Grie{\ss}bach}, {Grolleau}, {Grouffal}, {Grziwa}, {Guarcello}, {Gueguen}, {Guenther}, {Guilhem}, {Guillerot}, {Guiot}, {Guterman}, {Guti{\'e}rrez}, {Guti{\'e}rrez-Canales}, {Hagelberg}, {Haldemann}, {Hall}, {Handberg}, {Harrison}, {Harrison}, {Hasiba}, {Haswell}, {Hatalova}, {Hatzes}, {Haywood}, {H{\'e}brard}, {Heckes}, {Heiter}, {Hekker}, {Heller}, {Helling}, {Helminiak}, {Hemsley}, {Heng}, {Hermans}, {Hermes}, {Hidalgo Torres}, {Hinkel}, {Hobbs}, {Hodgkin}, {Hofmann}, {Hojjatpanah}, {Houdek}, {Huber}, {Huesler}, {Hui-Bon-Hoa}, {Huygen}, {Huynh}, {Iro}, {Irwin}, {Irwin}, {Izidoro}, {Jacquinod}, {Emborg Jannsen}, {Janson}, {Jeszenszky}, {Jiang}, {Jos{\'e} Jimenez Mancebo}, {Jofre}, {Johansen}, {Johnston}, {Jones}, {Kallinger}, {K{\'a}lm{\'a}n}, {Kanitz}, {Karjalainen}, {Karjalainen}, {Karoff}, {Kawaler}, {Kawata}, {Keereman}, {Keiderling}, {Kennedy}, {Kenworthy}, {Kerschbaum}, {Kidger}, {Kiefer}, {Kintziger}, {Kislyakova}, {Kiss}, {Klagyivik}, {Klahr}, {Klevas}, {Kochukhov}, {K{\"o}hler}, {Kolb}, {Koncz}, {Korth}, {Kostogryz}, {Kov{\'a}cs}, {Kov{\'a}cs}, {Kozhura}, {Krivova}, {Ku{\v{c}}inskas}, {Kuhlemann}, {Kupka}, {Laauwen}, {Labiano}, {Lagarde}, {Laget}, {Laky}, {Lam}, {Lambrechts}, {Lammer}, {Lanza}, {Lanzafame}, {Lares Martiz}, {Laskar}, {Latter}, {Lavanant}, {Lawrenson}, {Lazzoni}, {Lebre}, {Lebreton}, {Lecavelier des Etangs}, {Leinhardt}, {Leleu}, {Lendl}, {Leto}, {Levillain}, {Libert}, {Lichtenberg}, {Ligi}, {Lignieres}, {Lillo-Box}, {Linsky}, {Scige Liu}, {Loidolt}, {Longval}, {Lopes}, {Lorenzani}, {Ludwig}, {Lund}, {Sloth Lundkvist}, {Luri}, {Maceroni}, {Madden}, {Madhusudhan}, {Maggio}, {Magliano}, {Magrin}, {Mahy}, {Maibaum}, {Malac-Allain}, {Malapert}, {Malavolta}, {Maldonado}, {Mamonova}, {Manchon}, {Mann}, {Mantovan}, {Marafatto}, {Marconi}, {Mardling}, {Marigo}, {Marinoni}, {Marques}, {Marques}, {Marrese}, {Marshall}, {Mart{\'\i}nez Perales}, {Mary}, {Marzari}, {Masana}, {Mascher}, {Mathis}, {Mathur}, {Mattiuci Figueiredo}, {Maxted}, {Mazeh}, {Mazevet}, {Mazzei}, {McCormac}, {McMillan}, {Menou}, {Merle}, {Meru}, {Mesa}, {Messina}, {M{\'e}sz{\'a}ros}, {Meunier}, {Meunier}, {Micela}, {Michaelis}, {Michel}, {Michielsen}, {Michtchenko}, {Miglio}, {Miguel}, {Milligan}, {Mirouh}, {Mitchell}, {Moedas}, {Molendini}, {Moln{\'a}r}, {Mombarg}, {Montalban}, {Montalto}, {Monteiro}, {Morales}, {Morales-Calderon}, {Morbidelli}, {Mordasini}, {Moreau}, {Morel}, {Morello}, {Morin}, {Mortier}, {Mosser}, {Mourard}, {Mousis}, {Moutou}, {Mowlavi}, {Moya}, {Muehlmann}, {Muirhead}, {Munari}, {Musella}, {Mustill}, {Nardetto}, {Nardiello}, {Narita}, {Nascimbeni}, {Nash}, {Neiner}, {Nelson}, {Nettelmann}, {Nicolini}, {Nielsen}, {Niemi}, {Noack}, {Noels-Grotsch}, {Noll}, {Norazman}, {Norton}, {Nsamba}, {Ofir}, {Ogilvie}, {Olander}, {Olivetto}, {Olofsson}, {Ong}, {Ortolani}, {Oshagh}, {Ottacher}, {Ottensamer}, {Ouazzani}, {Paardekooper}, {Pace}, {Pajas}, {Palacios}, {Palandri}, {Palle}, {Paproth}, {Parro}, {Parviainen}, {Granado}, {Passegger}, {Pastor-Morales}, {P{\"a}tzold}, {Gade Pedersen}, {Pena Hidalgo}, {Pepe}, {Pereira}, {Persson}, {Pertenais}, {Peter}, {Petit}, {Petit}, {Pezzuto}, {Pichierri}, {Pietrinferni}, {Pinheiro}, {Pinsonneault}, {Plachy}, {Plasson}, {Plez}, {Poppenhaeger}, {Poretti}, {Portaluri}, {Portell}, {Frederico Porto de Mello}, {Poyatos}, {Pozuelos}, {Prada Moroni}, {Pricopi}, {Prisinzano}, {Quade}, {Quirrenbach160}, {Rabanal Reina6}, {Rabello Soares}, {Raimondo}, {Rainer}, {Ram{\'o}n Rod{\'o}n}, {Ram{\'o}n-Ballesta}, {Ramos Zapata}, {R{\"a}tz}, {Rauterberg}, {Redman}, {Redmer}, {Reese}, {Regibo}, {Reiners}, {Reinhold}, {Renie}, {Ribas}, {Ribeiro}, {Pereira Ricciardi}, {Rice}, {Richard}, {Riello}, {Rieutord}, {Ripepi}, {Rixon}, {Rockstein}, {Rodr{\'\i}guez}, {Rodr{\'\i}guez D{\'\i}az}, {Rodriguez Garcia}, {Rodriguez-Gomez}, {Roehlly}, {Roig}, {Rojas-Ayala}, {Rolf}, {Lysgaard R{\o}rsted}, {Rosado}, {Rosotti}, {Roth}, {Roth}, {Rousseau}, {Roxburgh}, {Roy}, {Royer}, {Ruane}, {Rufini Mastropasqua}, {Ruiz de Galarreta}, {Russi}, {Saar}, {Saillenfest}, {Salaris}, {Salmon}, {Saltas}, {Samadi}, {Samadi}, {Samra}, {Sanches da Silva}, {Andr{\'e}s S{\'a}nchez Carrasco}, {Santerne}, {Santoli}, {Santos}, {Sanz Mesa}, {Sarro}, {Scandariato}, {Sch{\"a}fer}, {Schlafly}, {Schmider}, {Schneider}, {Schou}, {Schunker}, {J{\"o}rg Schwarzkopf}, {Serenelli}, {Seynaeve}, {Shan}, {Shapiro}, {Shipman}, {Sicilia}, {Sierra Sanmartin}, {Sigot}, {Silliman}, {Silvotti}, {Simon}, {Simoyama Napoli}, {Skarka}, {Smalley}, {Smiljanic}, {Smit}, {Smith}, {Smith}, {Snellen}, {S{\'o}dor}, {Sohl}, {Solanki}, {Sortino}, {Sousa}, {Southworth}, {Souto}, {Sozzetti}, {Stamatellos}, {Stassun}, {Steller}, {Stello}, {Stelzer}, {Stiebeler}, {Stokholm}, {Storelvmo}, {Strassmeier}, {Str{\o}m}, {Strugarek}, {Sulis}, {{\v{S}}vanda}, {Szabados}, {Szab{\'o}}, {Szab{\'o}}, {Szuszkiewicz}, {Talens}, {Teti}, {Theisen}, {Th{\'e}venin}, {Thoul}, {Tiphene}, {Titz-Weider}, {Tkachenko}, {Tomecki}, {Tonfat}, {Tosi}, {Trampedach}, {Traven}, {Triaud}, {Tr{\o}nnes}, {Tsantaki}, {Tschentscher}, {Turin}, {Tvaruzka}, {Ulmer}, {Ulmer-Moll}, {Ulusoy}, {Umbriaco}, {Valencia}, {Valentini}, {Valio}, {Valverde Guijarro}, {Van Eylen}, {Van Grootel}, {van Kempen}, {Van Reeth}, {Van Zelst}, {Vandenbussche}, {Vasiliou}, {Vasilyev}, {Vaz de Mascarenhas}, {Vazan}, {Vela Nunez}, {Nunes Velloso}, {Ventura}, {Ventura}, {Venturini}, {Trallero}, {Veras}, {Verdugo}, {Verma}, {Vibert}, {Vicanek Martinez}, {Vida}, {Vigan}, {Villacorta}, {Villaver}, {Villaverde Aparicio}, {Viotto}, {Vorobyov}, {Vorontsov}, {Wagner}, {Walloschek}, {Walton}, {Walton}, {Wang}, {Waters}, {Watson}, {Wedemeyer}, {Weeks}, {Weingril}, {Weiss}, {Wendler}, {West}, {Westerdorff}, {Westphal}, {Wheatley}, {White}, {Whittaker}, {Wickhusen}, {Wilson}, {Windsor}, {Winter}, {Lykke Winther}, {Winton}, {Witteck}, {Witzke}, {Woitke}, {Wolter}, {Wuchterl}, {Wyatt}, {Yang}, {Yu}, {Zanmar Sanchez}, {Rosa Zapatero Osorio}, {Zechmeister},
  {Zhou}, {Ziemke}, \& {Zwintz}}]{2024arXiv240605447R}
{Rauer}, H., {Aerts}, C., {Cabrera}, J., {et~al.} 2024, arXiv e-prints, arXiv:2406.05447, \dodoi{10.48550/arXiv.2406.05447}

\bibitem[{{Reinhold} {et~al.}(2023){Reinhold}, {Shapiro}, {Solanki}, \& {Basri}}]{2023A&A...678A..24R}
{Reinhold}, T., {Shapiro}, A.~I., {Solanki}, S.~K., \& {Basri}, G. 2023, \aap, 678, A24, \dodoi{10.1051/0004-6361/202346789}

\bibitem[{{Richardson} {et~al.}(2011){Richardson}, {Morrison}, {Kryukova}, \& {Adelman}}]{2011AJ....141...17R}
{Richardson}, N.~D., {Morrison}, N.~D., {Kryukova}, E.~E., \& {Adelman}, S.~J. 2011, \aj, 141, 17, \dodoi{10.1088/0004-6256/141/1/17}

\bibitem[{{Ricker} {et~al.}(2015){Ricker}, {Winn}, {Vanderspek}, {Latham}, {Bakos}, {Bean}, {Berta-Thompson}, {Brown}, {Buchhave}, {Butler}, {Butler}, {Chaplin}, {Charbonneau}, {Christensen-Dalsgaard}, {Clampin}, {Deming}, {Doty}, {De Lee}, {Dressing}, {Dunham}, {Endl}, {Fressin}, {Ge}, {Henning}, {Holman}, {Howard}, {Ida}, {Jenkins}, {Jernigan}, {Johnson}, {Kaltenegger}, {Kawai}, {Kjeldsen}, {Laughlin}, {Levine}, {Lin}, {Lissauer}, {MacQueen}, {Marcy}, {McCullough}, {Morton}, {Narita}, {Paegert}, {Palle}, {Pepe}, {Pepper}, {Quirrenbach}, {Rinehart}, {Sasselov}, {Sato}, {Seager}, {Sozzetti}, {Stassun}, {Sullivan}, {Szentgyorgyi}, {Torres}, {Udry}, \& {Villasenor}}]{2015JATIS...1a4003R}
{Ricker}, G.~R., {Winn}, J.~N., {Vanderspek}, R., {et~al.} 2015, Journal of Astronomical Telescopes, Instruments, and Systems, 1, 014003, \dodoi{10.1117/1.JATIS.1.1.014003}

\bibitem[{{Rogers}(2015)}]{2015ApJ...815L..30R}
{Rogers}, T.~M. 2015, \apjl, 815, L30, \dodoi{10.1088/2041-8205/815/2/L30}

\bibitem[{{Rogers} {et~al.}(2013){Rogers}, {Lin}, {McElwaine}, \& {Lau}}]{2013ApJ...772...21R}
{Rogers}, T.~M., {Lin}, D.~N.~C., {McElwaine}, J.~N., \& {Lau}, H.~H.~B. 2013, \apj, 772, 21, \dodoi{10.1088/0004-637X/772/1/21}

\bibitem[{{Rogers} \& {McElwaine}(2017)}]{2017ApJ...848L...1R}
{Rogers}, T.~M., \& {McElwaine}, J.~N. 2017, \apjl, 848, L1, \dodoi{10.3847/2041-8213/aa8d13}

\bibitem[{{Salpeter}(1955)}]{1955ApJ...121..161S}
{Salpeter}, E.~E. 1955, \apj, 121, 161, \dodoi{10.1086/145971}

\bibitem[{{Samadi} {et~al.}(2010){Samadi}, {Belkacem}, {Goupil}, {Dupret}, {Brun}, \& {Noels}}]{2010Ap&SS.328..253S}
{Samadi}, R., {Belkacem}, K., {Goupil}, M.~J., {et~al.} 2010, \apss, 328, 253, \dodoi{10.1007/s10509-009-0215-3}

\bibitem[{{Sana} {et~al.}(2012){Sana}, {de Mink}, {de Koter}, {Langer}, {Evans}, {Gieles}, {Gosset}, {Izzard}, {Le Bouquin}, \& {Schneider}}]{2012Sci...337..444S}
{Sana}, H., {de Mink}, S.~E., {de Koter}, A., {et~al.} 2012, Science, 337, 444, \dodoi{10.1126/science.1223344}

\bibitem[{{Sana} {et~al.}(2014){Sana}, {Le Bouquin}, {Lacour}, {Berger}, {Duvert}, {Gauchet}, {Norris}, {Olofsson}, {Pickel}, {Zins}, {Absil}, {de Koter}, {Kratter}, {Schnurr}, \& {Zinnecker}}]{2014ApJS..215...15S}
{Sana}, H., {Le Bouquin}, J.~B., {Lacour}, S., {et~al.} 2014, \apjs, 215, 15, \dodoi{10.1088/0067-0049/215/1/15}

\bibitem[{{Schneider} {et~al.}(2014){Schneider}, {Langer}, {de Koter}, {Brott}, {Izzard}, \& {Lau}}]{2014A&A...570A..66S}
{Schneider}, F.~R.~N., {Langer}, N., {de Koter}, A., {et~al.} 2014, \aap, 570, A66, \dodoi{10.1051/0004-6361/201424286}

\bibitem[{{Schonhut-Stasik} \& {Stassun}(2023)}]{2023RNAAS...7...18S}
{Schonhut-Stasik}, J., \& {Stassun}, K. 2023, Research Notes of the American Astronomical Society, 7, 18, \dodoi{10.3847/2515-5172/acb936}

\bibitem[{{Schultz} {et~al.}(2022){Schultz}, {Bildsten}, \& {Jiang}}]{2022ApJ...924L..11S}
{Schultz}, W.~C., {Bildsten}, L., \& {Jiang}, Y.-F. 2022, \apjl, 924, L11, \dodoi{10.3847/2041-8213/ac441f}

\bibitem[{{Schultz} {et~al.}(2023){Schultz}, {Bildsten}, \& {Jiang}}]{2023ApJ...951L..42S}
---. 2023, \apjl, 951, L42, \dodoi{10.3847/2041-8213/acdf50}

\bibitem[{{Shen} {et~al.}(2023){Shen}, {Li}, {Abdusamatjan}, {Fu}, {Zhu}, {Yu}, {Zhang}, {L{\"u}}, {Zhai}, \& {Liu}}]{2023ApJ...955..123S}
{Shen}, D.-X., {Li}, G., {Abdusamatjan}, I., {et~al.} 2023, \apj, 955, 123, \dodoi{10.3847/1538-4357/acf197}

\bibitem[{{Shiode} {et~al.}(2013){Shiode}, {Quataert}, {Cantiello}, \& {Bildsten}}]{2013MNRAS.430.1736S}
{Shiode}, J.~H., {Quataert}, E., {Cantiello}, M., \& {Bildsten}, L. 2013, \mnras, 430, 1736, \dodoi{10.1093/mnras/sts719}

\bibitem[{{Sim{\'o}n-D{\'\i}az} {et~al.}(2011{\natexlab{a}}){Sim{\'o}n-D{\'\i}az}, {Castro}, {Garcia}, {Herrero}, \& {Markova}}]{2011BSRSL..80..514S}
{Sim{\'o}n-D{\'\i}az}, S., {Castro}, N., {Garcia}, M., {Herrero}, A., \& {Markova}, N. 2011{\natexlab{a}}, Bulletin de la Societe Royale des Sciences de Liege, 80, 514, \dodoi{10.48550/arXiv.1009.5824}

\bibitem[{{Sim{\'o}n-D{\'\i}az} {et~al.}(2011{\natexlab{b}}){Sim{\'o}n-D{\'\i}az}, {Castro}, {Herrero}, {Puls}, {Garcia}, \& {Sab{\'\i}n-Sanjuli{\'a}n}}]{2011JPhCS.328a2021S}
{Sim{\'o}n-D{\'\i}az}, S., {Castro}, N., {Herrero}, A., {et~al.} 2011{\natexlab{b}}, in Journal of Physics Conference Series, Vol. 328, Journal of Physics Conference Series (IOP), 012021, \dodoi{10.1088/1742-6596/328/1/012021}

\bibitem[{{Sim{\'o}n-D{\'\i}az} {et~al.}(2017){Sim{\'o}n-D{\'\i}az}, {Godart}, {Castro}, {Herrero}, {Aerts}, {Puls}, {Telting}, \& {Grassitelli}}]{2017A&A...597A..22S}
{Sim{\'o}n-D{\'\i}az}, S., {Godart}, M., {Castro}, N., {et~al.} 2017, \aap, 597, A22, \dodoi{10.1051/0004-6361/201628541}

\bibitem[{{Sim{\'o}n-D{\'\i}az} \& {Herrero}(2014)}]{2014A&A...562A.135S}
{Sim{\'o}n-D{\'\i}az}, S., \& {Herrero}, A. 2014, \aap, 562, A135, \dodoi{10.1051/0004-6361/201322758}

\bibitem[{{Sim{\'o}n-D{\'\i}az} {et~al.}(2014){Sim{\'o}n-D{\'\i}az}, {Herrero}, {Sab{\'\i}n-Sanjuli{\'a}n}, {Najarro}, {Garcia}, {Puls}, {Castro}, \& {Evans}}]{2014A&A...570L...6S}
{Sim{\'o}n-D{\'\i}az}, S., {Herrero}, A., {Sab{\'\i}n-Sanjuli{\'a}n}, C., {et~al.} 2014, \aap, 570, L6, \dodoi{10.1051/0004-6361/201424742}

\bibitem[{{Sota} {et~al.}(2014){Sota}, {Ma{\'\i}z Apell{\'a}niz}, {Morrell}, {Barb{\'a}}, {Walborn}, {Gamen}, {Arias}, \& {Alfaro}}]{2014ApJS..211...10S}
{Sota}, A., {Ma{\'\i}z Apell{\'a}niz}, J., {Morrell}, N.~I., {et~al.} 2014, \apjs, 211, 10, \dodoi{10.1088/0067-0049/211/1/10}

\bibitem[{{Stanishev} {et~al.}(2002){Stanishev}, {Kraicheva}, {Boffin}, \& {Genkov}}]{2002A&A...394..625S}
{Stanishev}, V., {Kraicheva}, Z., {Boffin}, H.~M.~J., \& {Genkov}, V. 2002, \aap, 394, 625, \dodoi{10.1051/0004-6361:20021163}

\bibitem[{{Stassun} {et~al.}(2019){Stassun}, {Oelkers}, {Paegert}, {Torres}, {Pepper}, {De Lee}, {Collins}, {Latham}, {Muirhead}, {Chittidi}, {Rojas-Ayala}, {Fleming}, {Rose}, {Tenenbaum}, {Ting}, {Kane}, {Barclay}, {Bean}, {Brassuer}, {Charbonneau}, {Ge}, {Lissauer}, {Mann}, {McLean}, {Mullally}, {Narita}, {Plavchan}, {Ricker}, {Sasselov}, {Seager}, {Sharma}, {Shiao}, {Sozzetti}, {Stello}, {Vanderspek}, {Wallace}, \& {Winn}}]{2019AJ....158..138S}
{Stassun}, K.~G., {Oelkers}, R.~J., {Paegert}, M., {et~al.} 2019, \aj, 158, 138, \dodoi{10.3847/1538-3881/ab3467}

\bibitem[{{Stevenson}(1979)}]{1979GApFD..12..139S}
{Stevenson}, D.~J. 1979, Geophysical and Astrophysical Fluid Dynamics, 12, 139, \dodoi{10.1080/03091927908242681}

\bibitem[{{Szewczuk} {et~al.}(2022){Szewczuk}, {Walczak}, {Daszy{\'n}ska-Daszkiewicz}, \& {Mo{\'z}dzierski}}]{2022MNRAS.511.1529S}
{Szewczuk}, W., {Walczak}, P., {Daszy{\'n}ska-Daszkiewicz}, J., \& {Mo{\'z}dzierski}, D. 2022, \mnras, 511, 1529, \dodoi{10.1093/mnras/stac168}

\bibitem[{{Talon} \& {Charbonnel}(2008)}]{2008A&A...482..597T}
{Talon}, S., \& {Charbonnel}, C. 2008, \aap, 482, 597, \dodoi{10.1051/0004-6361:20078620}

\bibitem[{{Tarczay-Neh{\'e}z} {et~al.}(2023){Tarczay-Neh{\'e}z}, {Moln{\'a}r}, {B{\'o}di}, \& {Szab{\'o}}}]{2023A&A...676A..28T}
{Tarczay-Neh{\'e}z}, D., {Moln{\'a}r}, L., {B{\'o}di}, A., \& {Szab{\'o}}, R. 2023, \aap, 676, A28, \dodoi{10.1051/0004-6361/202346094}

\bibitem[{{Thompson} {et~al.}(2024){Thompson}, {Herwig}, {Woodward}, {Mao}, {Denissenkov}, {Bowman}, \& {Blouin}}]{2024MNRAS.531.1316T}
{Thompson}, W., {Herwig}, F., {Woodward}, P.~R., {et~al.} 2024, \mnras, 531, 1316, \dodoi{10.1093/mnras/stae1162}

\bibitem[{{Torres} {et~al.}(2011){Torres}, {Fressin}, {Batalha}, {Borucki}, {Brown}, {Bryson}, {Buchhave}, {Charbonneau}, {Ciardi}, {Dunham}, {Fabrycky}, {Ford}, {Gautier}, {Gilliland}, {Holman}, {Howell}, {Isaacson}, {Jenkins}, {Koch}, {Latham}, {Lissauer}, {Marcy}, {Monet}, {Prsa}, {Quinn}, {Ragozzine}, {Rowe}, {Sasselov}, {Steffen}, \& {Welsh}}]{2011ApJ...727...24T}
{Torres}, G., {Fressin}, F., {Batalha}, N.~M., {et~al.} 2011, \apj, 727, 24, \dodoi{10.1088/0004-637X/727/1/24}

\bibitem[{{Triana} {et~al.}(2015){Triana}, {Moravveji}, {P{\'a}pics}, {Aerts}, {Kawaler}, \& {Christensen-Dalsgaard}}]{2015ApJ...810...16T}
{Triana}, S.~A., {Moravveji}, E., {P{\'a}pics}, P.~I., {et~al.} 2015, \apj, 810, 16, \dodoi{10.1088/0004-637X/810/1/16}

\bibitem[{{Vanon} {et~al.}(2023){Vanon}, {Edelmann}, {Ratnasingam}, {Varghese}, \& {Rogers}}]{2023ApJ...954..171V}
{Vanon}, R., {Edelmann}, P.~V.~F., {Ratnasingam}, R.~P., {Varghese}, A., \& {Rogers}, T.~M. 2023, \apj, 954, 171, \dodoi{10.3847/1538-4357/ace9db}

\bibitem[{{Varghese} {et~al.}(2024){Varghese}, {Ratnasingam}, {Vanon}, {Edelmann}, {Mathis}, \& {Rogers}}]{2024ApJ...970..104V}
{Varghese}, A., {Ratnasingam}, R.~P., {Vanon}, R., {et~al.} 2024, \apj, 970, 104, \dodoi{10.3847/1538-4357/ad54b5}

\bibitem[{{Varghese} {et~al.}(2023){Varghese}, {Ratnasingam}, {Vanon}, {Edelmann}, \& {Rogers}}]{2023ApJ...942...53V}
{Varghese}, A., {Ratnasingam}, R.~P., {Vanon}, R., {Edelmann}, P.~V.~F., \& {Rogers}, T.~M. 2023, \apj, 942, 53, \dodoi{10.3847/1538-4357/aca092}

\bibitem[{{Virtanen} {et~al.}(2020){Virtanen}, {Gommers}, {Oliphant}, {Haberland}, {Reddy}, {Cournapeau}, {Burovski}, {Peterson}, {Weckesser}, {Bright}, {van der Walt}, {Brett}, {Wilson}, {Millman}, {Mayorov}, {Nelson}, {Jones}, {Kern}, {Larson}, {Carey}, {Polat}, {Feng}, {Moore}, {VanderPlas}, {Laxalde}, {Perktold}, {Cimrman}, {Henriksen}, {Quintero}, {Harris}, {Archibald}, {Ribeiro}, {Pedregosa}, {van Mulbregt}, \& {SciPy 1. 0 Contributors}}]{2020NatMe..17..261V}
{Virtanen}, P., {Gommers}, R., {Oliphant}, T.~E., {et~al.} 2020, Nature Methods, 17, 261, \dodoi{10.1038/s41592-019-0686-2}

\bibitem[{{Ziegler} {et~al.}(2018){Ziegler}, {Law}, {Baranec}, {Howard}, {Morton}, {Riddle}, {Duev}, {Salama}, {Jensen-Clem}, \& {Kulkarni}}]{2018AJ....156...83Z}
{Ziegler}, C., {Law}, N.~M., {Baranec}, C., {et~al.} 2018, \aj, 156, 83, \dodoi{10.3847/1538-3881/aace59}

\bibitem[{{Zorec} {et~al.}(2023){Zorec}, {Hubert}, {Martayan}, \& {Fr{\'e}mat}}]{2023A&A...676A..81Z}
{Zorec}, J., {Hubert}, A.~M., {Martayan}, C., \& {Fr{\'e}mat}, Y. 2023, \aap, 676, A81, \dodoi{10.1051/0004-6361/202346018}

\end{thebibliography}
\bibliographystyle{aasjournal}


\end{document}